\documentclass[twocolumn]{aastex701}
\usepackage{chemformula}
\usepackage{multirow}

\begin{document}

\title{The SOFIA Massive (SOMA) Radio Survey. III. Radio Emission from Intermediate-Mass Protostars}

\author[orcid=0000-0001-8169-1437, gname='Francisco', sname='Sequeira-Murillo']{Francisco Sequeira-Murillo}
\affiliation{Department of Astronomy, University of Wisconsin-Madison, 475 N. Charter St., Madison, WI 53703, USA}
\email[show]{sequeiramuri@wisc.edu}

\author[orcid=0000-0001-8596-1756, gname='Viviana', sname='Rosero']{Viviana Rosero}
\affiliation{Cahill Center for Astronomy and Astrophysics, MC 249-17, California Institute of Technology, Pasadena, CA 91125, USA}
\affiliation{Space Science Institute, 4750 Walnut Street, Suite 205, Boulder, CO 80301, USA}
\email[show]{vrosero@caltech.edu}   

\author[orcid=0000-0003-1111-8066, gname='Joshua', sname='Marvil']{Joshua Marvil}
\affiliation{National Radio Astronomy Observatory, 1003 Lópezville Rd., Socorro, NM 87801, USA}
\email{jmarvil@nrao.edu}

\author[orcid=0000-0003-4040-4934, gname='Ruben', sname='Fedriani']{Ruben Fedriani}
\affiliation{Instituto de Astrofísica de Andalucía, CSIC, Glorieta de la Astronomía s/n, 18008 Granada, Spain}
\affiliation{Department of Space, Earth \& Environment, Chalmers University of Technology, 412 93 Gothenburg, Sweden} 
\email{fedriani@iaa.es}  

\author[orcid=0000-0001-7511-0034, gname='Yichen', sname='Zhang']{Yichen Zhang}
\affiliation{State Key Laboratory of Dark Matter Physics, School of Physics and Astronomy, Shanghai Jiao Tong University, Shanghai 200240, People’s Republic of China}
\affiliation{Department of Astronomy, University of Virginia, Charlottesville, Virginia 22904, USA}
\email{yichen.zhang@sjtu.edu.cn}

\author[orcid=0000-0003-1602-6849, gname='Prasanta', sname='Gorai']{Prasanta Gorai}
\affiliation{Rosseland Centre for Solar Physics, University of Oslo, PO Box 1029 Blindern, 0315, Oslo, Norway}
\affiliation{Institute of Theoretical Astrophysics, University of Oslo, PO Box 1029 Blindern, 0315, Oslo, Norway}
\email{prasanta.gorai@astro.uio.no}

\author[orcid=0000-0001-7378-4430, gname='James M.', sname='De Buizer']{James M. De Buizer}
\affiliation{Carl Sagan Center for Research, SETI Institute, Mountain View, CA, USA}
\email{jdebuizer@seti.org}

\author[orcid=0000-0003-3315-5626, gname='Maria T.', sname='Beltrán']{Maria T. Beltrán}
\affiliation{INAF-Osservatorio Astrofisico di Arcetri, Largo E. Fermi 5, I-50125 Firenze, Italy}
\email{maria.beltran@inaf.it}

\author[orcid=0000-0002-3389-9142, gname='Jonathan C.', sname='Tan']{Jonathan C. Tan}
\affiliation{Department of Physics \& Astronomy, Chalmers University of Technology, 412 93 Gothenburg, Sweden}
\affiliation{Department of Astronomy, University of Virginia, Charlottesville, Virginia 22904, USA} 
\email{jonathan.tan@chalmers.se}  

\defcitealias{Bruizer_2017}{SOMA I}
\defcitealias{Liu_2019}{SOMA II}
\defcitealias{Liu_2020}{SOMA III}
\defcitealias{Fedriani_2023}{SOMA IV}
\defcitealias{Telkamp_2025}{SOMA V}
\defcitealias{Rosero_2019}{SOMA Radio I}
\defcitealias{Sequeira-Murillo_2025}{SOMA Radio II}
\defcitealias{Zhang_2018}{ZT18}
\defcitealias{Tanaka_2016}{TTZ16}

\begin{abstract}
We present results from Very Large Array (VLA) radio continuum observations of twelve intermediate-mass (IM) protostars, as part of the \textit{SOFIA} Massive Star Formation Survey. Using these observations, we studied their morphology, multiplicity and radio spectral energy distributions (SEDs). Across our target regions, we resolve multiple compact sources and report eight new detections, four of which are entirely new and four that have counterparts at other wavelengths, but are detected here for the first time at radio frequencies. 
Based on radio morphologies and spectral indices, we assess the nature of the detected sources, highlighting seven that display jet-like structures and spectral indices consistent with ionized jets. Combining our results with the SOMA Radio I and II results, we expand the overall sample to 29 protostars, covering a bolometric luminosity range from $L_{\rm bol}\sim 10^2$ to $10^6\:L_\odot$. These sources help define a potential evolutionary sequence in the radio versus bolometric luminosity diagram. IM protostars have radio luminosities that are lower than expected from a simple power law extrapolation from low-mass protostars. However, this result is consistent with theoretical expectations from protostellar evolution models, which show low levels of photoionization and reduced shock ionization emission due to expanded stellar radii during this phase.
Overall our expanded SOMA Radio sample provides new constraints on theoretical models of massive protostellar evolution, especially the connection to ionized gas structures.
\end{abstract}


\section{Introduction} \label{sec:intro}

Intermediate-mass (IM) protostars are classified as those with masses in the range $2<m_*/M_\odot<8$, i.e., linking the low- and high-mass regimes. IM stars produce more UV photons and are thought to form in more densely clustered environments, compared to the low-mass protostars \citep[e.g.,][]{2015Ap&SS.355..283B}. These IM star-forming regions are generally closer and less extincted than massive star-forming regions, making them easier to study.

Both massive and IM protostars form in dusty environments where they are typically deeply embedded within their natal clouds. High-sensitivity centimeter-continuum observations provide valuable insights into the earliest and most obscured stages of intermediate-mass and high-mass star formation. Moreover, free-free radio emission at centimeter wavelengths has long been a well-established method for estimating important physical properties of YSO jets, as high-resolution observations can help characterize proper motions, variability, and outflow rotation. In addition, the radio luminosity of these jets is well correlated with both the bolometric luminosity and the outflow momentum rate, making radio continuum observations a powerful tracer of protostellar activity \citep{2018A&ARv..26....3A}.

In particular, radio continuum emission plays a crucial role in constraining the ionizing luminosity of protostars and probing their surrounding environments. \citet{2016ApJS..227...25R, 2019ApJ...880...99R} showed that many centimeter continuum sources exhibit morphologies and properties characteristic of ionized jets, consistent with the free–free emission models of massive protostars developed by \cite{Tanaka_2016}. These results highlight the importance of radio continuum observations for testing theoretical models and refining our understanding of the evolutionary sequence of star formation across the mass spectrum.

The SOFIA Massive (SOMA) Star Formation Survey (PI: J. Tan) aims to characterize a sample of $\geq$ 50 high and intermediate-mass protostars over a range of evolutionary stages, environments, and core masses, using SOFIA-FORCAST $\sim$ 10–40 $\mu$m data. In Paper I of the survey \citep{Bruizer_2017} (hereafter \citetalias{Bruizer_2017}), the first eight sources were presented, consisting primarily of massive protostars. In Paper II \citep{Liu_2019} (hereafter \citetalias{Liu_2019}), seven high-luminosity sources were presented, which are among the most massive protostars in the survey. In Paper III \citep{Liu_2020} (hereafter \citetalias{Liu_2020}), 14 intermediate-mass sources were presented. In Paper IV \citep{Fedriani_2023} (hereafter \citetalias{Fedriani_2023}), 11 isolated sources, based on the 37 $\mu$m imaging, were presented. In Paper V \citep{Telkamp_2025} (hereafter \citetalias{Telkamp_2025}), seven regions of relatively clustered massive star formation were studied.

Fitting infrared-only radiative transfer (RT) models of massive protostars \citep[e.g.,][hereafter \citetalias{Zhang_2018}]{Zhang_2018} typically leads to significant degeneracies in key parameters. However, incorporating centimeter continuum data, along with models for free-free emission such as the \citet[][hereafter \citetalias{Tanaka_2016}]{Tanaka_2016} models, is expected to help break these degeneracies and thus provide tighter constraints on massive protostellar properties.

This study is the third in a series of centimeter–continuum follow-ups to the SOMA papers. 
\citet[][hereafter \citetalias{Rosero_2019}]{Rosero_2019} presented and analyzed the first eight sources from \citetalias{Bruizer_2017}. 
Subsequently, \citet[][hereafter \citetalias{Sequeira-Murillo_2025}]{Sequeira-Murillo_2025} presented and analyzed nine sources from \citetalias{Liu_2019}, including two IM protostars G305.20+0.21~A and IRAS~16562$-$3959~N from \citetalias{Liu_2020}. 
Here, we present high-sensitivity Karl~G.~Jansky Very Large Array (VLA) observations at 1.3 and 6~cm for seven regions containing a total of 12 sources, all of which were analyzed in \citetalias{Liu_2020}.

Our approach follows the general methodology described in \citetalias{Rosero_2019} and \citetalias{Sequeira-Murillo_2025} to analyze the data and construct the radio spectral energy distributions (SEDs). This includes reducing the centimeter–continuum observations, measuring flux densities from the images, and analyzing the morphology and multiplicity of each source. These steps ensure that all protostars are analyzed in a uniform and systematic manner. In addition, we present an analysis of the source properties across the SOMA sample, which comprises 29 sources studied in \citetalias{Bruizer_2017}, \citetalias{Liu_2019}, and \citetalias{Liu_2020}, as well as in their corresponding centimeter–continuum follow-ups (\citetalias{Rosero_2019}, \citetalias{Sequeira-Murillo_2025}, and this work).

Building on the methods outlined in \citetalias{Sequeira-Murillo_2025}, we incorporate the updated SED fitting methods from \citetalias{Telkamp_2025}, which introduced several improvements to the general SOMA analysis framework. Accordingly, we adopt the aperture sizes, as well as the infrared data and results (bolometric luminosities, fluxes, and models), from \citetalias{Telkamp_2025} in place of those from \citetalias{Liu_2020}.

Our paper is organized as follows: methodology and information regarding the observations are presented in \S\ref{sec:methods}. The observational results for each source are presented in \S\ref{subsec:results}, while the analysis and discussion of the sample are presented in \S\ref{analysis}. The summary and conclusions are presented in \S\ref{summary}.

\section{Methods} \label{sec:methods}

The SOMA Star Formation Survey sample was constructed from SOFIA-FORCAST observations (i.e., from $\sim$10 to 40 $\mu$m). The sample analyzed in this paper is presented in \citetalias{Liu_2020}, except for  sources  G305.20$+$0.21 A and IRAS 16562$-$3959 N that were presented in \citetalias{Sequeira-Murillo_2025}. Thus, a total of twelve protostars in seven target regions will be analyzed: S235, IRAS 22198$+$6336, NGC 2071, Cepheus E, L1206 (A and B), IRAS 22172$+$5549 (MIR1, MIR2, and MIR3), and IRAS 21391$+$5802 (BIMA2, BIMA3, and MIR48). 

The data analyzed in this study are summarized in Table \ref{tab:SOMA_Sources}. The first column presents the region names, while columns 2, 3, and 4 provide the band frequency, R.A., and Decl., respectively. Synthesized beam size and position angle (PA), along with the rms of the resulting images, are detailed in columns 5 and 6. The distance to every region, as adopted by \citetalias{Liu_2020}, as well as the isotropic bolometric luminosities and the bolometric luminosities evaluated by \citetalias{Telkamp_2025}, are shown in columns 7, 8, and 9, respectively. Furthermore, Table \ref{tab:VLA_Calibrators} provides a list of phase calibrators utilized in the observations at 6 and 1.3 cm.

\startlongtable
\begin{deluxetable*}{ccccccccc}
    \tabletypesize{\scriptsize}
    \tablewidth{0pt}
    \tablecaption{SOMA Sources: Radio Continuum Data. \label{tab:SOMA_Sources}}
    \tablehead{\colhead{Region} & \colhead{Frequency Band} & \colhead{R.A} & \colhead{Decl.} & \colhead{Beam Size} & \colhead{rms} & \colhead{D\small\textsuperscript{a}} & \colhead{$L_{\rm bol, iso}$\small\textsuperscript{b}} & \colhead{$L_{\rm bol}$\small\textsuperscript{b}} \\
    \colhead{} & \colhead{(GHz)} & \colhead{(J2000)} & \colhead{(J2000)} & \colhead{($^{\prime\prime}$ x $^{\prime\prime}$, degree)} & \colhead{($\mu$Jy beam$^{-1}$)} & \colhead{(kpc)} & \colhead{($L_{\odot}$)} & \colhead{($L_{\odot}$)}}
    \startdata
    S235 & 4.0 $-$ 8.0 & 05 40 52.40 & +35 41 30.0 & 0.30 $\times$ 0.28, $+$80.8 & 4.9 & 1.8 & $2.2^{+2.3}_{-1.1} \times 10^{3}$ & $2.6^{+6.9}_{-1.9} \times 10^{4}$  \\
         & 18.0 $-$ 26.0 & ... & ... & 0.39 $\times$ 0.24, $+$77.1 & 9.7 & ... & ... & ... \\
    IRAS 22198+6336 & 4.0 $-$ 8.0 & 22 21 26.68& +63 51 38.2 & 0.59 $\times$ 0.28, $-$71.6 & 4.2 & 0.764 & $3.9^{+17.8}_{-3.2} \times 10^{2}$ & $8.0^{+20.2}_{-5.7} \times 10^{2}$ \\
                    & 18.0 $-$ 26.0 & ... & ... & 0.42 $\times$ 0.24, $-$87.6 & 8.9 & ... & ... & ... \\
    NGC 2071 & 4.0 $-$ 8.0 & 05 47 04.74 & +00 21 42.9 & 0.29 $\times$ 0.27, $-$59.6 & 7.6  & 0.43 & $3.6^{+9.1}_{-2.6} \times 10^{2}$ & $6.9^{+12.0}_{-4.4} \times 10^{2}$  \\
             & 18.0 $-$ 26.0 & ... & ... & 0.31 $\times$ 0.25, $+$16.2 & 13.7 & ... & ... & ... \\
    Cepheus E & 4.0 $-$ 8.0 & 23 03 12.8 & +61 42 26.0 & 0.33 $\times$ 0.28, $+$8.05 & 3.5 & 0.73 & $3.3^{+10.1}_{-2.5} \times 10^{2}$ & $7.4^{+14.3}_{-4.9} \times 10^{2}$  \\
              & 18.0 $-$ 26.0 & ... & ... & 0.63 $\times$ 0.30, $-$73.9 & 9.7 & ... & ... & ... \\
    L1206 & 4.0 $-$ 8.0 & 22 28 51.41 & +64 13 41.1 & 0.39 $\times$ 0.37, $+$48.1 & 4.5 & 0.776 & $1.7^{+0.7}_{-0.5} \times 10^{3}$ & $3.7^{+3.3}_{-1.7} \times 10^{3}$ \\
          & 18.0 $-$ 26.0 & ... & ... & 0.55 $\times$ 0.30, $-$85.7 & 8.2 & ... & ... & ... \\
    IRAS 22172+5549 & 4.0 $-$ 8.0 & 22 19 09.48 & +56 05 00.4 & 0.41 $\times$ 0.32, $-$75.6 & 6.6 & 2.4 & $1.5^{+1.9}_{-0.8} \times 10^{3}$ & $6.2^{+26.1}_{-5.0} \times 10^{4}$ \\
                    & 18.0 $-$ 26.0 & ... & ... & 0.46 $\times$ 0.24, $-$78.2 & 8.7 & ... & ... & ... \\
    IRAS 21391+5802 & 4.0 $-$ 8.0 & 21 40 41.90 & +58 16 12.3 & 0.66 $\times$ 0.28, $-$67.1 & 4.0 & 0.75 & $9.9^{+9.8}_{-4.9} \times 10^{1}$ & $5.6^{+7.0}_{-3.1} \times 10^{2}$  \\
                    & 18.0 $-$ 26.0 & ... & ... & 0.41 $\times$ 0.24, $-$84.7 & 9.1 & ... & ... & ... \\
    \enddata
    \tablecomments{ The source position are the field centers for the radio observations and come from \citetalias{Liu_2020}.\\
    Units of R.A. are hours, minutes, and seconds. Units of decl. are degrees, arcminutes, and arcseconds. \\
    \textsuperscript{a} References cited in \citetalias{Liu_2020}, except for NGC 2071, where we use the values from \citet{Tobin2020} with Gaia observations. \\
    \textsuperscript{b} Average and dispersion of the bolometric luminosities of the good models from \citetalias{Telkamp_2025}, these values and other intrinsic properties are reported in Table C1.}
\end{deluxetable*}

\begin{deluxetable}{cccc}
    \tabletypesize{\scriptsize}
    \tablewidth{0pt}
    \tablecaption{VLA Calibrators. \label{tab:VLA_Calibrators}}
    \tablehead{
    \colhead{Calibrator} & \colhead{Astrometry Precision\small\textsuperscript{a}} & \colhead{Source Calibrated} & \colhead{Band}}
    \startdata
    J0555+3948 & A & S235            & C, K \\
    J2230+6946 & A & IRAS 22198+6336 & C \\
    J2148+6107 & C & IRAS 22198+6336 & K \\ 
    J0541-0541 & A & NGC 2071        & C, K \\
    J2230+6946 & A & Cepheus E       & C \\
    J2148+6107 & C & Cepheus E       & K \\
    J2230+6946 & A & L1206           & C \\
    J2148+6107 & C & L1206           & K \\
    J2202+4216 & A & IRAS 22172+5549 & C \\
    J2148+6107 & C & IRAS 22172+5549 & K \\
    J2022+6136 & B & IRAS 21391+5802 & C \\
    J2148+6107 & C & IRAS 21391+5802 & K \\
    \enddata
    \tablecomments{\\
    \textsuperscript{a} Astrometric precisions of A, B and C correspond to positional accuracies of $<$0.002 arcsec, 0.002\textendash0.01 arcsec and 0.01\textendash0.15 arcsec, respectively.}
\end{deluxetable}

\subsection{VLA Data}\label{VLA_data}

\subsubsection{The 6 cm Data}\label{6cm}

The 6 cm (C-Band) observations were made in the A configuration, providing angular resolutions of $\sim$0$^{\prime\prime}$.3 - 0$^{\prime\prime}$.6. The data for regions S235, IRAS 22198+6336, NGC 2071, Cepheus E, and IRAS 21391+5802, consist of two $\sim$2~GHz wide basebands (3-bit samplers) centered at 5.03 and 6.98~GHz. The data were recorded in 30 unique spectral windows (SPWs), each comprised of 64 channels and each channel being 2~MHz wide, resulting in a total bandwidth of 3842~MHz (before “flagging”). Source 3C48 was used as flux density and bandpass calibrator for regions S235, NGC~2071, and Cepheus E, and 3C286 was used as flux density and bandpass calibrator for regions IRAS~22198+6336 and IRAS~21391+5802. Meanwhile, the data for regions L1206 and IRAS~22172+5549, consist of two 1~GHz wide basebands (8-bit samplers) centered at 5.3 and 6.3~GHz. The data were recorded in 16 unique SPWs, each comprised of 64 channels and each channel being 2~MHz wide, resulting in a total bandwidth of 2048~MHz (before “flagging”). Source 3C48 was used as a flux density and bandpass calibrator.

Table \ref{tab:VLA_Phase_Times} shows approximations of the observation times for each source at both 6 and 1.3~cm bands. All the observations were made by alternating between the target source and the phase calibrator. Column 3 shows the project code of the observations, columns 4 and 5 show these alternating times, and column 6 shows the total integration time for each observation.

Data reduction was carried out in a manner similar to the VLA observations from the \citetalias{Sequeira-Murillo_2025}. We used calibrated data from the 41154 (Pipeline-CASA51-P2-B) version of the VLA Calibration Pipeline\footnote{science.nrao.edu/facilities/vla/data-processing/pipeline} for regions IRAS 22198+6336, Cepheus E, and IRAS 21391+5802. For region NGC 2071, we used the VLA Calibration Pipeline version 2021.2.0.128, and for region S235, we used the VLA Calibration Pipeline version 42536 (Pipeline-CASA54-P3-B-SRDP). The differences in the data reduction pipelines arise from the year of the observations and the version of the VLA Calibration Pipeline available at that time (see column 3 in Table \ref{tab:VLA_Phase_Times}). The images were made in CASA \citep{2022PASP..134k4501C} using the \textit{tclean} task and a Briggs \textit{Robust} = 0.5 weighting \citep{1995PhDT.......238B}.

For regions S235, IRAS~22198+6336, NGC~2071, Cepheus E, and IRAS~21391+5802, we made two images, each of a $\sim$2 GHz baseband composed of 15 SPWs, and also a combined image using data from both basebands with a total of 30 SPWs. For regions L1206 and IRAS 22172+5549, we made two images, each of a  $\sim$1 GHz baseband composed of 8 SPWs, and also a combined image using data from both basebands with a total of 16 SPWs. All maps were primary beam-corrected. Columns 5 and 6 of Table \ref{tab:SOMA_Sources} show the synthesized beam (size and position angle) and the rms of the combined images.

\subsubsection{The 1.3 cm Data}\label{1.3 cm}

The 1.3 cm (K-Band) observations were made in the B configuration, providing angular resolutions of $\sim$0$^{\prime\prime}$.3 - 0$^{\prime\prime}$.6. The data consist of two $\sim$4 GHz wide basebands (3-bit samplers) centered at 20.4 and 24.4 GHz, except for regions S235 and NGC 2071, centered at 20.5 and 24.5 GHz. The data were recorded in 60 (62 for source L1206) unique SPWs, comprised of 64 channels and each channel being 2 MHz wide, resulting in a total bandwidth of 7680 MHz (7936 MHz for source L1206), before “flagging”. Source 3C286 was used as a flux density and bandpass calibrator for all the regions except for NGC 2071, where source 3C48 was used as a flux density and bandpass calibrator. Additional information about the observations can be found in Table \ref{tab:VLA_Phase_Times}.

The data reduction was done in the same manner as that for the C-Band observations and using the VLA Calibration Pipeline version 2021.2.0.128 for regions S235 and NGC 2071, and the VLA Calibration Pipeline version 42270 (Pipeline-CASA54-P2-B) for the other regions. The images were made using the \textit{tclean} task and a Briggs \textit{Robust} = 0.5 weighting. We made two images, each of a $\sim$4 GHz baseband composed of 30 SPWs, and also a combined image using data from both basebands with a total of 60 SPWs for all sources except L1206, that have a total of 62 SPWs. All maps were primary beam-corrected. Columns 5 and 6 of Table \ref{tab:SOMA_Sources} show the synthesized beam (size and position angle) and the rms of the combined images.

\begin{deluxetable}{cccccc}
    \tabletypesize{\footnotesize}
    \tablewidth{0pt}
    \tablecaption{Observations times for each source in minutes. \label{tab:VLA_Phase_Times}}
    \tablehead{
    \colhead{Source} & \colhead{Band} & \colhead{Project\scriptsize\textsuperscript{a}} &  \colhead{Target} & \colhead{Phase} & \colhead{Total}\\
    \colhead{} & \colhead{} & \colhead{Code} &  \colhead{Source\scriptsize\textsuperscript{b}} & \colhead{Calibrator\scriptsize\textsuperscript{b}} & \colhead{Integration}\\
    \colhead{} & \colhead{} & \colhead{} & \colhead{(min)} & \colhead{(min)} & \colhead{(min)}}
    \startdata
    S235 & C & 19A-216 & 9.3 & 0.5 & 37.2 \\
         & K & 21B-229 & 2.4 & 0.6 & 21.8 \\
    IRAS 22198+6336 & C & 18A-294 & 8.4 & 0.5 & 42.1 \\
                    & K & 19A-216 & 1.3 & 0.5 & 23.3 \\
    NGC 2071 & C & 22A-125 & 7.9 & 0.5 & 39.5 \\
             & K & 21B-229 & 2.3 & 0.5 & 20.9 \\
    Cepheus E & C & 18A-294 & 8.2 & 0.4 & 40.9 \\
         & K & 19A-216 & 2.2 & 0.4 & 19.6 \\
    L1206 & C & 12B-140\scriptsize\textsuperscript{c} & 6.4 & 0.6 & 31.8 \\
         & K & 19A-216 & 1.2 & 0.5 & 21.8 \\
    IRAS 22172+5549 & C & 12B-140\scriptsize\textsuperscript{c} & 6.4 & 0.6 & 31.8 \\
         & K & 19A-216 & 1.3 & 0.6 & 22.6 \\
    IRAS 21391+5802 & C & 18A-294 & 8.2 & 0.4 & 40.8 \\
         & K & 19A-216 & 1.4 & 0.6 & 24.4 \\
    \enddata
    \tablecomments{\\
    \textsuperscript{a} Project code from the VLA observations. The first two numbers in the project code denotes the year in which the observations were made, e.g., 19A corresponds to observations taken in 2019. \\
    \textsuperscript{b} Alternating time from source to calibrator.\\
    \textsuperscript{c} Observations with project code 12B-140 were Jansky VLA public archival data (PI: M. Hoare), the rest of the observations were our own (PI: V. Rosero).}
\end{deluxetable}

\section{Results}\label{subsec:results}

We follow the methodology described in \citetalias{Rosero_2019} and \citetalias{Sequeira-Murillo_2025} (see Section~4 of \citetalias{Rosero_2019} for further details). In brief, a radio detection is defined when the peak intensity $I_\nu$ is $\geq 5$ times the image rms ($\sigma$) in either of the baseband-combined images at a given band (i.e., C~band or K~band). For non-detections in one of the combined images, we report a $3\sigma$ upper limit for the flux density at the corresponding frequency. Figure~\ref{fig:VLA_Contours} shows VLA contour plots of the combined images at C~band (6~cm; red) and K~band (1.3~cm; cyan) for all the radio sources, overlaid on SOFIA/FORCAST 37~$\mu$m images. The infrared data from \citet{Liu_2020} (including SOFIA, \textit{Herschel}, and \textit{Spitzer} observations) and the VLA data presented here have astrometric accuracies better than $1\farcs5$ and $0\farcs1$ (see Table~\ref{tab:VLA_Calibrators}), respectively.

The spatial scales and the procedures used to measure flux densities and their uncertainties follow those in \citetalias{Rosero_2019} and \citetalias{Sequeira-Murillo_2025}. For the {\it SOMA} and \textit{Intermediate} scales (see below for definitions), we used the CASA task \texttt{imstat} with the aperture sizes listed in Table~\ref{tab:SOMA_scales}. For the \textit{Inner} scale, flux densities were obtained with the CASA task \texttt{imfit}, and we report the image component size (deconvolved from the beam). In cases where we are unable to deconvolved the beam, we report the size (major and minor axes) of the ellipse used to measure the source's flux density. A detailed discussion of the uncertainty calculations is provided in Section~4 of \citetalias{Rosero_2019}.

\begin{figure*}[ht!]
\figurenum{1}
\begin{center}
\includegraphics[width=0.49\linewidth]{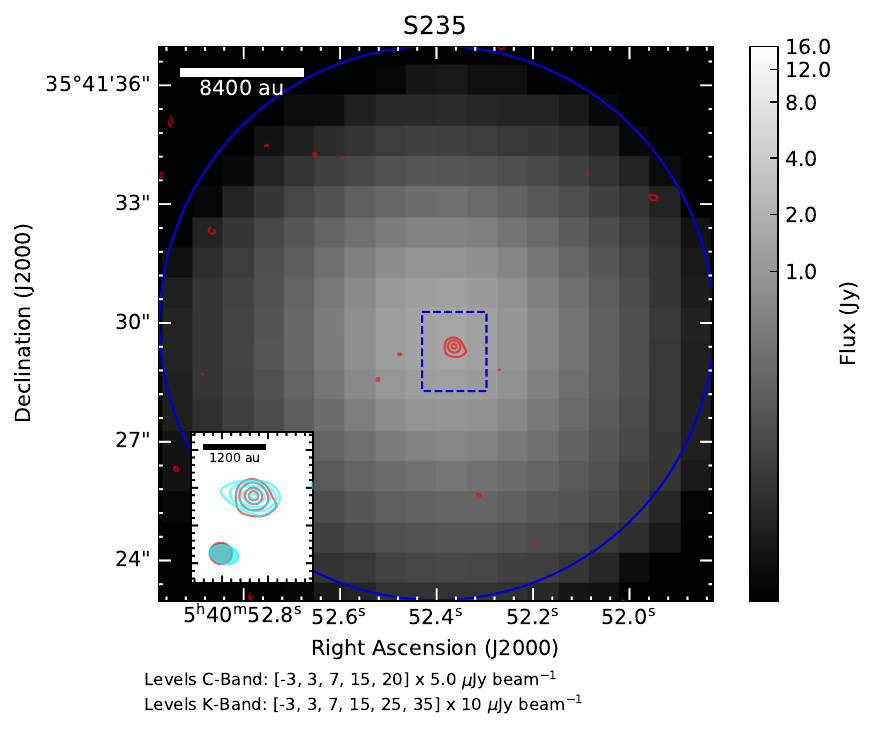}\quad\includegraphics[width=0.49\linewidth]{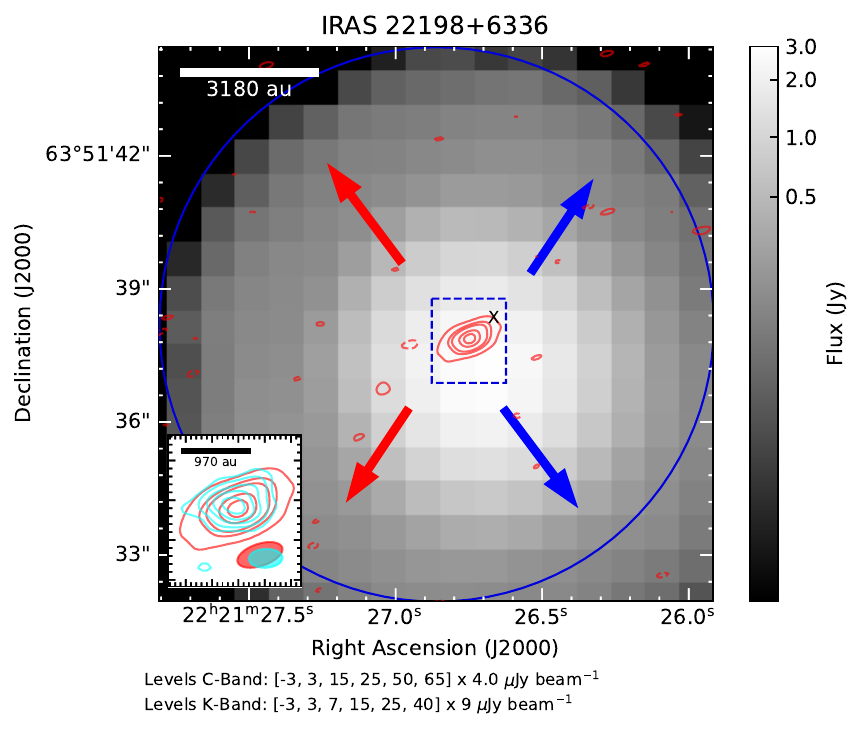} \\
\includegraphics[width=0.49\linewidth]{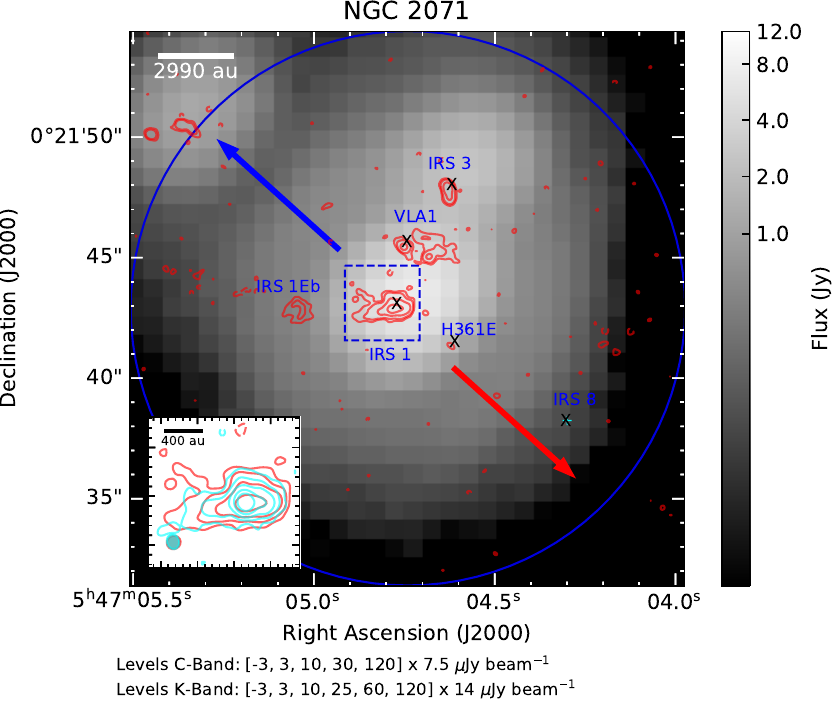}\quad\includegraphics[width=0.49\linewidth]{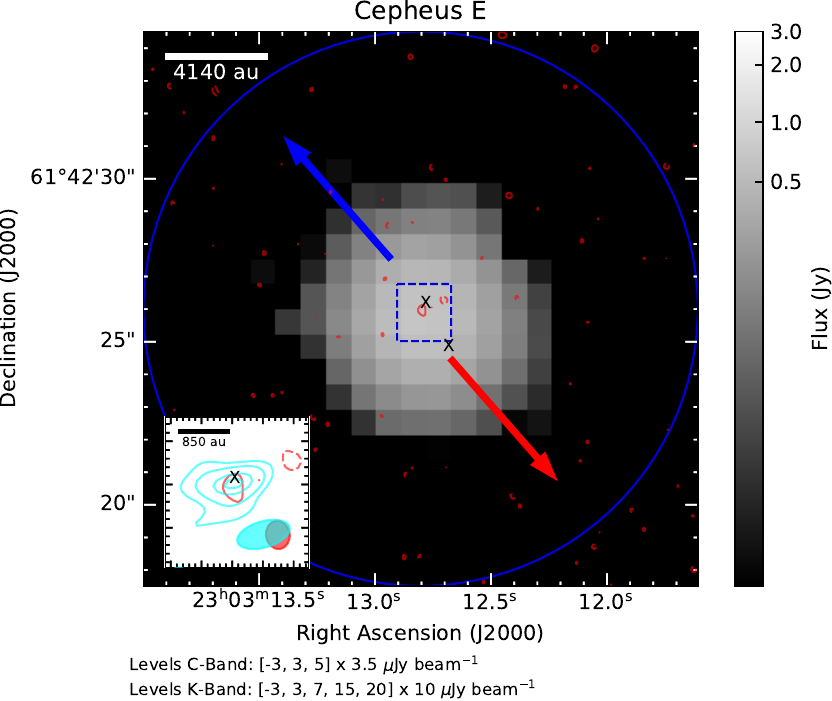} \\
\caption{Images are SOFIA-FORCAST 37 $\mu$m with VLA contours $-$ red: C-band (6 cm); cyan: K-band (1.3 cm) $-$ of the combined radio maps overlaid. The blue dashed squares correspond to the area and the location of the inset image showing a zoom-in of the central region, and the synthesized beams are shown in the lower corners of these insets. The cyan crosses in NGC 2071 denotes the position of detections only at K-band (cyan contours). The black $\times$ denotes the position of the mm core on the regions with millimeter observations, references as follows: IRAS 22198+6336: \cite{Sanchez-Monge2010IRASCore}, NGC 2071: \cite{Cheng2022}, Cepheus E: \cite{2018A&A...618A.145O}, L1206 A: \cite{Beltran2006}, IRAS 21391+5802 BIMA2 and BIMA3: \cite{2007A&A...468L..33N}. 
The blue circles are the SOMA apertures used by \citetalias{Telkamp_2025} and reported in Table C1.
The blue and red arrows represent the direction of a molecular outflow detected toward the region, only for regions with know association with molecular outflows (references are given in section \ref{Morph}). A scale bar in units of au is shown in the upper left of the figures. \label{fig:VLA_Contours}}
\end{center}
\end{figure*}

\begin{figure*}[ht!]
\figurenum{1}
\begin{center}
\includegraphics[width=0.49\linewidth]{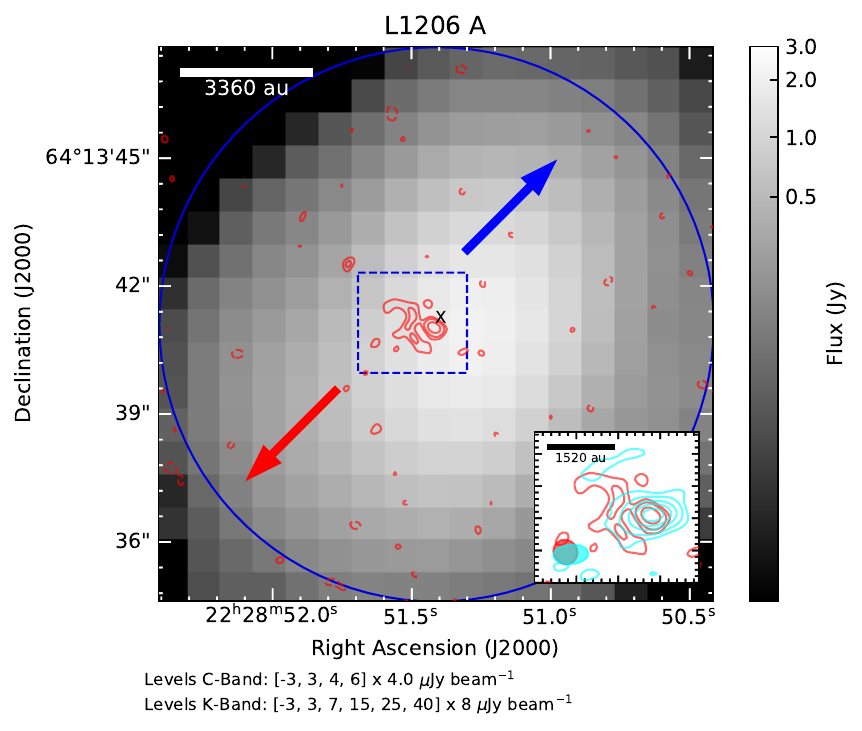}\quad\includegraphics[width=0.49\linewidth]{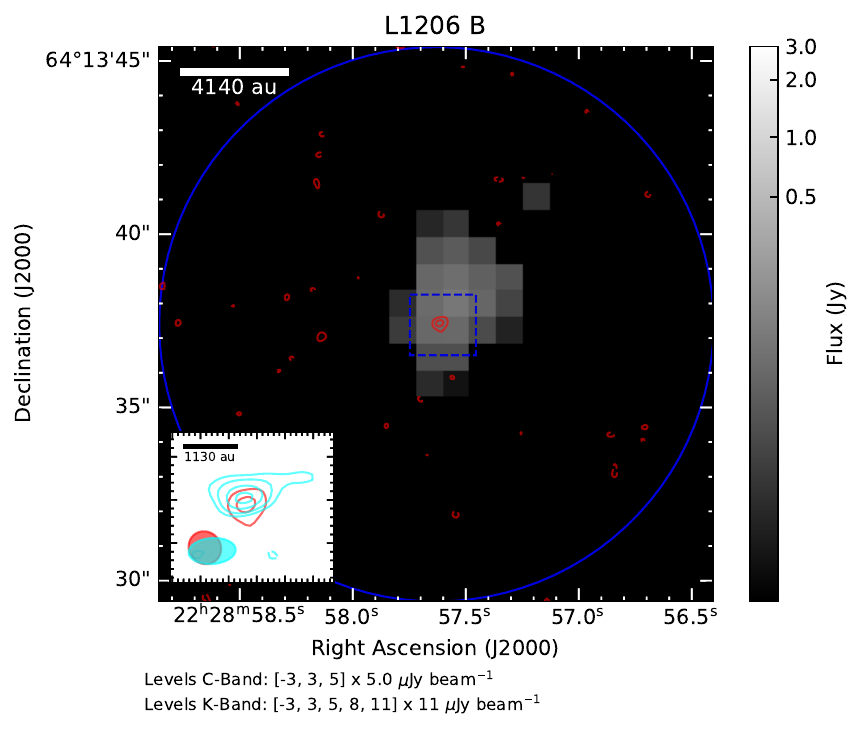} \\
\includegraphics[width=0.49\linewidth]{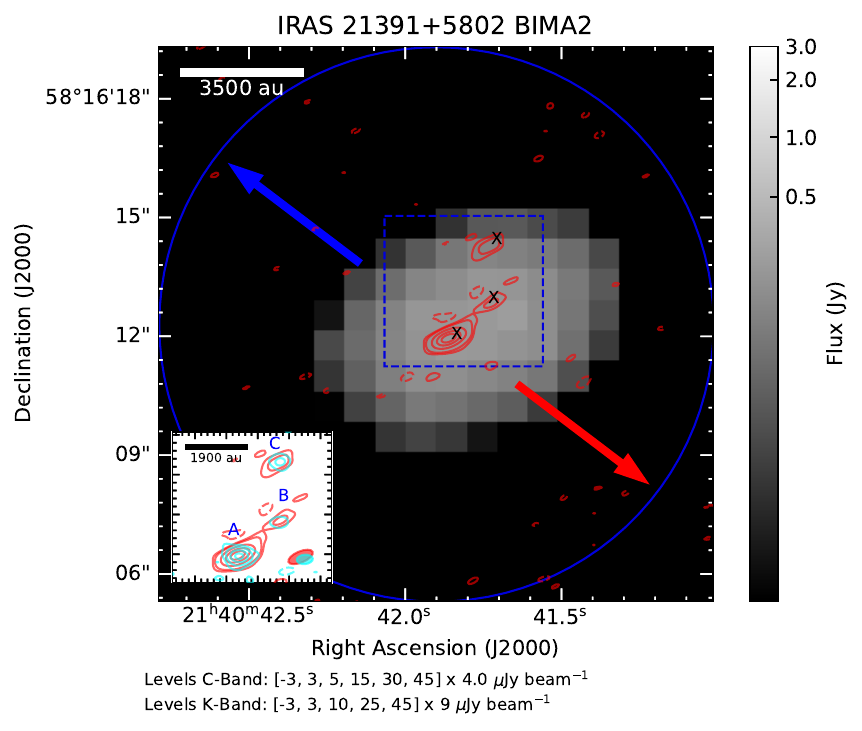}\quad\includegraphics[width=0.49\linewidth]{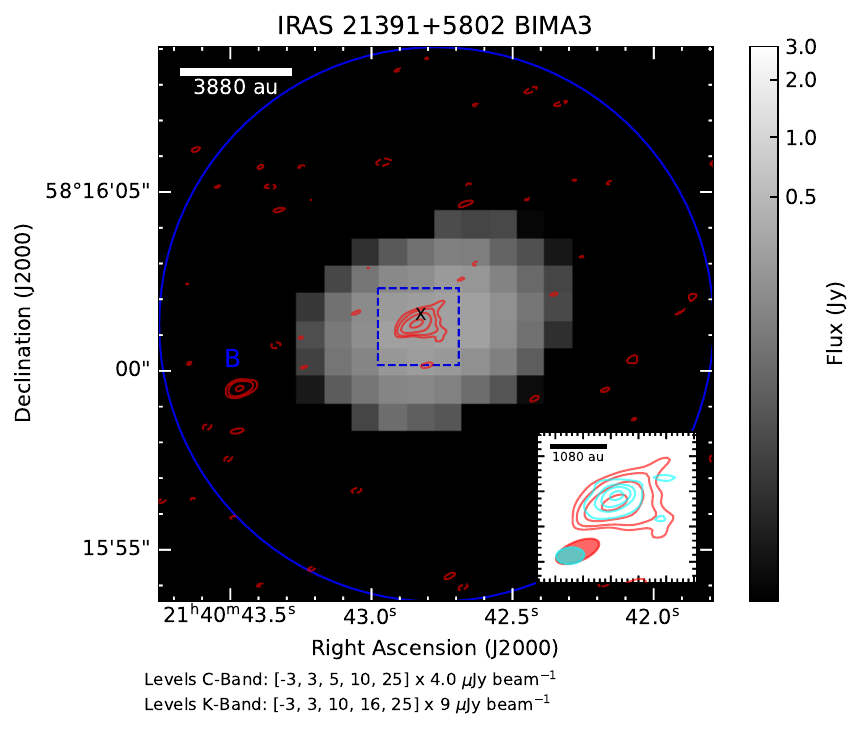} \\
\caption{(Continued)}
\end{center}
\end{figure*}

Table~\ref{tab:Parameters_Radio} lists the radio parameters for each of the 12 protostars studied in this paper. These parameters are measured at three spatial scales, defined as follows. The {\it SOMA} scale corresponds to the aperture radius used in 
\citetalias{Telkamp_2025} to measure the infrared fluxes. The \textit{Intermediate} scale is based on the morphology of the radio source, specifically whether the detections exhibit jet-like structures. In the sample analyzed here, only NGC~2071 display such morphology. The \textit{Inner} scale corresponds to the size of the central radio emission that is most likely associated with the driving protostar.

Columns~1 and 2 of Table~\ref{tab:Parameters_Radio} list the region and spatial scale. For each scale, Columns~3 and 4 give the R.A. and Decl. coordinates. For the {\it SOMA} scale, these values correspond to the SOFIA/FORCAST pointing centers reported in \citet[][Table~\ref{tab:SOMA_Sources}]{Liu_2020}. For the \textit{Intermediate} scale, they represent the midpoint of the jet-like detection, while for the \textit{Inner} scale, they correspond to the peak intensity position of the central detected object. The subsequent columns list the flux densities ($S_\nu$) at various frequencies, with the corresponding uncertainties given in parentheses. The final column in Table~\ref{tab:Parameters_Radio} reports the spectral indices and their uncertainties for each scale (see Section~\ref{Radio_SEDs}).

\begin{deluxetable}{ccccc}
    \tabletypesize{\footnotesize}
    \tablewidth{0pt}
    \tablecaption{SOMA, \textit{Intermediate} and \textit{Inner} Scales. \label{tab:SOMA_scales}}
    \label{tab:Sources_Scales}
    \tablehead{
    \colhead{Region} & \colhead{Frequency} & \colhead{SOMA} & \colhead{\textit{Intermediate}} & \colhead{Inner}\\
    \colhead{} & \colhead{Band} & \colhead{$R\;(\arcsec)$} & \colhead{\textit{$w\;(\arcsec)$ $\times$ $h\;(\arcsec)$}} & \colhead{\textit{$a\;(\arcsec)$ $\times$ $b\;(\arcsec)$}}}
    \startdata
    \multirow{2}{*}{S235}                  & C & \multirow{2}{*}{7.0}  & \nodata & 0.19 $\times$ 0.09 \\
                                           & K &                       & \nodata & 0.09 $\times$ 0.04 \\
    \multirow{2}{*}{IRAS 22198+6336}       & C & \multirow{2}{*}{6.25} & \nodata & 0.41 $\times$ 0.20 \\
                                           & K &                       & \nodata & 0.28 $\times$ 0.25 \\
    \multirow{2}{*}{NGC 2071}              & C & \multirow{2}{*}{11.5} & \multirow{2}{*}{2.83 $\times$ 1.65} & \multirow{2}{*}{1.34 $\times$ 0.75\small\textsuperscript{*}} \\
                                           & K &                       &                    & \\
    \multirow{2}{*}{Cepheus E}             & C & \multirow{2}{*}{7.5}  & \nodata & \nodata \\
                                           & K &                       & \nodata & 0.48 $\times$ 0.11 \\
    \multirow{2}{*}{L1206 A}               & C & \multirow{2}{*}{6.5}  & \nodata & 1.41 $\times$ 0.46 \\
                                           & K &                       & \nodata & 0.23 $\times$ 0.14 \\
    \multirow{2}{*}{L1206 B}               & C & \multirow{2}{*}{8.0}  & \nodata & 0.50 $\times$ 0.25 \\
                                           & K &                       & \nodata & 0.39 $\times$ 0.11 \\
    \multirow{2}{*}{IRAS 22172+5549 MIR1}  & C & \multirow{2}{*}{8.0}  & \nodata & \nodata \\
                                           & K &                       & \nodata & \nodata \\
    \multirow{2}{*}{IRAS 22172+5549 MIR2}  & C & \multirow{2}{*}{3.8}  & \nodata & \nodata \\
                                           & K &                       & \nodata & \nodata \\
    \multirow{2}{*}{IRAS 22172+5549 MIR3}  & C & \multirow{2}{*}{7.75} & \nodata & \nodata \\
                                           & K &                       & \nodata & \nodata \\
    \multirow{2}{*}{IRAS 21391+5802 BIMA2} & C & \multirow{2}{*}{7.0}  & \nodata & \multirow{2}{*}{0.32 $\times$ 0.25\small\textsuperscript{*}} \\
                                           & K &                       & \nodata & \\
    \multirow{2}{*}{IRAS 21391+5802 BIMA3} & C & \multirow{2}{*}{7.75} & \nodata & 0.53 $\times$ 0.33 \\
                                           & K &                       & \nodata & 0.31 $\times$ 0.09 \\
    \multirow{2}{*}{IRAS 21391+5802 MIR48} & C & \multirow{2}{*}{7.75} & \nodata & \nodata \\
                                           & K &                       & \nodata & \nodata \\
    \enddata
    \tablecomments{ The reported values correspond to a circle of radius \textit{R} for the SOMA scale and a box of height \textit{h} and width \textit{w} for the \textit{Intermediate} scale. The inner scale corresponds to image component (deconvolved with beam) size from the task imfit, of major axis \textit{a} and minor axis \textit{b}. Except for the scales that show the ($^*$), in this case the scale corresponds to an ellipse of major axis \textit{a} and minor axis \textit{b}. In the cases that the inner scale between the C and K images are different, we report both values.}
\end{deluxetable}

\clearpage
\startlongtable
\begin{deluxetable*}{ccccccccc}
    \tabletypesize{\footnotesize}
    \tablewidth{0pt}
    \tablecaption{Parameters from Radio Continuum. \label{tab:Parameters_Radio}}
    \tablehead{
    \colhead{Region} & \colhead{Scale} & \colhead{R.A} & \colhead{Decl.} & \colhead{$S_{5.0}$ $_{GHz}$} & \colhead{$S_{7.0}$ $_{GHz}$} & \colhead{$S_{20.6}$ $_{GHz}$} & \colhead{$S_{24.5}$ $_{GHz}$} & \colhead{Spectral} \\
    \colhead{} & \colhead{} & \colhead{(J2000)} & \colhead{(J2000)} & \colhead{(mJy)} & \colhead{(mJy)} & \colhead{(mJy)} & \colhead{(mJy)} & \colhead{Index}}
    \startdata
    & SOMA & 05:40:52.40 & +35.41.30.00 & 1.33(0.24) & 0.31(0.28) & 0.46(0.42) & $<$1.76  & $<$-0.3 \\
    S235 & \textit{Intermediate} & \nodata & \nodata & \nodata & \nodata & \nodata & \nodata & \nodata \\
    & \textit{Inner} & 05:40:52.37 & +35.41.29.39 & 0.11(0.01) & 0.12(0.02) & 0.32(0.03) & 0.42(0.05) & 0.9(0.1) \\
    \hline
    & SOMA & 22:21:26.68 & +63.51.38.20 & 0.47(0.12) & 0.45(0.20) & 0.68(0.33) & 0.86(0.45) & 0.3(0.5) \\
    IRAS 22198+6336 & \textit{Intermediate} & \nodata & \nodata & \nodata & \nodata & \nodata & \nodata & \nodata \\
    & \textit{Inner} & 22:21:26.75 & +63.51.37.92 & 0.24(0.09) & 0.53(0.06) & 0.78(0.09) & 0.80(0.13) & 0.5(0.1) \\
    \hline
    & SOMA & 05:47:04.74 & +00.21.42.96 & 8.21(1.15) & 7.01(0.98) & 10.47(1.41) & 10.87(1.90) & 0.3(0.1) \\
    NGC 2071\small\textsuperscript{a} & \textit{Intermediate} & 05:47:04.70 & +00.21.45.41 & 1.76(0.20) & 1.29(0.15) & 2.39(0.26) & 2.65(0.31) & 0.4(0.1) \\
    & \textit{Inner} & 05:47:04.79 & +00.21.42.96 & 4.88(0.49) & 4.51(0.45) & 6.11(0.61) & 6.81(0.68) & 0.2(0.1) \\
    \hline
    & SOMA & 23:03:12.80 & +61.42.26.00 & 0.02(0.17) & $<$0.80  & 0.03(0.39) & $<$1.48  & $<$0.9  \\
    Cepheus E & \textit{Intermediate} & \nodata & \nodata & \nodata & \nodata & \nodata & \nodata & \nodata \\
    & \textit{Inner} & 23:03:12:80 & +61.42.26.04 & $<$0.01 & $<$0.01 & 0.19(0.03) & 0.27(0.04) & $>$2.1  \\
    \hline
    & SOMA & 22:28:51.41 & +64.13.41.10 & 0.04(0.18) & 0.23(0.21) & 1.36(0.29) & 1.60(0.41) & 1.7(1.2) \\
    L1206 A & \textit{Intermediate} & \nodata & \nodata & \nodata & \nodata & \nodata & \nodata & \nodata \\
    & \textit{Inner} & 22:28:51.43 & +64.13.41.00 & 0.15(0.03) & 0.08(0.03) & 0.43(0.05) & 0.24(0.09) & 0.9(0.2) \\
    \hline
    & SOMA & 22:28:51.41 & +64.13.41.10 & 0.05(0.21) & $<$0.80  & 0.56(0.41) & 0.22(0.64) & $>$1.2 \\
    L1206 B & \textit{Intermediate} & \nodata & \nodata & \nodata & \nodata & \nodata & \nodata & \nodata \\
    & \textit{Inner} & 22:28:51.43 & +64.13.41.00 & 0.05(0.01) & 0.03(0.01) & 0.16(0.02) & 0.17(0.04) & 1.1(0.1) \\
    \hline
    & SOMA & 22:19:08.33 & +56.05.10.52 & 0.17(0.31) & $<$1.23  & $<$1.25  & $<$1.63  & $<$0.7 \\
    IRAS 22172+5549 MIR1 & \textit{Intermediate} & \nodata & \nodata & \nodata & \nodata & \nodata & \nodata & \nodata \\
    & \textit{Inner} & \nodata & \nodata & $<$0.03 & $<$0.03 & $<$0.03 & $<$0.04 & $<$0.2 \\
    \hline
    & SOMA & 22:19:09.48 & +56.05.00.37 & $<$0.44  & 0.38(0.21) & $<$0.68  & $<$0.90  & $<$0.5 \\
    IRAS 22172+5549 MIR2 & \textit{Intermediate} & \nodata & \nodata & \nodata & \nodata & \nodata & \nodata & \nodata \\
    & \textit{Inner} & \nodata & \nodata & $<$0.03 & $<$0.03 & $<$0.04 & $<$0.04 & $<$0.3 \\
    \hline
    & SOMA & 22:19:09.43 & +56.04.45.58 & 0.05(0.31) & 0.05(0.44) & 0.03(0.50) & 0.09(0.59) & 0.1(2.7) \\
    IRAS 22172+5549 MIR3 & \textit{Intermediate} & \nodata & \nodata & \nodata & \nodata & \nodata & \nodata & \nodata \\
    & \textit{Inner} & \nodata & \nodata & $<$0.03 & $<$0.03 & $<$0.04 & $<$0.04 & $<$0.3 \\
    \hline
    & SOMA & 21:40:41.90 & +58.16.12.30 & 0.27(0.13) & 0.36(0.18) & 0.52(0.41) & 0.73(0.51) & 0.5(0.8)  \\
    IRAS 21391+5802 BIMA2\small\textsuperscript{b} & \textit{Intermediate} & \nodata & \nodata & \nodata & \nodata & \nodata & \nodata & \nodata \\
    & \textit{Inner} & 21:40:41.86 & +58.16.11.93 & 0.23(0.03) & 0.26(0.03) & 0.64(0.07) & 0.58(0.07) & 0.7(0.1) \\
    \hline
    & SOMA & 21:40:42.77 & +58.16.01.28 & 0.14(0.16) & 0.22(0.20) & 0.26(0.44) & 0.42(0.57) & 0.5(1.7)  \\
    IRAS 21391+5802 BIMA3 & \textit{Intermediate} & \nodata & \nodata & \nodata & \nodata & \nodata & \nodata & \nodata \\
    & \textit{Inner} & 21:40:42.85 & +58.16.01.46 & 0.01(0.02) & 0.18(0.03) & 0.30(0.04) & 0.20(0.07) & 0.9(0.2) \\
    \hline
    & SOMA & 21:40:41.43 & +58.16.38.09 & 0.07(0.16) & $<$0.61  & $<$1.43  & $<$1.93  & $<$1.3 \\
    IRAS 21391+5802 MIR48 & \textit{Intermediate} & \nodata & \nodata & \nodata & \nodata & \nodata & \nodata & \nodata \\
    & \textit{Inner} & \nodata & \nodata & $<$0.02 & $<$0.02 & $<$0.04 & $<$0.04 & $<$0.6 \\
    \enddata
    \tablecomments{ 
    Units of R.A. are hours, minutes, and seconds. Units of decl. are degrees, arcminutes, and arcseconds. \\
    \textsuperscript{a} Inner scale detection corresponds to the location of source IRS1 (more details in section \ref{NGC2071_info}). \\
    \textsuperscript{b} Inner scale detection corresponds to the location of source VLA 2A (more details in section \ref{I21391_info}).\\ }
\end{deluxetable*}

\clearpage
\subsection{Morphology and Multiplicity}\label{Morph}

All  target regions presented in this paper have been observed in the centimeter continuum; we describe their morphology as ``compact'' if the detection shows no structure on the scale of a few synthesized beams, or ``extended'' otherwise. Next, we describe the centimeter wavelength detections toward each target. For a detailed background on each of these regions, see \citetalias{Liu_2019, Liu_2020}.

\subsubsection{S235} \label{S235_info}

S235~A and S235~B are regions located at a distance of 1.8~kpc \citep{1986PASJ...38..531N,2016ApJ...819...66D} and form part of the extended star-forming region S235 within the giant molecular cloud G174$+$2.5 in the Perseus spiral arm \citep{1996ApJ...463..630H}. For our analysis, we focus on S235~B, classified by \citet{2009MNRAS.399..778B} as an early-type (B1~V) Herbig~Be star based on low- and medium-resolution spectra, whose spectral type, emission lines, nebulosity, and environment meet the criteria for HBe stars defined by \citet{1960ApJS....4..337H}.

In our observations, we detect a single compact source at both 6 cm and 1.3 cm. This detection is consistent with the results of \citet{2006A&A...453..911F}, who also observed the source at these wavelengths, naming it VLA-2 and reporting a flux density of 0.47~mJy with a resolution of 1$^{\prime\prime}$ at 1.3 cm. At 6 cm, they reported an upper limit of 0.27 mJy with a resolution of 3$^{\prime\prime}$. The flux density measured in our 6 cm observations is within this upper limit.

\cite{2006A&A...453..911F} reported an observed spectral index smaller than 0.6 for VLA-2. In our observations, we report an estimated value of the spectral index of $\alpha \sim$ 0.9. This result may indicate that the source is a point source associated with thermal emission. This interpretation is consistent with the observed morphology, which appears compact and not elongated in any direction.

Our results, showing a compact morphology and a spectral index consistent with thermal emission, reinforce the Herbig~Be star classification proposed by \citet{2009MNRAS.399..778B} and support the conclusion regarding the nature of S235~B.

\subsubsection{IRAS 22198+6336} \label{I22198_info}

IRAS 22198+6336 is an intermediate-mass protostar located at 764 $\pm$ 27 pc \citep{Hirota2008Astrometry1204G}. Due to the presence of a strong and compact submillimeter dust condensation \citep{Jenness1995Embedded} and no near-infrared observation, \cite{Sanchez-Monge2008SurveyWavelengths} proposed that IRAS 22198+6336 is a deeply embedded intermediate-mass YSO, classified as an early-type B protostar in an evolutionary stage similar to Class 0 source in the low-mass regime. CO(1-0) and CO(2-1) maps from \cite{Sanchez-Monge2010IRASCore} reveal an outflow with a quadrupolar morphology clearly centered on the position of the dust condensation. 

In our observations, we detected a single source at both 6~cm and 1.3~cm, slightly elongated along the NW--SE direction. This source, previously reported as VLA~2 by \citet{Sanchez-Monge2008SurveyWavelengths}, was observed at 3.6, 1.3, and 0.7~cm with angular resolutions of $\sim$9.6$^{\prime\prime}$, $\sim$3.2$^{\prime\prime}$, and $\sim$1.7$^{\prime\prime}$, respectively. Their spectral index of $\sim$0.5 agrees with our results and is typical of ionized jets associated with YSOs \citep[e.g.,][]{Reynolds1986Continuum,Anglada1998SpectralSources,Tanaka_2016}. Furthermore, the source’s elongated morphology and likely alignment with a larger-scale NW-SE molecular outflow support this interpretation.

\subsubsection{NGC 2071} \label{NGC2071_info}
 
NGC~2071 is an intermediate-mass star-forming region in the L1630 molecular cloud of Orion~B, located at a distance of 430~pc \citep{Tobin2020}. A bipolar molecular outflow has been observed toward NGC~2071, oriented in the northeast–southwest direction, extending $\sim$15$^{\prime}$ in length and reaching velocities of $\sim$120~km~s$^{-1}$. This outflow has been studied in \ch{CO} \citep{1982ApJ...261..558B}, \ch{HI} \citep{1983ApJ...266L..61B}, \ch{OH} \citep{1992ApJ...398..139R}, as well as \ch{SiO} and \ch{CH3OH} \citep{2000ApJ...545..861G}.  

In our observations, we detected 7 sources at both 1.3 and 6 cm, and an additional detection only at 1.3 cm. Several of these sources were previously detected using VLA observations, such as the studies of \cite{Trinidad2009} at 1.3 cm  (K-band) with an angular resolution of $\sim 0.1\arcsec$ and a  sensitivity of 160$\mu$Jy/beam, and \cite{Carrasco2012} at 20 cm (L), 6 cm (C), 3.6 cm (X) and 2 cm (Ku), with resolutions of $\sim 1.44\arcsec$, $\sim 0.71\arcsec$, $\sim 0.35\arcsec$ and $\sim 0.18\arcsec$, respectively. 
In Table \ref{tab:Multiplicity_NGC2071} we present more information about the multiplicity towards NGC 2071, including the R.A. and Decl. positions, the estimated spectral index, association with other wavelengths, and whether each source is a new radio detection or not. 
We present below an individual discussion of each detected source.

\startlongtable
\begin{deluxetable*}{ccccccc}
    \tabletypesize{\footnotesize}
    \tablewidth{0pt}
    \tablecaption{Multiplicity in NGC 2071.}
    \label{tab:Multiplicity_NGC2071}
    \tablehead{
    \colhead{Detection} & \colhead{R.A (J2000)} & \colhead{Decl. (J2000)} & \colhead{Spectral Index} & \colhead{Association} & \colhead{New Detection}}
    \startdata
    IRS 1    & 05:47:04.78 & +00.21.42.93 & 0.2 (0.08)    & IR, mm & No \\
    IRS 1Eb   & 05:47:05.04 & +00.21.42.82 & 0.1 (0.5)   & \nodata & Yes\small\textsuperscript{a} \\
    VLA 1    & 05:47:04.75 & +00.21.45.43 & 0.4 (0.11)   & X-ray, mm & No \\
    VLA 1-B1 & 05:47:04.68 & +00.21.45.07 & 0.1 (0.10)   & \nodata & Yes\small\textsuperscript{a} \\
    VLA 1-B2 & 05:47:04.65 & +00.21.45.20 & 0.1 (0.19)   & \nodata & Yes\small\textsuperscript{a} \\
    HOPS-361-E-cm  & 05:47:04.31 & +00.21.38.05 & $<$1.1   & mm & Yes\small\textsuperscript{b} \\
    IRS 8-cm   & 05:47:04.62 & +00.21.41.32 & $<$0.7    & IR, mm & Yes\small\textsuperscript{b} \\
    IRS 3    & 05:47:04.63 & +00.21.47.84 & 0.4 (0.10)    & IR, mm & No \\
    \enddata
    \tablecomments{ Units of R.A. are hours, minutes, and seconds. Units of Decl. are degrees, arcminutes, and arcseconds. \\
    \textsuperscript{a} These sources have been observed but not detected. \\
    \textsuperscript{b} These sources have been previously detected at different wavelengths but not in radio continuum.}
\end{deluxetable*}

\textit{IRS 1:} From our results, we find this source to have an extended morphology elongated in the E–W direction, consistent with previous reports by \citet{Trinidad2009} and \citet{Carrasco2012}. Using higher-resolution 1.3 cm observations ($\sim 0.1\arcsec$, which is $\sim$3 times higher than our observations), \citet{Trinidad2009} resolved the source into three continuum peaks, IRS 1C, IRS 1W, and IRS 1E, and proposed that IRS 1W and IRS 1E are condensations ejected by IRS 1C, the latter being an ionized jet associated with a YSO. All three sources are associated with water masers, consistent with the jet scenario. In addition, \citet{Trinidad2009} detected an ionized source, IRS 1Wb, located $\sim1\arcsec.8$ southwest of IRS 1C, which is not associated with water masers. Interestingly, we do not detect this source despite our observations being $\sim$10 times more sensitive.

Given its complex morphology and spectral index ($\alpha\sim$ 0.45) between 20-1.3 cm,
\cite{Carrasco2012} suggested that the emission from  IRS1 is arising from the  superposition of two jets due to a binary system. Our estimated spectral index for the inner scale of IRS 1 is slightly flatter (see Table \ref{tab:Multiplicity_NGC2071}) than their reported one, but it is still consistent with thermal emission.

\begin{figure}
\figurenum{2}
\begin{center}
\includegraphics[width=0.95\linewidth]{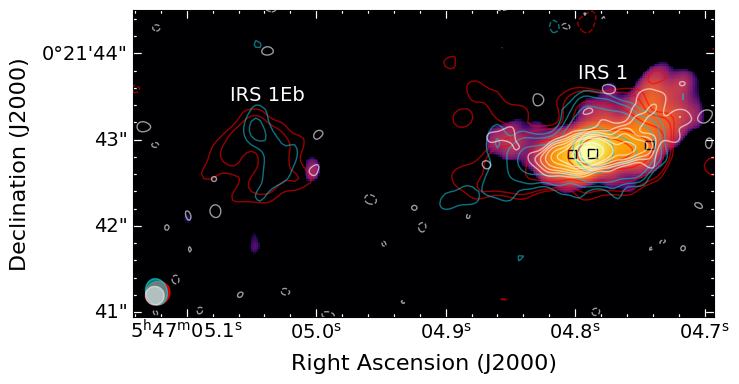}
\caption{For region NGC 2071. VLA 3.6 cm continuum map and contours (white) from Fig. 1 in \cite{Carrasco2012} (see more details on their paper regarding the radio maps), of the radio sources IRS 1 and IRS 1Eb in the NGC 2071 region 
with VLA contours overlaid (1.3 cm--cyan, 3.6 cm--white, and 6 cm--red). The contour levels are at [-3, 3, 5, 10, 50, 100] $\times$ 13.7 $\mu$Jy/beam for 1.3 cm (K-band), at [-3, 3, 5, 7, 10, 15, 20, 40, 60] $\times$ 30 $\mu$Jy/beam for 3.6 cm, and at [-3, 3, 5, 15, 60, 120] $\times$ 7.6 $\mu$Jy/beam for 6 cm (C-band). The beam sizes are indicated in the lower left-hand corner. \label{fig:NGC_IRS1}}
\end{center}
\end{figure}

\textit{IRS 1Eb:} We detected this source at both C and K bands, and to the best of our knowledge it has not been previously reported. However, a point source at a close position is visible in Fig. 1 of \citet{Carrasco2012}, which shows 3.6 cm VLA data at comparable resolution, but was not discussed by the authors. 
Figure \ref{fig:NGC_IRS1} shows the VLA 3.6 continuum map from \cite{Carrasco2012} with VLA contours overlaid from our observations at both C and K band, of sources IRS1 and IRS 1Eb. From this figure we can see the compact source and how it is close to the position of the contours from our observations. This may indicate that the point source from \cite{Carrasco2012} is related to source IRS 1Eb.

Moreover, no millimeter or infrared counterpart is found in the literature. The source shows two peaks, slightly elongated in the NE–SW direction, and lies $\sim4\arcsec$ east of IRS 1C, roughly aligned with the centimeter emission from IRS 1. Its spectral index of $\alpha\sim0.1$ is consistent with thermal emission. Given these characteristics, we speculate that IRS 1Eb may trace ejected material from IRS 1C.

\textit{IRS 3:} In our data, this source appears slightly elongated in the N-S direction, displaying a jet-like morphology and a spectral index consistent with previous studies \citep{1998ApJ...505..756T, Carrasco2012, Trinidad2009}. \citet{Trinidad2009} resolved it into three components, IRS 3C, IRS 3N, and IRS 3S. However, we do not resolve these three components at the lower resolution of our data.  

Our results are consistent with previous studies that describe this source as an ionized jet. Furthermore, it is thought to be part of a jet–disk system, with the disk-like structure traced by 3 mm continuum emission and water masers distributed along the disk plane \citep{Carrasco2012}.

\begin{figure}
\figurenum{3}
\begin{center}
\includegraphics[width=0.95\linewidth]{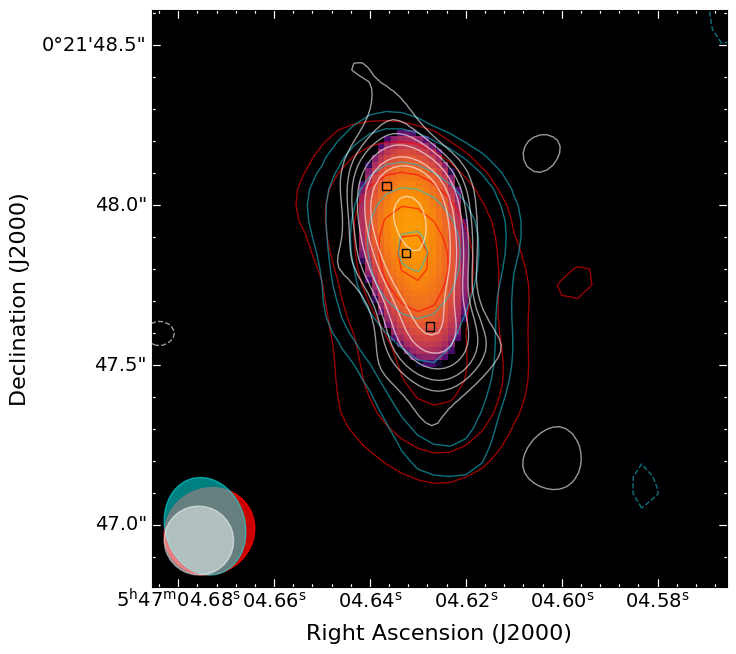}
\caption{For region NGC 2071. VLA 3.6 cm continuum map and contours (white) from Fig. 1 in \cite{Carrasco2012} (see more details on their paper regarding the radio maps), of the radio source IRS 3 in the NGC 2071 region with VLA contours overlaid (1.3 cm--cyan, 3.6 cm--white, and 6 cm--red). The contour levels are at [-3, 3, 5, 10, 50, 100] $\times$ 13.7 $\mu$Jy/beam for 1.3 cm (K-band), at [-3, 3, 5, 7, 10, 15, 20, 40, 60] $\times$ 30 $\mu$Jy/beam for 3.6 cm, and at [-3, 3, 5, 15, 60, 120] $\times$ 7.6 $\mu$Jy/beam for 6 cm (C-band). The beam sizes are indicated in the lower left-hand corner. \label{fig:NGC_IRS3}}
\end{center}
\end{figure}

Figure \ref{fig:NGC_IRS3} shows the VLA 3.6 continuum map from \cite{Carrasco2012} with VLA contours overlaid from our observations at both 6 cm (C-band) and 1.3 cm (K-band), of source IRS 3. \cite{Carrasco2012} presented several epochs of 3.6 cm continuum maps toward this source between 1998 and 1999 in Fig. 5 of their paper. They interpret the difference in the size of the source due to an ejection of ionized material together with the precession of the jet. We see an increase in the size of the emission from IRS3 when comparing the 1.3 cm and 6 cm contours with the 3.6 cm contours from \cite{Carrasco2012}.

\textit{VLA 1:} From our observations, we detect a compact source previously reported as VLA~1 by \citet{Trinidad2009} and \citet{Carrasco2012}. Our estimated spectral index, along with the overall morphology of the system, suggests that this source is likely an ionized jet. Moreover, \citet{Trinidad2009} and \citet{Carrasco2012} proposed that VLA 1 is a younger and more deeply embedded YSO compared to other sources in the region, as it is associated with millimeter and X-ray emission, but remains undetected at infrared wavelengths. \\

\begin{figure}
\figurenum{4}
\begin{center}
\includegraphics[width=0.95\linewidth]{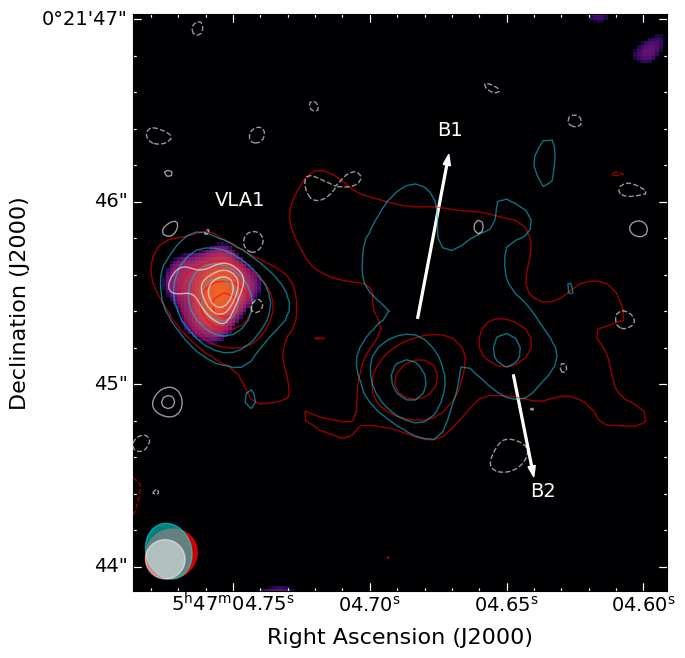}
\caption{For region NGC 2071. VLA 3.6 continuum map and contours (white) from Fig. 1 in \cite{Carrasco2012} (see more details on their paper regarding the radio maps), of the radio source VLA 1 in the NGC 2071 region with VLA contours overlaid (1.3 cm--cyan, 3.6 cm--white, and 6 cm--red). The contour levels are at [-3, 3, 5, 10, 50, 100] $\times$ 13.7 $\mu$Jy/beam for 1.3 cm (K-band), at [-3, 3, 5, 7, 10, 15, 20, 40, 60] $\times$ 30 $\mu$Jy/beam for 3.6 cm, and at [-3, 3, 5, 15, 60, 120] $\times$ 7.6 $\mu$Jy/beam for 6 cm (C-band). The beam sizes are indicated in the lower left-hand corner. \label{fig:NGC_VLA1}}
\end{center}
\end{figure}

\textit{VLA 1-B1 and VLA 1-B2:} We detect two new point sources at both 1.3 and 6 cm, labeled VLA1-B1 and VLA1-B2, which, to the best of our knowledge, have not been previously reported (see Figure~\ref{fig:NGC_VLA1}). These sources are separated by $\sim$0$\arcsec$.5 and located $\sim$1$\arcsec$ and $\sim$1$\arcsec$.5 southwest of VLA1, respectively. Both are aligned with the centimeter emission from VLA1 and appear slightly extended, with an additional peak north of VLA1-B1. Their estimated spectral indices of $\sim$0.1 are consistent with optically thin free-free emission. Neither source appears to be associated with 3mm emission based on Figure 1 of \citet{Carrasco2012}.

It is noteworthy that VLA1-B1, the brighter of the two sources, with peak intensities of 380$\mu$Jy and 320~$\mu$Jy at 1.3 and 6 cm, respectively, should have been detectable in previous studies. Based on the reported sensitivities of 40~$\mu$Jy/beam at 6~cm \citep{Carrasco2012} and 70$\mu$Jy/beam at 1.3~cm \citep{Trinidad2009}, VLA1-B1 would correspond to an $\sim$8$\sigma$ and $\sim$5$\sigma$ detection, respectively. However, it was not detected in those studies, nor was it detected at 3.6cm (X-band) in \cite{Carrasco2012}, which had a sensitivity of 19~$\mu$Jy/beam.

While the nature of these sources remains uncertain, their  spectral indices, alignment with the jet axis, and lack of millimeter counterparts suggest they may be ionized ejecta from VLA~1. 

\begin{figure}
\figurenum{5}
\begin{center}
\includegraphics[width=0.95\linewidth]{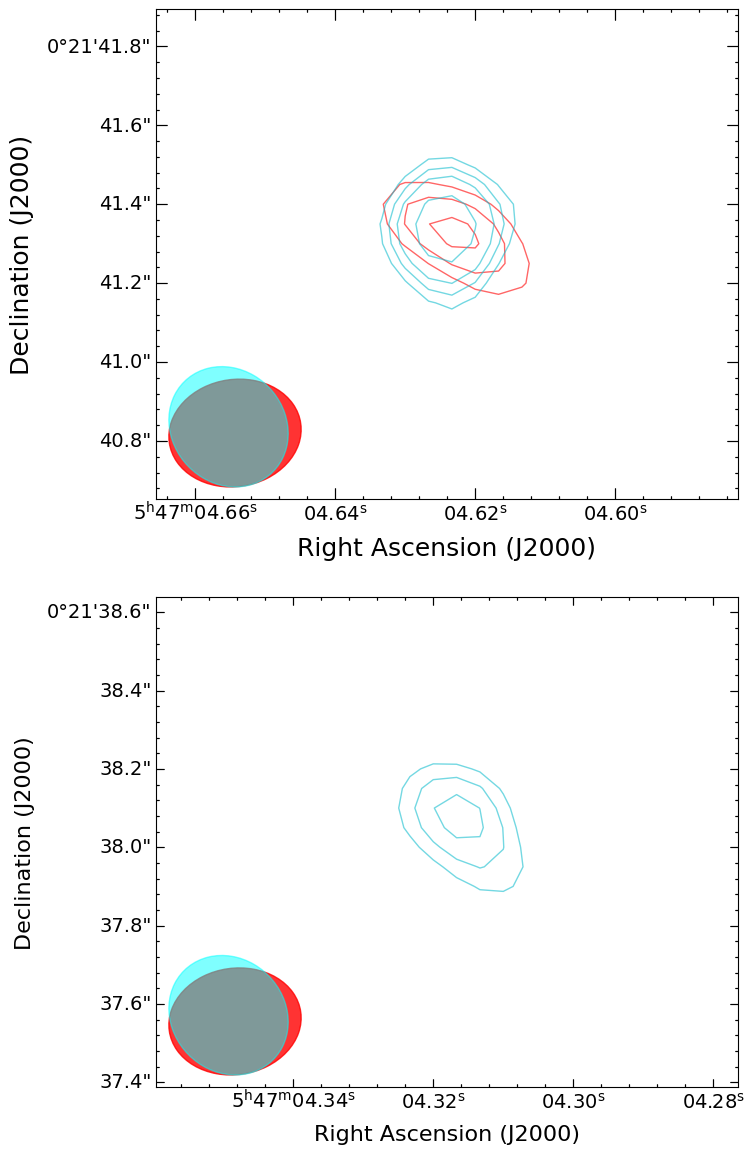}
\caption{For region NGC 2071. VLA continuum maps of the radio sources: HOPS-361-E-cm (top) and IRS 8 (bottom) in the NGC 2071 region observed at 1.3 cm (cyan) and 6 cm (red). The contour levels are at [-3, 3, 4, 5] $\times$ 7.6 $\mu$Jy/beam for 1.3 cm (K-band), and at [-3, 3, 4, 5, 7] $\times$ 13.7 $\mu$Jy/beam for 6 cm (C-band). The beam sizes are indicated in the lower left-hand corner. \label{fig:NGC_IRS8}}
\end{center}
\end{figure}

\textit{IRS 8-cm and HOPS-361-E-cm:} We report two newly detected weak, compact centimeter sources, HOPS-361-E-cm and IRS 8-cm, both identified only at 1.3 cm with 5$\sigma$ significance. HOPS-361-E-cm shows a marginal 4$\sigma$ detection at 6 cm (see Figure~\ref{fig:NGC_IRS8}), while IRS 8-cm is undetected at this frequency, with its position marked by an ``x'' in Figure~\ref{fig:VLA_Contours}. Both sources are spatially associated with 1.3 mm continuum emission from \cite{Cheng2022}. Due to their non-detection at 6 cm, we estimate upper limits on their spectral indices: $\alpha \lesssim 1.1$ for HOPS-361-E-cm and $\alpha \lesssim 0.7$ for IRS 8-cm. Given their spectral properties and millimeter associations, we suggest these are likely individual YSOs within the NGC~2071 cluster.

\subsubsection{Cepheus E} \label{CepE_info}

Cepheus~E is a molecular cloud at a distance of 730~pc \citep{1977ApJ...218..736S} whose central source, the isolated intermediate-mass Class~0 protostar Cep~E-mm \citep{1996A&A...313L..17L,2001ApJ...555..146M}, drives a molecular outflow and jet, as revealed by CO(2-1) and CO(3-2) maps \citep{LeflochCepE1,LeflochCepE2}. The southern jet terminates at the bright Herbig–Haro object HH~377 \citep{2000AJ....120..909A,Gusdorf_CepE}. \citet{2018A&A...618A.145O} detected a hot corino in Cep~E-mm and resolved it into a binary system (Cep~E-A and Cep~E-B), using NOEMA observations at 1.3 and 3.3 mm with resolutions of $\sim$1.4$^{\prime\prime}$ and $\sim$2.4$^{\prime\prime}$, respectably. With their mm data, \citet{2018A&A...618A.145O} concluded that both millimeter components drive high-velocity molecular jets, with Cep~E-A powering the jet associated with HH~377, and Cep~E-B powering another jet almost perpendicular to the Cep~E-A jet. \citet{2019MNRAS.490.2679O} identified several molecular species in the Cep~E-A jet, including \ch{SO}, \ch{SiO}, \ch{H2CO}, \ch{CS}, \ch{HCO+}, and \ch{HCN}.

From our observations, we detect a compact source at K~band (1.3~cm) and only a $3\sigma$ detection at C~band, representing, to our knowledge, the first reported centimeter detection toward this source. The emission is coincident with Cep~E-mm, previously detected in the infrared and millimeter \citep{1996A&A...313L..17L,LeflochCepE2,2001ApJ...555..146M}, but at our resolution the binary components are not resolved and it appears we are only resolving Cep~E-A. We estimate a lower limit to the spectral index of $\alpha>2.1$.

This source exhibits several star-formation tracers, such as a mm counterpart, molecular outflows, and a rich variety of molecular species associated with the jet powered by Cep~E-A. Since this source is detected only at K~band and we report a lower limit for its spectral index, the current evidence is insufficient to draw a firm conclusion about its nature. Further observations are needed to determine it with more certainty. 

\subsubsection{L1206} \label{L1206_info}

L1206, also known as IRAS 22272+6358, is a region located at a distance of 776~pc \citep{2010Rygl}. In \citetalias{Liu_2020}, two MIR sources are presented in the field of view separated at distance of about 40$^{\prime\prime}$. Consequently, our analysis focuses on the L1206 A and L1206 B regions, where we present 6 and 1.3 cm observations.

\textit{L1206 A:} Region L1206 A presents a CO molecular outflow at the position of the source OVRO 2 (IRAS 22272+6358A hereafter referred to as L1206 A) \citep{Beltran2006, 2021ApJS..253...15L}. Moreover, \cite{Surcis2013II} with VLBI observations, detected 26 6.7-GHz methanol (\ch{CH3OH}) masers features towards L1206 A and aligned in the direction of the molecular outflow. 

The source L1206 A, has no optical counterpart, but has been observed at infrared, millimeter (2 and 2.7 mm) and radio continuum wavelengths in the Q (0.7~cm) and C bands \citep{2000AJ....119..323S, Beltran2006, 2021Purser}. \cite{Beltran2006} suggested that the intermediate-mass protostar, L1206 A is in a transition between a Class 0 and I object, giving the dust emission morphology and properties of OVRO 2. Meanwhile, \cite{2021Purser} suggested that L1206 A traces an ionized jet (candidate) at 6 cm (C-band), but is dominated by disk emission at Q-band. 

In our 6 and 1.3 cm observations, we report a single detection that is coincident with the source A and OVRO 2 reported by \cite{2021Purser} and \cite{Beltran2006} respectively. This source presents an interesting morphology. At 6 cm (C-band), it shows an elongation in the N-S direction, but at higher sigma levels and at 1.3 cm (K-band), it displays a compact morphology. We report an estimate spectral index of $\alpha\sim0.9$, which is similar to the value reported by \cite{2021Purser} of $\alpha \sim 1.05$. Given the morphology of the source, the weak centimeter emission and a spectral index of $\alpha \sim 1$, we are in accordance with \cite{2021Purser} and conclude that L1206 A appears to be ionized jet. We also report that source L1206 A presents a 6.7 GHz methanol (\ch{CH3OH}) maser within the synthesized beam, with a flux density around 2.88~mJy.

\textit{L1206 B:} Source IRAS 22272+6358 B (hereafter referred to as L1206 B) is a less luminous object that lies 40$^{\prime\prime}$ to the east of L1206 A and is clearly visible at near infrared wavelengths. \cite{1991AJ....102.1398R} suggested that L1206 B is a late Class I object or an early Class II object.

For source L1206 B, we present the first radio observations (to the knowledge of the authors) of this region and we report a single compact detection with an estimated spectral index of $\alpha \sim1.1$ that may indicate free-free emission from an ionized jet or an optically thick HC HII region \citep{2019ApJ...880...99R}, although higher resolution observations are needed to confirm either scenario. 

\subsubsection{IRAS 22172+5549} \label{I22172_info}

IRAS~22172$+$5549 is located at 2.4~kpc \citep{2002Molinari} and hosts a range of YSOs, from intermediate- to high-mass protostars deeply embedded in massive dense cores \citep{2013ApJ...762..120P}. Our analysis focuses on the MIR1, MIR2, and MIR3 regions identified by \citet{2002Molinari}. \citet{2004Fontani} reported a CO(1-0) bipolar outflow centered on MIR2 (also known as IRS1), whose parameters suggest a relatively massive protostar. \citet{Liu_2020} also detected extended emission along the blue-shifted lobe of this outflow using SOFIA observations. For MIR1 and MIR3, no mm emission or molecular outflows have been reported \citep{2004Fontani,2013ApJ...762..120P}.  

In our VLA observations, none of the MIR sources from \citet{2002Molinari} are detected, consistent with \citet{2021Purser}, whose similar-resolution and sensitivity data revealed only one C-band (6 cm) source, but which was not coincident with any MIR positions, and no Q-band (0.7~cm) detections.

\subsubsection{IRAS 21391$+$5802} \label{I21391_info}

IRAS~21391$+$5802 is a young, intermediate-mass object deeply embedded in the bright-rimmed globule IC~1396N at a distance of 750~pc \citep{1979A&A....75..345M}. This source shows strong submillimeter and millimeter continuum dust emission \citep{1993AJ....106..250W}, high-density gas \citep{2001A&A...376..271C}, and water maser emission \citep{2005A&A...443..535V}.  

\citet{Beltran2002} resolved the source into three millimeter cores—BIMA~1, BIMA~2, and BIMA~3—associated with centimeter sources VLA~1, VLA~2, and VLA~3, respectively, detected using 3.6 cm observations. BIMA~2 is an IM protostar surrounded by the less massive BIMA~1 and BIMA~3, and it drives a strong east–west molecular outflow. \citet{2010ApJ...717.1067C} identified $\sim$45 YSOs in IC~1396N from NIR observations, with MIR~50 and MIR~54 corresponding to BIMA~2 and BIMA~3, and MIR~48 located $\sim$45$^{\prime\prime}$ north of BIMA~2.

Our analysis focuses on BIMA~2, BIMA~3, and MIR~48. We detect multiple radio sources toward BIMA~2 and BIMA~3, but no emission toward MIR~48. The latter result is consistent with the results shown by \citet{Beltran2002}, who did not detect any source around MIR~48 at 3.6 cm with a resolution of 17$^{\prime\prime}$.9 and a sensitivity of 32 $\mathrm{\mu}$Jy. Table~\ref{tab:Multiplicity_I21391} summarizes the multiplicity in BIMA~2 and BIMA~3, including the R.A. and Decl. positions, estimated spectral indices, associations at other wavelengths, whether each source represents a new radio detection, and the estimated spectral type (more details about this are presented in Sect.~\ref{multiplicity_detections}).

\startlongtable
\begin{deluxetable*}{ccccccccc}
    \tabletypesize{\footnotesize}
    \tablewidth{0pt}
    \tablecaption{Multiplicity in IRAS 21391 BIMA 2 and BIMA 3.}
    \label{tab:Multiplicity_I21391}
    \tablehead{
    \colhead{Region} & \colhead{Source} & \colhead{R.A (J2000)} & \colhead{Decl. (J2000)} & \colhead{Spectral Index} & \colhead{Association} & \colhead{New Detection} & \colhead{Spectral Type}}
    \startdata
    BIMA 2 & VLA 2A    & 21:40:41.86 & +58.16.11.95 & 0.7 (0.09) & mm & Yes\small\textsuperscript{a} & B3 \\
           & VLA 2B & 21:40:41.73 & +58.16.12.81 & 0.9 (0.66) & mm & Yes\small\textsuperscript{a} & B3 \\
           & VLA 2C & 21:40:41.73 & +58.16.14.34 & 1.2 (0.18) & mm & Yes\small\textsuperscript{a} & B4 \\
    BIMA 3 & VLA 3     & 21:40:42.84 & +58.16.01.36 & 0.9 (0.16) & IR, mm & No & B3 \\
           & VLA 3B          & 21:40:43.46 & +58.15.59.49 & 0.0 (0.20) & \nodata & Yes\small\textsuperscript{b} & B4 \\
    \enddata
    \tablecomments{ Units of R.A. are hours, minutes, and seconds. Units of decl. are degrees, arcminutes, and arcseconds. \\
    \textsuperscript{a} These sources have been previously detected in radio continuum but we are resolving into multiple detections \\
    \textsuperscript{b} These sources have been observed but not detected.}
\end{deluxetable*}

\textit{IRAS 21391+5802 BIMA2:} To our knowledge, this is the first time source VLA~2 has been resolved into three components. We detect emission at both C and K bands, labeling the components VLA~2A, VLA~2B, and VLA~2C. These ionized sources are associated with the millimeter cores IRAM~2A, 47.73$+$14.3, and 47.73$+$12.8 reported by \citet{2007A&A...468L..33N} using 3.3 and 1.3 mm observations with resolutions of $\sim$0.45$^{\prime\prime}$ and $\sim$1.0$^{\prime\prime}$, respectably. Their association supports the interpretation that each source is ionized by an independent protostellar object.

The central source, VLA~2A, is relatively compact, as its north–south elongation aligns with the beam direction. Sources VLA~2B and VLA~2C appear as point sources. The reported positive spectral indices, consistent with thermal emission, together with their spectral types (see Table~\ref{tab:Multiplicity_I21391}) and association with millimeter emission, provide further evidence that these three sources are independent protostellar objects.

\textit{IRAS 21391+5802 BIMA3:} We detect two  sources at both 1.3~cm and 6~cm. The first, VLA~3, was previously reported by \citet{Beltran2002}, who classified it as an evolved low-mass object based on its compact morphology and small dust emissivity index. In our data, VLA~3 appears slightly elongated in the NE--SW direction.  

We also identify a second compact source, located east of VLA~3 at around 10$^{\prime\prime}$ and labeled VLA~3B, which, to our knowledge, has not been previously reported. This source has no known counterparts at other wavelengths.

We report an estimated spectral index of $\alpha \sim 0.9$ for VLA~3, which denotes the presence of thermal emission and is in the range of expected spectral index for (0.2 $\leq$ $\alpha$ $\leq$ 1.2) ionized jets \citep{2019ApJ...880...99R, 1995RMxAC...1...67A}. For source VLA~3B we report a flat spectral index of $\alpha \sim 0.0$. 

Given the slightly extended morphology, the mm counterpart, and the estimated spectral index of VLA~3, we suggest that the nature of this source is an ionized jet; however, more information and additional observations are needed to confirm this assessment. Another possibility is that this source could be a protostellar object with a B3 spectral type, as shown in Table \ref{tab:Multiplicity_I21391}

Meanwhile, source VLA~3B is too far from VLA~3 to be considered associated with it. Furthermore, there is no mm counterpart in the high-resolution 1.3 and 3.3 mm observations reported by \citet{2007A&A...468L..33N}. Considering all this and given the compact morphology of this source, its nature could be extragalactic, but more information and observations are still required to confirm this assessment.

\subsection{Radio SEDs}\label{Radio_SEDs}

In Figure \ref{fig:Radio_SEDs} we present the extended (E-)SEDs, i.e., radio + IR SEDs, for our twelve sources. The dashed lines correspond to the best fit to the data using a power law of the form $S_{\nu} \propto \nu^{\alpha}$, where $\alpha$ is the spectral index at the different scales: {\it SOMA}, \textit{Intermediate}, and \textit{Inner}, as described above. The spectral index was calculated using the flux density at the central frequencies of the images, so $\alpha$ is calculated over a wide frequency range ($\sim$20 GHz). The uncertainty in the spectral index was calculated with a Monte Carlo simulation that bootstrapped the flux density uncertainties. We estimated an upper limit in the spectral index for non-detections at higher frequencies using a value of $S_{\nu}$ of 3$\sigma$. 


\begin{figure*}[ht!]
\figurenum{6}
\begin{center}
\includegraphics[width=0.32\linewidth]{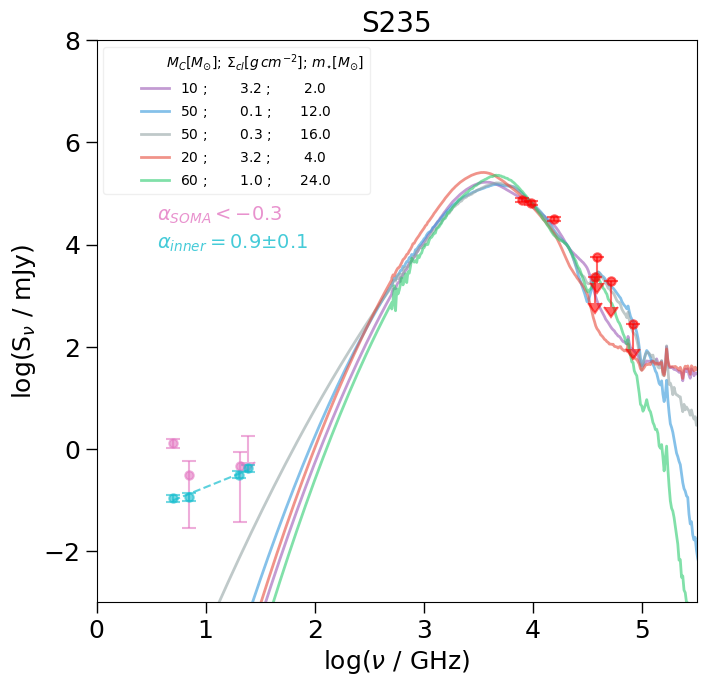}\quad\includegraphics[width=0.32\linewidth]{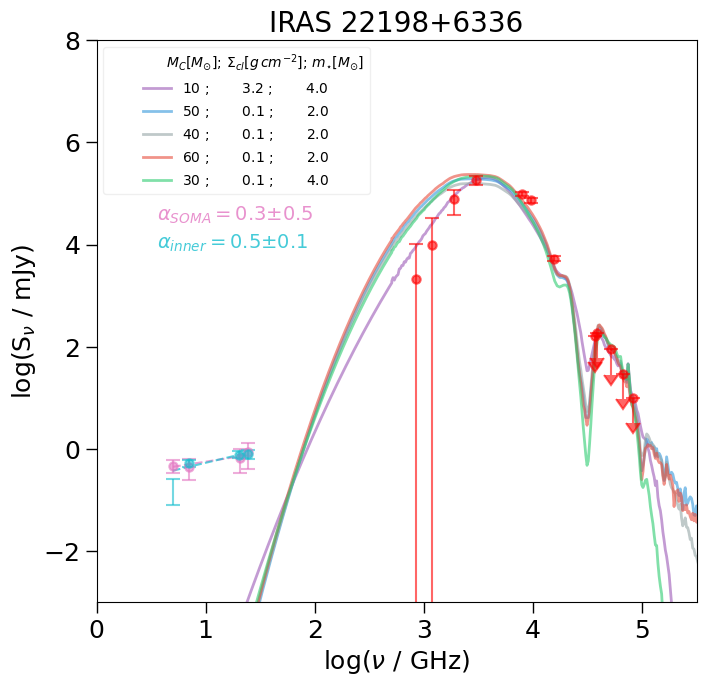}\quad\includegraphics[width=0.32\linewidth]{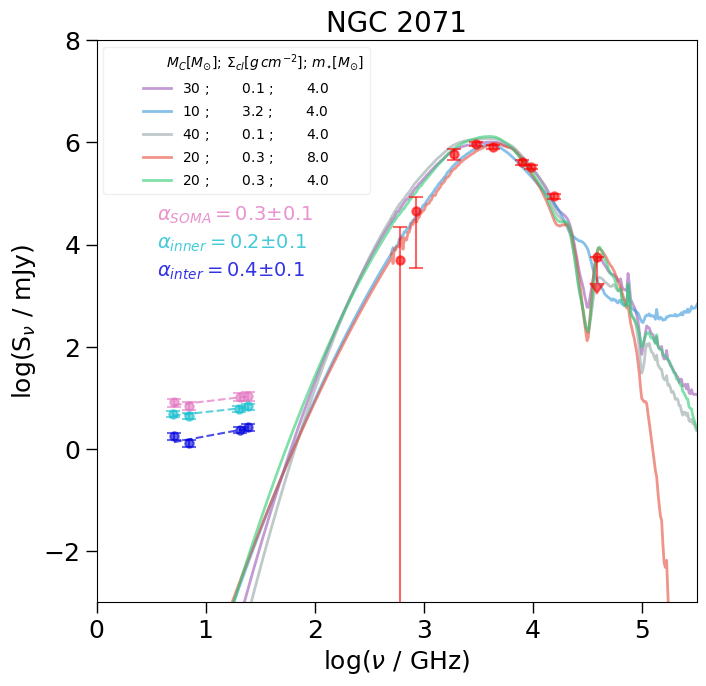}\\
\includegraphics[width=0.32\linewidth]{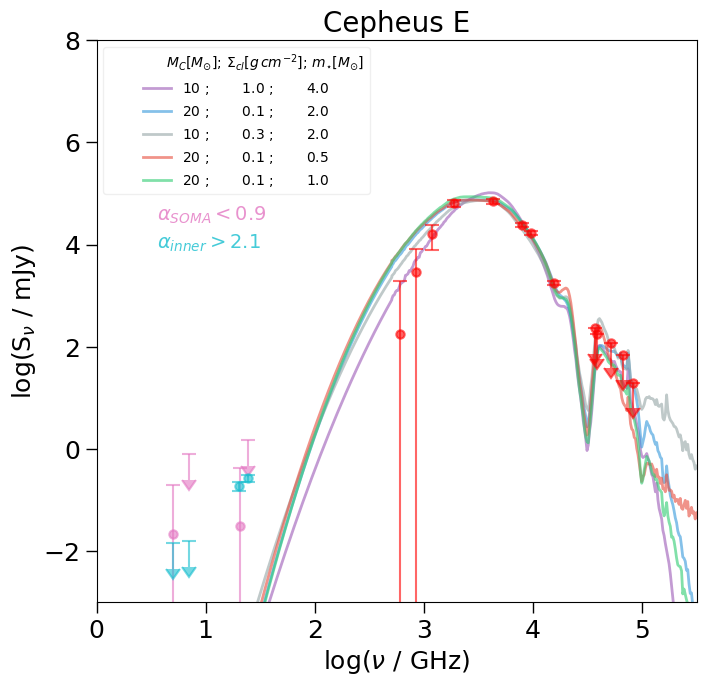}\quad\includegraphics[width=0.32\linewidth]{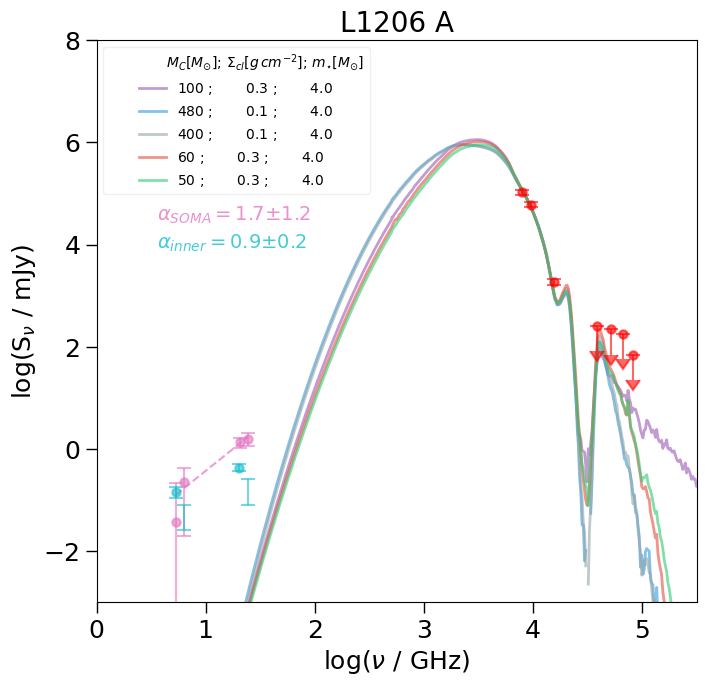}\quad\includegraphics[width=0.32\linewidth]{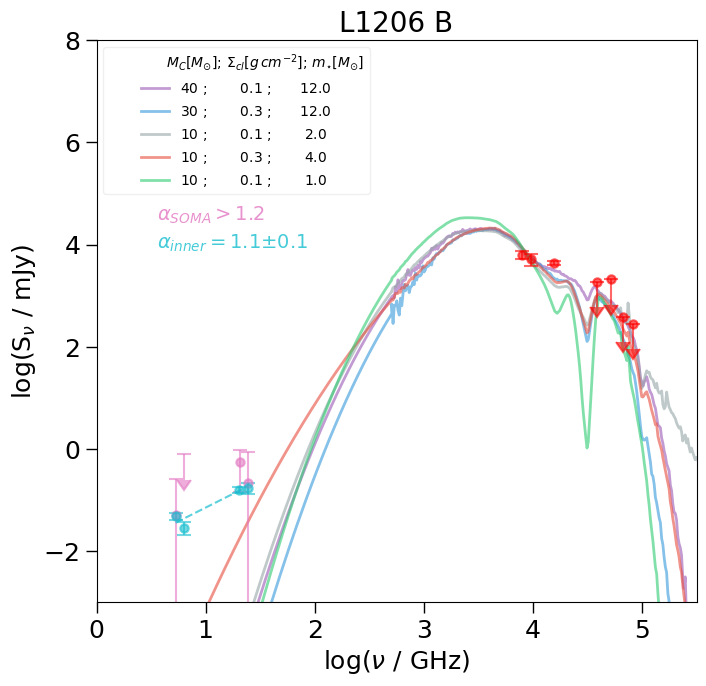}\\ 
\caption{Extended spectral energy distributions (E-SEDs) of SOMA protostars, consisting of radio and IR SEDs. The circles correspond to the flux density as a function of the frequency at each scale (magenta: \textit{SOMA}; blue: \textit{Intermediate}; cyan: \textit{Inner}). Error bars are explained in \S\ref{Radio_SEDs}. The solid lines show the five best fits (see legend) to the IR data SED from the \cite{Zhang_2018} models as fit by \citetalias{Telkamp_2025} (see table C1, Appendix in \citetalias{Telkamp_2025}), and the dashed lines are the best fit of the radio data using a power law of the form $S_{\nu} \propto \nu^{\alpha}$. \label{fig:Radio_SEDs}}
\end{center}
\end{figure*}

\begin{figure*}[ht!]
\figurenum{6}
\begin{center}
\includegraphics[width=0.32\linewidth]{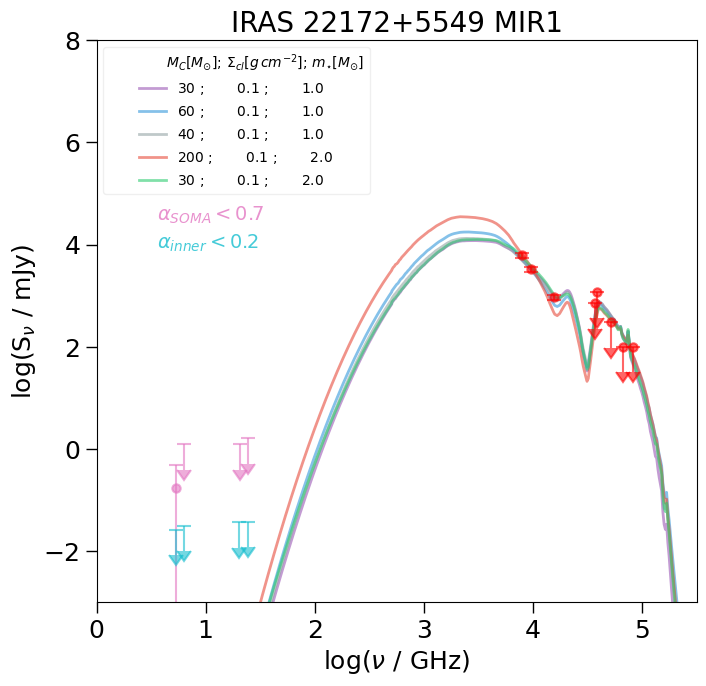}\quad\includegraphics[width=0.32\linewidth]{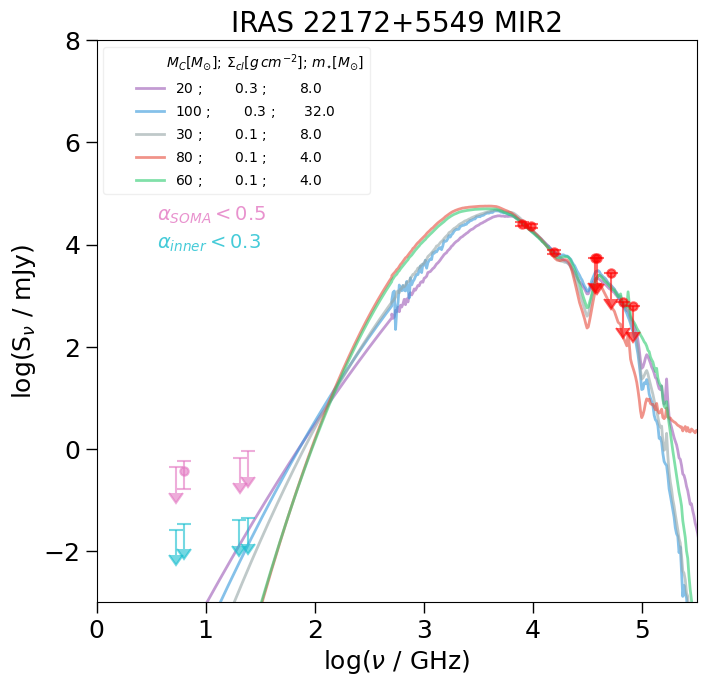}\quad\includegraphics[width=0.32\linewidth]{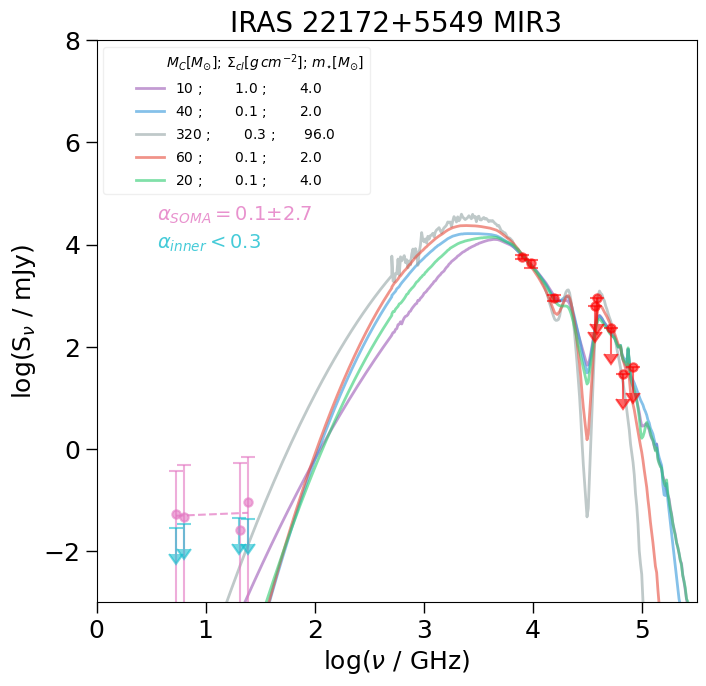}\\ 
\includegraphics[width=0.32\linewidth]{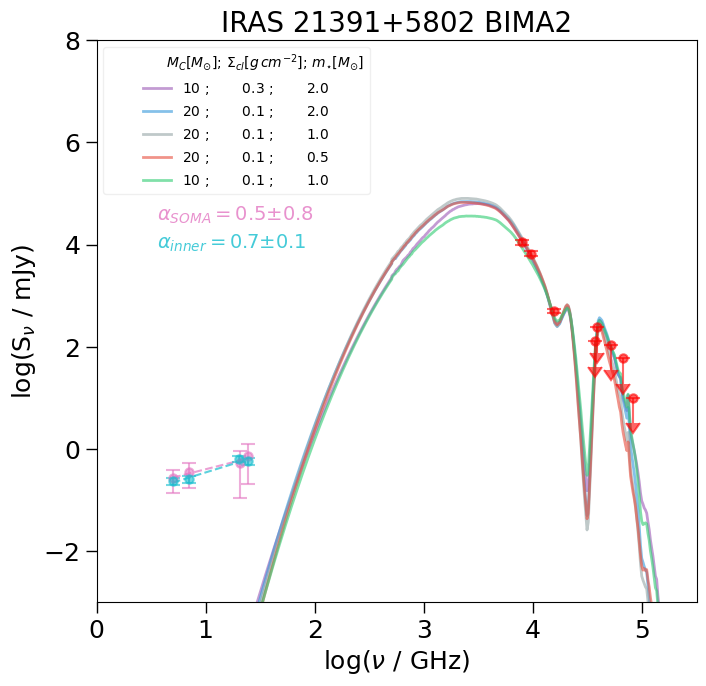}\quad\includegraphics[width=0.32\linewidth]{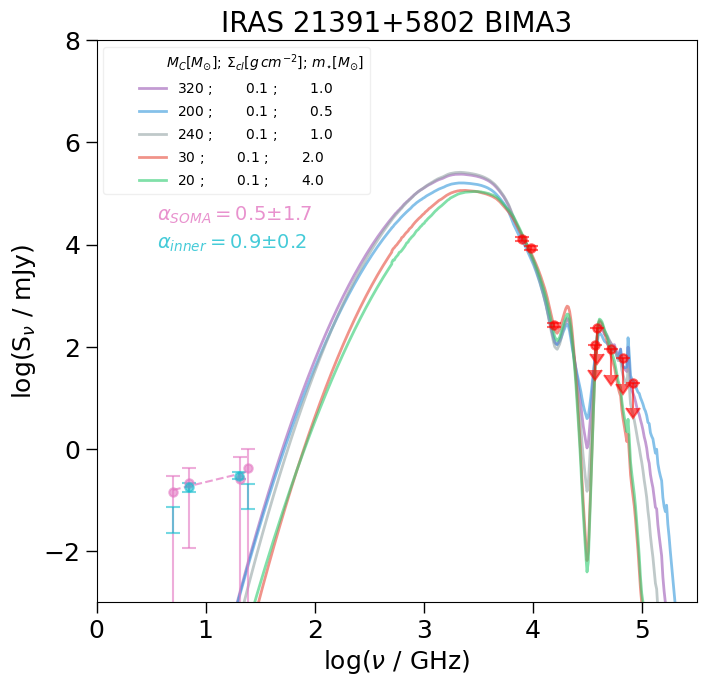}\quad\includegraphics[width=0.32\linewidth]{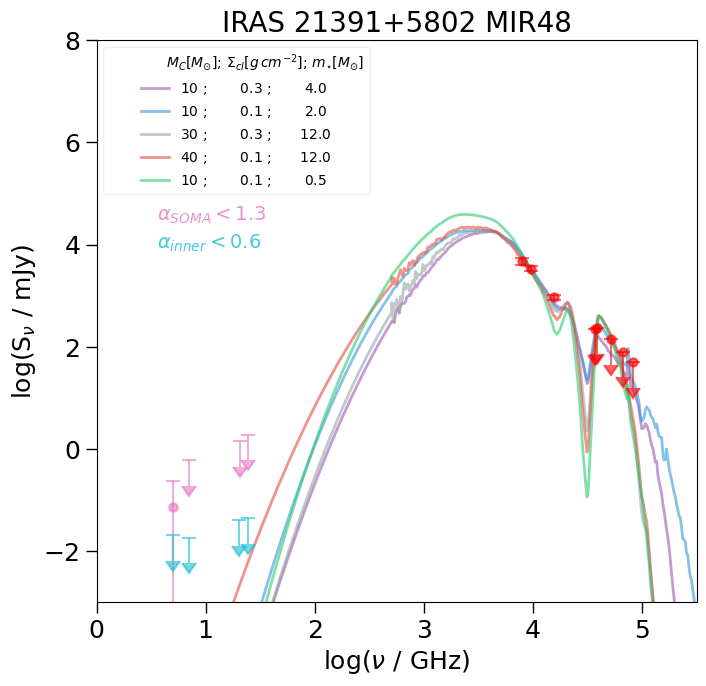}\\
\caption{(Continued)}
\end{center}
\end{figure*}

\section{Analysis and Discussion}\label{analysis}

To better understand the nature of the sources from radio observations, we analyze the morphology, multiplicity, and spectral indices of the SOMA~III sample. Unlike the sources in \citetalias{Rosero_2019} and \citetalias{Sequeira-Murillo_2025}, our targets are mostly located within 1~kpc, except for S235 and IRAS~22172$+$5549, and generally exhibit low radio flux densities. Among the twelve sources studied, only NGC~2071 shows flux densities above 3~mJy.

Following the approach of Section~4.2 in \citetalias{Sequeira-Murillo_2025}, we use VLA observations to assess the nature of each source based on morphology, spectral index and their association with any tracers of star formation. For objects with previous studies, we draw on earlier characterizations, comparing them with our measurements (see \S\ref{Morph} for individual analyzes). This comparison refines our understanding of the underlying radio emission mechanisms in each case.

\subsection{Multiplicity and source characterization}\label{multiplicity_detections}

Across the 12 regions analyzed, we detected 18 sources in total. Multiplicity is present in only three cases: NGC~2071 (eight sources), IRAS~21391$+$5802~BIMA2 (three sources), and IRAS~21391$+$5802~BIMA3 (two sources). Compared to \citetalias{Rosero_2019, Sequeira-Murillo_2025}, the sample of intermediate-mass protostars analyzed in this work shows a significantly lower multiplicity, with only 3 of 12 regions (25\%) showing multiple sources, while the earlier studies reported multiple detections in 14 of 17 regions ($\sim$82\%).

Of the 18 detections, four are entirely new, with no counterparts at any wavelength, and four are new radio detections of sources previously identified at other wavelengths. We also resolve IRAS~21391$+$5802~BIMA2 into three distinct components, each with a millimeter counterpart, which we consider new radio detections.

We identify four regions—IRAS~22172$+$5549~MIR1, MIR2, and MIR3, and IRAS~21391$+$5802~MIR48, where no sources are detected at the $5\sigma$ level. Following the methodology of \citetalias{Sequeira-Murillo_2025}, and in order to determine if we are missing any UCHII regions, we estimate the physical properties implied by the 5~GHz sensitivity limit (rms) using equations~(1) and (3) from \citet{1994ApJS...91..659K} and equation~(A.2.3) from \citet{Panagia1978}. 

These calculations assume spherical symmetry and optically thin emission from a uniform-density plasma with \( T_e = 10^4\,\mathrm{K} \). The results are presented in Table~\ref{tab:ZAMs_Calc}, where column~1 lists the region name, column~2 gives the logarithm of the Lyman continuum flux (\( \log N'_{\mathrm{Ly}} \)) required for ionization, and column~3 shows the corresponding spectral type from Table~II of \citet{1973AJ.....78..929P}, assuming a single ZAMS star is responsible for the ionization. For cases where we obtain B4 spectral types, we use an extrapolation from the values in Table~II of \citet{1973AJ.....78..929P}, since the minimum value of the logarithm of the Lyman continuum flux (\( \log N'_{\mathrm{Ly}} \)) in that table is 43.69 (for a B3 spectral type). The distances used in these calculations are taken from Table~\ref{tab:SOMA_Sources}. Column~4 lists the stellar mass associated with each spectral type, from \citet{2013ApJS..208....9P}.

\begin{deluxetable}{cccc}
    \tabletypesize{\footnotesize}
    \tablewidth{0pt}
    \tablecaption{Parameters from Radio Continuum at the sensitivity limit at 5 GHz. \label{tab:ZAMs_Calc}}
    \tablehead{
    \colhead{Region} & \colhead{$\mathrm{log\:N'_{Ly}}$\small\textsuperscript{a}} & \colhead{Sp\small\textsuperscript{b}} & \colhead{$M_{\odot}$\small\textsuperscript{c}}}
    \startdata
    S235                    & 42.93 & B3 & 5.4 \\
    IRAS 22198+6336         & 42.14 & B4 & 5.1 \\ 
    NGC 2071                & 41.99 & B4 & 5.1 \\
    Cepheus E               & 42.04 & B4 & 5.1 \\
    L1206 A                 & 42.25 & B4 & 5.1 \\
    L1206 B                 & 42.23 & B4 & 5.1 \\
    IRAS 22172+5549 MIR1    & 43.34 & B3 & 5.4 \\
    IRAS 22172+5549 MIR2    & 43.33 & B3 & 5.4 \\
    IRAS 22172+5549 MIR3    & 43.37 & B3 & 5.4 \\
    IRAS 21391+5802 BIMA2   & 42.14 & B4 & 5.1 \\
    IRAS 21391+5802 BIMA3   & 42.20 & B4 & 5.1 \\
    IRAS 21391+5802 MIR48   & 42.22 & B4 & 5.1 \\
    \enddata
    \tablecomments{\\
    \textsuperscript{a} Values calculated using equations from \cite{1994ApJS...91..659K, Panagia1978}.\\
    \textsuperscript{b} Spectral type estimation made from extrapolating the values in table II in \cite{1973AJ.....78..929P}.\\
    \textsuperscript{c} The estimate of the mass was taken from \cite{2013ApJS..208....9P}}
\end{deluxetable}

From Table \ref{tab:ZAMs_Calc} we can derive an estimation of the lowest mass ZAMS star that we can detect with the sensitivity of our observations. Therefore, we use this estimate as a completeness parameter on the multiplicity at the different regions, especially for the regions where we do not report any detection.

\subsection{Nature of the radio emission}

By leveraging previous studies on the sources in this sample, as well as the results from our radio observations, we have assessed the nature of each individual source (see \S\ref{Morph} for more details). From the 18 sources we report in this paper, it is important to highlight that 6 are radio jets given their morphology and the estimated spectral index. IRAS 22198+6336 and the sources IRS1 and IRS3 in NGC 2071, are examples of previously studied ionized jets. All three exhibit a jet-like morphology and a spectral index in the range for typical ionized jets associated with YSOs \citep{Reynolds1986Continuum, Anglada1998SpectralSources, Tanaka_2016}. Additionally, we suggest that sources L1206 A, L1206 B and IRAS 21391+5802 BIMA3 VLA-3 could be ionized jets given their morphology, weak centimeter emission and spectral index values consisted with ionized jets associated with YSOs. However, further studies are needed for a definitive interpretation of their nature.

Similarly, we have NGC 2071 VLA1, which was previously classified as a younger and more embedded YSO than the other sources associated in the region. From our results, and based on previous studies,  we agree with the scenario that this source has a radio jet given the jet-like morphology, the spectral index and two thermal components (VLA 1-B1 and VLA 1-B2) that are aligned in the direction of the jet. More observations are needed to fully resolve the sources and confirm the nature of this proposed radio jet.

Regarding the remaining sources in NGC 2071, IRS~1Eb exhibits a relatively flat spectral index, indicating thermal emission. Its proximity to the radio jet IRS~1 suggests that this source might be part of the jet emission. The two new radio detections, HOPS-361-E-cm and IRS~8~cm, have mm counterparts, and their compact morphology and positive spectral indices suggest that they are likely individual YSOs within the NGC~2071 cluster. Similarly, the three sources toward IRAS~21391+5802~BIMA2 (VLA~1, 2, and 3) show evidence of being individual YSOs. In this case, the most likely scenario is that source VLA~2 is the most massive and is driving the molecular outflow. Source S235 was classified as a Herbig Be star by \citet{2009MNRAS.399..778B}. Our results, which show a compact morphology and a spectral index consistent with thermal emission, support this conclusion regarding the nature of the source. Table~\ref{tab:nature} summarizes the detections and the corresponding assessments of the nature of each source already discussed in the individual subsections on \S\ref{Morph}.

\begin{deluxetable}{ccc}
    \tabletypesize{\footnotesize}
    \tablewidth{0pt}
    \tablecaption{Summary of the nature of the sources reported in the  \citetalias{Liu_2020} sample. \label{tab:nature}}
    \tablehead{\colhead{Region} & \colhead{Source} & \colhead{Nature}}
    \startdata
    S235                    & VLA-2 & Herbig Be star \\
    IRAS 22198+6336         & VLA2 & Ionized Jet \\ 
    NGC 2071                & IRS 1 & Jet Candidate \\ 
    NGC 2071                & VLA 1 & Jet Candidate \\ 
    NGC 2071                & VLA 1-B1 & Jet Knot (Candidate) \\ 
    NGC 2071                & VLA 1-B2 & Jet Knot (Candidate) \\ 
    NGC 2071                & IRS 1Eb & Jet Knot (Candidate) \\ 
    NGC 2071                & IRS 8-cm & YSOs \\ 
    NGC 2071                & HOPS-361-E-cm & YSOs \\ 
    NGC 2071                & IRS 3 & Jet Candidate \\ 
    Cepheus E               & VLA1 & Undetermined \\ 
    L1206 A                 & A & Jet Candidate \\ 
    L1206 B                 & B & Jet Candidate or HC HII Region\\ 
    IRAS 21391+5802 BIMA2   & VLA 2A & YSOs \\ 
    IRAS 21391+5802 BIMA2   & VLA 2B & YSOs \\
    IRAS 21391+5802 BIMA2   & VLA 2C & YSOs \\
    IRAS 21391+5802 BIMA3   & VLA 3 & Jet Candidate or YSOs\\ 
    IRAS 21391+5802 BIMA3   & VLA 3B & Undetermined \\ 
    \enddata
\end{deluxetable}

In \citetalias{Rosero_2019} and \citetalias{Sequeira-Murillo_2025}, we reported different evolutionary stages and nature of the sample, which included radio jets, UC HII regions, and variable stars. In general, 9 of the 17 regions previously analyzed presented characteristics that indicated a radio jet scenario. In this work, we present a slightly lower distribution with only 4 (IRAS 22198+6336, NGC 2071, L1206 A and L1206 B) of the 12 regions exhibiting characteristics consistent with radio jets, since 3 (NGC 2071: IRS1, IRS3, and VLA1) of the 6 radio jet scenarios that we report are from sources associated with region NGC 2071. Given our sensitivity limit, we also identify 4 regions where we do not report a 5$\sigma$ detection (IRAS~22172$+$5549: MIR1, MIR2, and MIR3, and IRAS~21391$+$5802~MIR48), in contrast to the first 17 regions analyzed, where only one, IRAS 16562-3959 N from \citetalias{Sequeira-Murillo_2025}, was a non-detections.

\subsection{Radio - bolometric luminosity of IM protostars}

\begin{figure*}[ht!]
\figurenum{7}
\begin{center}
\includegraphics[width=0.488\linewidth]{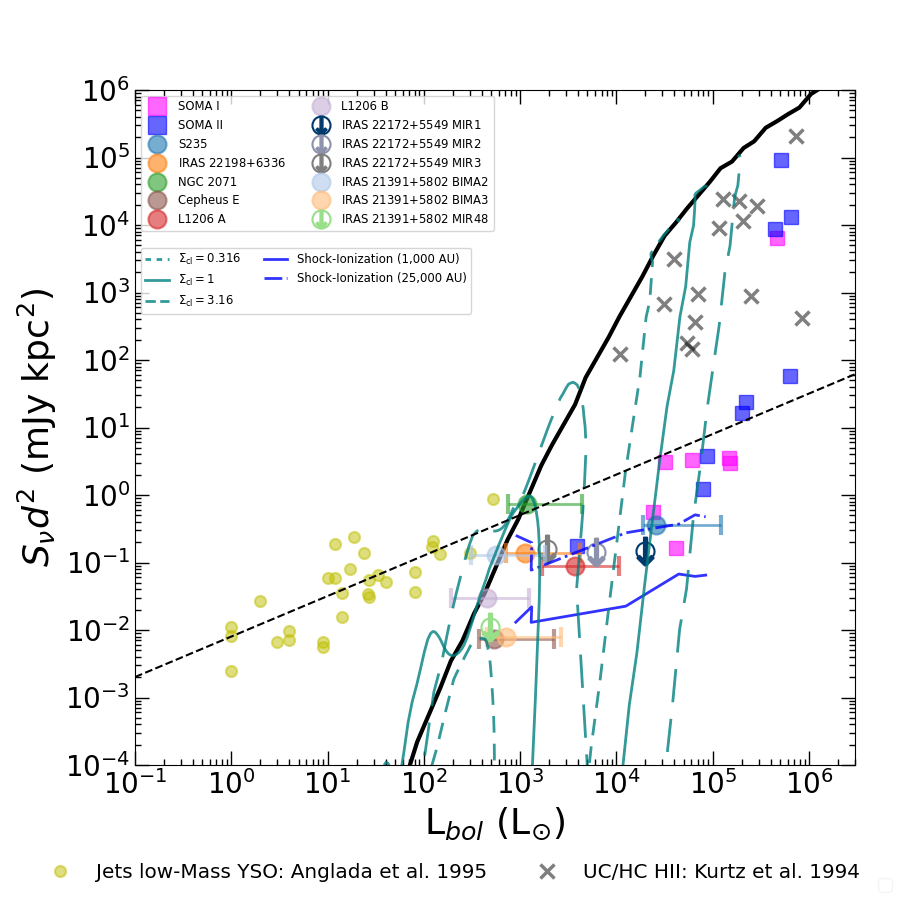} \quad \includegraphics[width=0.488\linewidth]{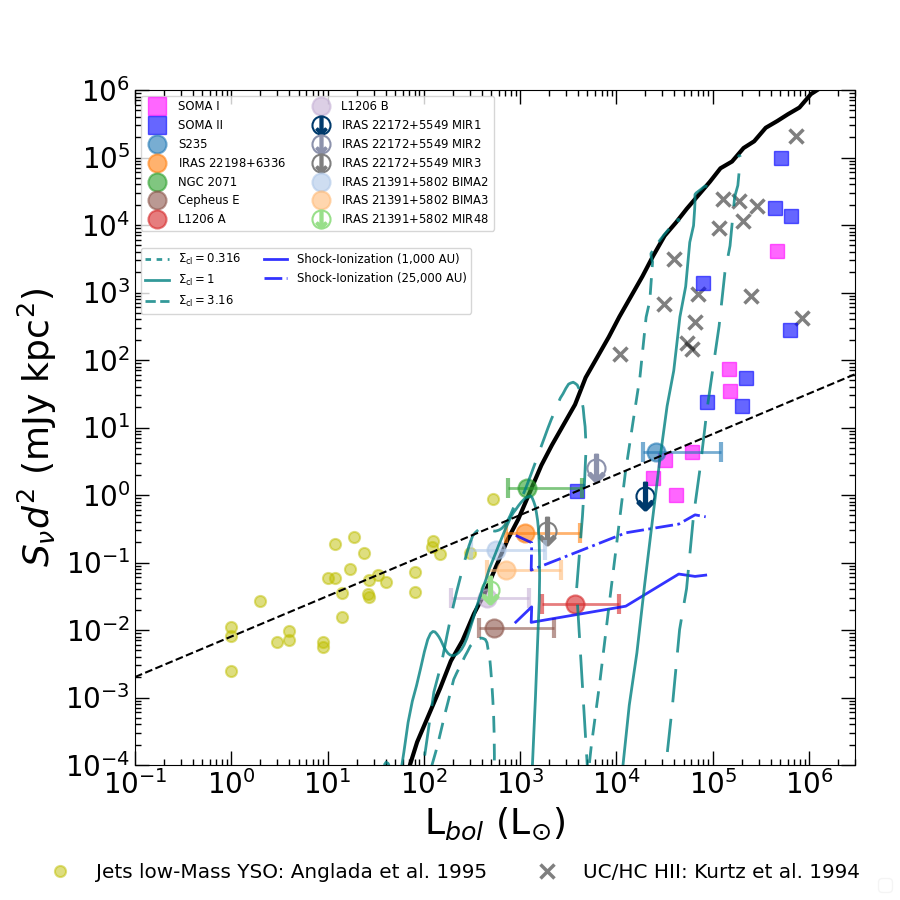}\\
\caption{Radio luminosity at 5 GHz for the Inner scale (left) and for the SOMA scale (right) as a function of the bolometric luminosity of for 29 SOMA sources from SOMA Radio I and II (squares) and III (circles) (see legend). In each panel we also show lower-mass YSOs from \cite{1995RMxAC...1...67A} (small yellow circles). The dashed line shows a power-law fit to these lower-mass YSOs \citep{2015aska.confE.121A}: $(S_{\nu}d^{2}/[{\rm mJy\:kpc^2}] ) = 8 \times 10^{-3}(L_{\rm bol}/L_{\odot})^{0.6}$. The $\times$ symbols are HC and UC HII regions from \cite{1994ApJS...91..659K}. The black solid line shows the radio emission expected from optically thin HII regions powered by ZAMS stars \citep{1984ApJ...283..165T}. The light blue lines show models for HII regions powered by TCA model protostars \citep{Tanaka_2016} (with $M_{c} = 60\:M_{\odot}$ and $\Sigma_{\rm cl} =$ 0.316, 1 and 3.16~g~cm$^{-2}$, as labeled). Note that these models, which do account for radiative transfer effects of the free-free emission, assume that all of the ionizing photons are reprocessed by the HII region, i.e., with zero escape fraction. The dark blue lines show radio emission from shock ionization in simulations of a fiducial TCA protostar, i.e., forming from a $60\:M_\odot$ core in a $1\:{\rm g\:cm}^{-2}$ clump environment \citep{2024ApJ...967..145G}, with the solid and dot-dashed lines showing emission from within 1,000 and 25,000~au of the protostar, respectively. \label{fig:Anglada_plot}}
\end{center}
\end{figure*}

Following the results from \citetalias{Sequeira-Murillo_2025}, in Figure \ref{fig:Anglada_plot} we present the radio luminosity at 5 GHz from the \textit{Inner} (left panel) and \textit{SOMA} (right panel) scales versus the bolometric luminosity. For the bolometric luminosity of SOMA sources (detections from SOMA I and II samples are magenta and blue squares, respectively) we report the average of the ``good'' models from Table C1 of \citetalias{Telkamp_2025} and the error bar for the \citetalias{Liu_2020} sources corresponds to the dispersion of these good models. We also show data for lower-mass YSOs associated with ionized jets from \cite{1995RMxAC...1...67A} (small yellow dots). We scaled their fluxes from 3.6~cm, using a factor of 0.74, assuming that these sources have a spectral index $\alpha =$ 0.6, which is the expected value of ionized jets. A power-law fit to these data of $(S_{\nu}d^{2}/[{\rm mJy\:kpc^2}] ) = 8 \times 10^{-3}(L_{\rm bol}/L_{\odot})^{0.6}$ is shown with a dashed line. UC/HC HII regions from \cite{1994ApJS...91..659K} are represented with $\times$ symbols. Note, the bolometric luminosities of the low-mass YSOs and the UC/HC HII regions are not measured in exactly the same way as the SED-fitting method of the SOMA sources, for which we are reporting intrinsic bolometric luminosities of the fitted SED models. Similar to \citetalias{Sequeira-Murillo_2025}, a plot showing the results based on isotropic bolometric luminosities is presented in Appendix~\ref{sec:appA}.

Several theoretical models for the radio luminosity of massive protostars are plotted in Figure~\ref{fig:Anglada_plot}. The black dotted line is the radio emission expected from an optically thin HII region, given the Lyman continuum luminosity of a single zero age main sequence (ZAMS) star at a given bolometric luminosity \citep{1984ApJ...283..165T}. The light blue lines correspond to the models from \citetalias{Tanaka_2016}, of the expected radio emission that arises from photoionization of a massive protostar forming via Turbulent Core Accretion (TCA), also for optically thin conditions at 5~GHz. These models have an initial core mass of $M_{c} = 60\:M_{\odot}$ and the three cases correspond to clump environment mass surface densities of $\Sigma_{\rm cl} = 0.316, 1$ and 3.16 g cm$^{-2}$. The lower mass surface densities correspond to lower accretion rates, for which protostellar contraction towards the ZAMS occurs sooner, i.e., at lower masses and lower values of $L_{\rm bol}$. Note, these models do not include any contribution from shock ionization. Similar plots with an additional sets of theoretical models based on optically thin radio emission from the ionizing photon output of the TCA protostellar evolution models of \citet{Zhang_2018} are shown in Appendix~\ref{sec:appB}.

\citet{2024ApJ...967..145G} developed a model for estimating the radio emission from free-free emission from ionized gas produced by collisional ionization in shocks for a massive protostar forming from a $60\:M_\odot$ core in a $1\:{\rm g\:cm}^{-2}$ clump environment. This radio emission is also shown in Figure~\ref{fig:Anglada_plot} with dark blue lines. We note that the contribution from shock ionization is quite spatially extended, with the emission from a 25,000~au radius aperture (more closely matching the SOMA scale) being about a factor of 10 higher than that from a 1,000~au aperture (more closely matching the inner scale).

As in \citetalias{Sequeira-Murillo_2025}, it is important to note that, at the SOMA scale, we may miss radio flux due to interferometric filtering. The SOMA scale (right plot) captures all the emission in the region, including contaminating contributions from jet knots and other nearby sources, further enhancing the radio luminosity. The artificial flux added by these sources, along with the differing spatial scales relative to the inner scale, makes the SOMA scale less reliable for assessing the intrinsic nature of individual sources. However, the SOMA scale still holds importance since it allows direct comparison with theoretical models with bigger scales that take into account jet knots and other multiple stellar sources in proto-clusters.

The differences between the results shown in the two plots of Figure~\ref{fig:Anglada_plot} (as well as Figures~\ref{fig:Anglada_plot2}, \ref{fig:YichenModels}, and \ref{fig:YichenModels2}), particularly for the sources we classified as having jet-like morphologies (IRAS 22198+6336, NGC 2071, L1206 A and L1206 B), arise from the different scales used to determine the radio luminosity \citep{2021A&A...645A..29K}. 

With the addition of the \citetalias{Liu_2020} sample, which is composed of intermediate-mass protostars, we now present a total of 29 SOMA protostars that expand the range of the radio versus bolometric luminosity relation. Our data points now cover values from $L_{\rm bol}\sim 10^2$ to $10^6\:L_\odot$. This large range of observations is important for studying the relationship between protostellar models and the observations.

Comparing the distribution of the \citetalias{Liu_2020} radio sources with the models, we find that our sample is considerably less luminous than the \citetalias{Bruizer_2017} and \citetalias{Liu_2019} sources. As expected, the sources analyzed in this work (intermediate-mass protostars) are positioned between the lower-mass YSOs from \citet{1995RMxAC...1...67A} and the more massive sources from our current sample, except for S235, which is closer to the massive sources but is still at the lower end of the high-mass stars in our sample. For a more in-depth analysis of the \citetalias{Bruizer_2017} and \citetalias{Liu_2019} sources, please refer to Sections 6 and 4 of \citetalias{Rosero_2019} and \citetalias{Sequeira-Murillo_2025}, respectively. It is important to note that most of our intermediate-mass protostars have lower radio luminosities than would be expected from an extrapolation of the power law relation defined by the low-mass protostars reported by \citet{2015aska.confE.121A}, i.e., most of our points fall below the dashed line at both the inner and the SOMA scale.

By expanding the SOMA protostar sample to lower luminosity sources, we see that the behavior described in \citetalias{Sequeira-Murillo_2025} still holds, namely that the sources exhibit relatively faint radio luminosities compared to the predictions of ZAMS models and of the power law extrapolation of radio luminosity from low-mass protostars. However, they overlap with the parameter space covered by TCA protostellar models. In particular, the radio faintness of IM protostars is consistent with predictions of protostellar evolution models with $\Sigma_{\rm cl}\lesssim 1\:{\rm g\:cm}^{-2}$: in the range of luminosities from $\sim 3\times10^2\:L_\odot$ to $\sim 10^4\:L_\odot$ the protostars produce very little photoionization since they are in an expanded, relatively cool state. Additionally, we see that the shock ionization model of \citet{2024ApJ...967..145G}, which falls below the extrapolated lower-mass YSO relation, is generally consistent with most of the sources in the SOMA III sample. Thus shock ionization is still expected to dominate in this regime, but the models of \citet{2024ApJ...967..145G} indicate this emission may be relatively weak, likely because outflow speeds are also reduced given the expanded protostellar radii.

The addition of the intermediate-mass protostars strengthens the suggestion that the rise in ionizing luminosity may occur only at relatively higher protostellar masses and luminosities. Moreover, we find that the radio luminosity measured on the SOMA scale rises steeply, by about three dex, as the bolometric luminosity increases by one dex. This rise is smaller than that found in the more massive protostars (SOMA II), but is still considered a steep increase, consistent with the basic expectations of TCA massive protostar models for the onset of Kelvin-Helmholz contraction to the ZAMS. However, more detailed testing of the TCA models via the SOMA Radio data will require independent estimates of other protostellar properties, such as protostellar masses and accretion rates.

\section{Summary and Conclusions}\label{summary}

Following the work done in \citetalias{Rosero_2019} and \citetalias{Sequeira-Murillo_2025}, we present results from radio continuum VLA follow-up observations of the \citetalias{Liu_2020} sample, which is primarily composed of intermediate-mass protostars. These observations allow us to precisely locate the protostars, investigate multiplicity, and identify ionized gas emission. The latter provides insight into outflow morphology, particularly through the orientation of radio jet axes, and serves as a potential evolutionary indicator for massive protostars, especially as they approach the zero-age main sequence, when photoionization is expected to play an increasingly dominant role. With the addition of the 12 sources in the 7 target region, our SOMA Radio sample now is composed of 29 sources, covering values from $L_{\rm bol}\sim 10^2$ to $10^6\:L_\odot$. Below, we summarize our main findings:

\begin{itemize}
    \item We detected 18 sources in the 12 regions that are part of the \citetalias{Liu_2020} sample, but we find multiplicity (e.g. two or more detections 5$\sigma$ within the SOMA scale) in only 3 of these 12 regions (NGC~2071, IRAS~21391+5802 BIMA2, and IRAS~21391+5802 BIMA3). Moreover, we present 4 regions where we do not report a detection above our 5$\sigma$ threshold (IRAS 22172+5549: MIR1, MIR2, and MIR3, and IRAS 21391+5802 MIR48).
    \item Similar to \citetalias{Rosero_2019} and \citetalias{Sequeira-Murillo_2025}, we report a variety of nature and evolutionary stages for the sources in the sample, with 7 out of the 18 sources being classified as ionized jet candidates given their morphologies and spectral indices.
    \item We find the \citetalias{Liu_2020} sample to be considerably less luminous than the \citetalias{Bruizer_2017} and \citetalias{Liu_2019} sources, and, as expected, these IM protostars are positioned between the lower-mass YSOs from \citet{1995RMxAC...1...67A} and the more massive sources from the \citetalias{Bruizer_2017} and \citetalias{Liu_2019} sample. 
    \item The IM protostars present lower radio luminosities than would be expected from a power law extrapolation from the low-mass protostars reported by \citet{2015aska.confE.121A}, which is consistent with the theoretically expected low photoionization rate. The shock ionization model of \citet{2024ApJ...967..145G}, which also falls below the extrapolated lower-mass YSO relation, is generally consistent with most of the sources in the \citetalias{Liu_2020} sample.
    \item The broad range of observations enables us to directly connect protostellar models with observations, providing new constraints on theories of massive protostar ionization and offering deeper insight into how ionization is tied to protostellar structure and evolutionary stage.
\end{itemize}

\begin{acknowledgments}
The authors would like to thank Prof. Carlos Carrasco-González for providing the additional VLA images of source NGC 2071 originally presented in \citet{Carrasco2012}. F. S. M. acknowledges support from the NRAO NINE (National and International Non-traditional Exchange) program. V. R. acknowledges support from NSF grant AST–2206437 to the Space Science Inst. J.C.T. acknowledges support from NSF grant AST–2206437 and ERC Advanced Grant 788829 (MSTAR). R.F. acknowledges support from the grants PID2023-146295NB-I00, and from the Severo Ochoa grant CEX2021-001131-S funded by MCIN/AEI/ 10.13039/501100011033 and by ``European Union NextGenerationEU/PRTR''.The National Radio Astronomy Observatory and Green Bank Observatory are facilities of the U.S. National Science Foundation operated under cooperative agreement by Associated Universities, Inc. 
 
\end{acknowledgments}

\facility{VLA} 

\software{CASA \citep{2022PASP..134k4501C}, Astropy \citep{astropy:2013, astropy:2018, astropy:2022}, APLpy \citep{aplpy2012, aplpy2019}}

\section{ORCID iDs}

\raggedright
F. Sequeira-Murillo \href{https://orcid.org/0000-0001-8169-1437}{https://orcid.org/0000-0001-8169-1437} \\
V. Rosero \href{https://orcid.org/0000-0001-8596-1756}{https://orcid.org/0000-0001-8596-1756} \\
J. Marvil \href{https://orcid.org/0000-0003-1111-8066}{https://orcid.org/0000-0003-1111-8066} \\
J. C. Tan \href{https://orcid.org/0000-0002-3389-9142}{https://orcid.org/0000-0002-3389-9142} \\
R. Fedriani \href{https://orcid.org/0000-0003-4040-4934}{https://orcid.org/0000-0003-4040-4934} \\
Y. Zhang \href{https://orcid.org/0000-0001-7511-0034}{https://orcid.org/0000-0001-7511-0034} \\
P. Gorai \href{https://orcid.org/0000-0003-1602-6849}{https://orcid.org/0000-0003-1602-6849} \\
J. M. De Buizer \href{https://orcid.org/0000-0001-7378-4430}{https://orcid.org/0000-0001-7378-4430} \\
M. T. Beltran \href{https://orcid.org/0000-0003-3315-5626}{https://orcid.org/0000-0003-3315-5626} \\

\appendix

\section{Full radio continuum parameters of the sources with multiplicity}

Table \ref{tab:Parameters_Radio_multiplicity} presents additional parameters related to the radio continuum detections listed in Tables \ref{tab:Multiplicity_NGC2071} and \ref{tab:Multiplicity_I21391}. We report the source sizes, as well as the flux densities and their associated uncertainties at each frequency.

\renewcommand{\thetable}{A\arabic{table}}
\setcounter{table}{0}
\startlongtable
\begin{deluxetable*}{cccccccccc}
    \tabletypesize{\footnotesize}
    \tablewidth{0pt}
    \tablecaption{Parameters from Radio Continuum of the sources with multiplicity in NGC 2071, IRAS 21391 BIMA 2 and BIMA 3. \label{tab:Parameters_Radio_multiplicity}}
    \tablehead{
    \colhead{Source} & \colhead{Detection} & \colhead{Scale} & \colhead{R.A} & \colhead{Decl.} & \colhead{$S_{5.0}$ $_{GHz}$} & \colhead{$S_{7.0}$ $_{GHz}$} & \colhead{$S_{20.6}$ $_{GHz}$} & \colhead{$S_{24.5}$ $_{GHz}$} & \colhead{Spectral} \\
    \colhead{} & \colhead{} & \colhead{\textit{$a\;(\arcsec)$ $\times$ $b\;(\arcsec)$}} & \colhead{(J2000)} & \colhead{(J2000)} & \colhead{(mJy)} & \colhead{(mJy)} & \colhead{(mJy)} & \colhead{(mJy)} & \colhead{Index}}
    \startdata
    NGC 2071 & IRS 1          & 1.34 $\times$ 0.75\small\textsuperscript{*} & 05:47:04.78 & +00.21.42.93 & 4.88 (0.49) & 4.51 (0.45) & 6.11 (0.61) & 6.81 (0.68) & 0.2 (0.08) \\
             & IRS 1Eb        & 1.91 $\times$ 0.63 & 05:47:05.04 & +00.21.42.82 & 0.30 (0.13) & 0.19 (0.07) & 0.27 (0.12) & 0.26 (0.11) & 0.1 (0.5)  \\
             & VLA 1          & 0.21 $\times$ 0.16 & 05:47:04.75 & +00.21.45.43 & 0.52 (0.07) & 0.44 (0.05) & 0.82 (0.13) & 0.82 (0.12) & 0.4 (0.11) \\
             & VLA 1-B1       & 0.22 $\times$ 0.14 & 05:47:04.68 & +00.21.45.07 & 0.48 (0.05) & 0.31 (0.04) & 0.44 (0.05) & 0.51 (0.06) & 0.1 (0.10) \\
             & VLA 1-B2       & 0.65 $\times$ 0.37 & 05:47:04.65 & +00.21.45.20 & 0.45 (0.05) & 0.11 (0.03) & 0.29 (0.04) & 0.26 (0.06) & 0.1 (0.19) \\
             & HOPS-361-E-cm  & 0.22 $\times$ 0.16 & 05:47:04.31 & +00.21.38.05 & $<$0.04     & $<$0.03     & 0.08 (0.01) & 0.14 (0.02) & $<$1.1      \\
             & IRS 8-cm       & 0.15 $\times$ 0.14 & 05:47:04.62 & +00.21.41.32 & $<$0.04     & $<$0.03     & 0.03 (0.02) & 0.14 (0.03) & $<$0.7      \\
             & IRS 3          & 0.50 $\times$ 0.12 & 05:47:04.63 & +00.21.47.84 & 1.28 (0.13) & 1.24 (0.16) & 2.02 (0.45) & 2.44 (0.25) & 0.4 (0.10)   \\
    BIMA 2  & VLA 2A          & 0.32 $\times$ 0.25\small\textsuperscript{*} & 21:40:41.86 & +58.16.11.95 & 0.23 (0.03) & 0.26 (0.03) & 0.64 (0.07) & 0.58 (0.07) & 0.7 (0.09) \\
            & VLA 2B          & 0.35 $\times$ 0.22 & 21:40:41.73 & +58.16.12.81 & $<$0.02     & 0.02 (0.01) & 0.04 (0.01) & 0.09 (0.01) & 0.9 (0.66) \\
            & VLA 2C          & 0.29 $\times$ 0.19\small\textsuperscript{*} & 21:40:41.73 & +58.16.14.34 & 0.02 (0.01) & 0.03 (0.01) & 0.11 (0.01) & 0.11 (0.02) & 1.2 (0.18) \\
    BIMA 3  & VLA 3           & 0.53 $\times$ 0.33 & 21:40:42.84 & +58.16.01.36 & 0.01 (0.02) & 0.18 (0.03) & 0.30 (0.04) & 0.20 (0.07) & 0.9 (0.16)  \\
            & VLA 3B          & 0.41 $\times$ 0.13 & 21:40:43.46 & +58.15.59.49 & 0.04 (0.02) & 0.05 (0.01) & 0.04 (0.01) & 0.07 (0.02) & 0.0 (0.20) \\
    \enddata
    \tablecomments{ Units of R.A. are hours, minutes, and seconds. Units of decl. are degrees, arcminutes, and arcseconds. \\
    The scale corresponds to image component (deconvolved with beam) size from the task imfit, of major axis \textit{a} and minor axis \textit{b}. Except for the scales that show the ($^*$), in this case the scale corresponds to an ellipse of major axis \textit{a} and minor axis \textit{b}.}
\end{deluxetable*}

\section{Radio luminosity versus isotropic bolometric luminosity}\label{sec:appA}

\begin{figure*}[ht!]
\figurenum{B1}
\begin{center}
\includegraphics[width=0.488\linewidth]{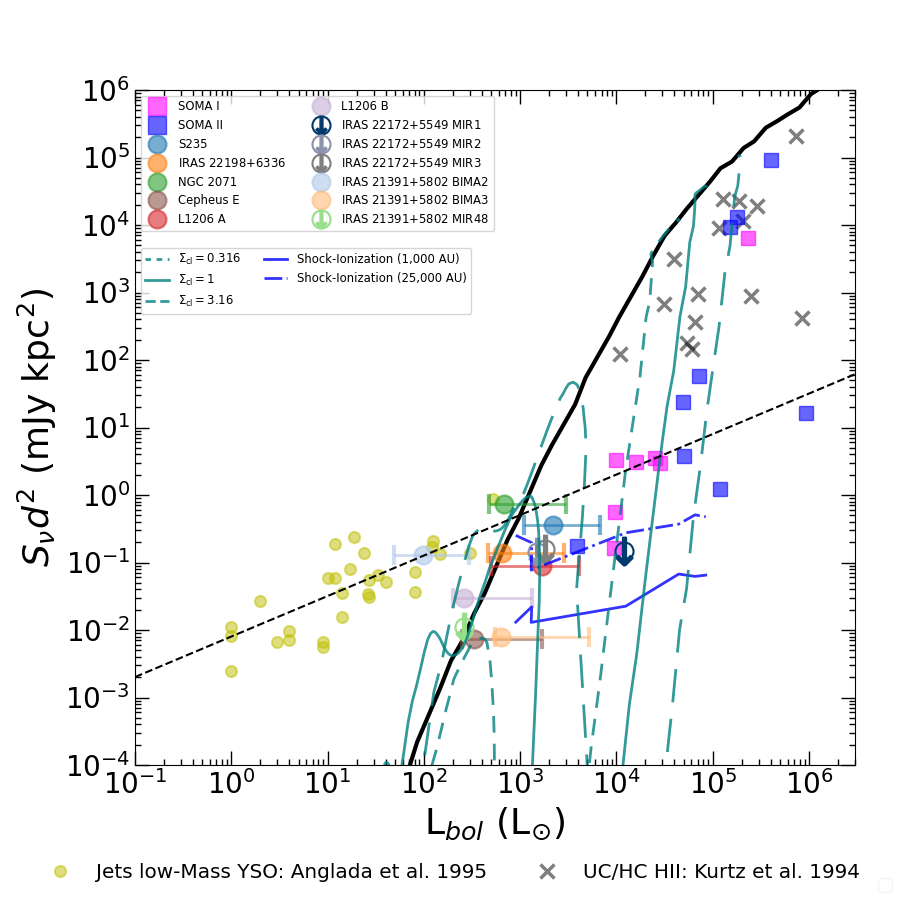}\quad \includegraphics[width=0.488\linewidth]{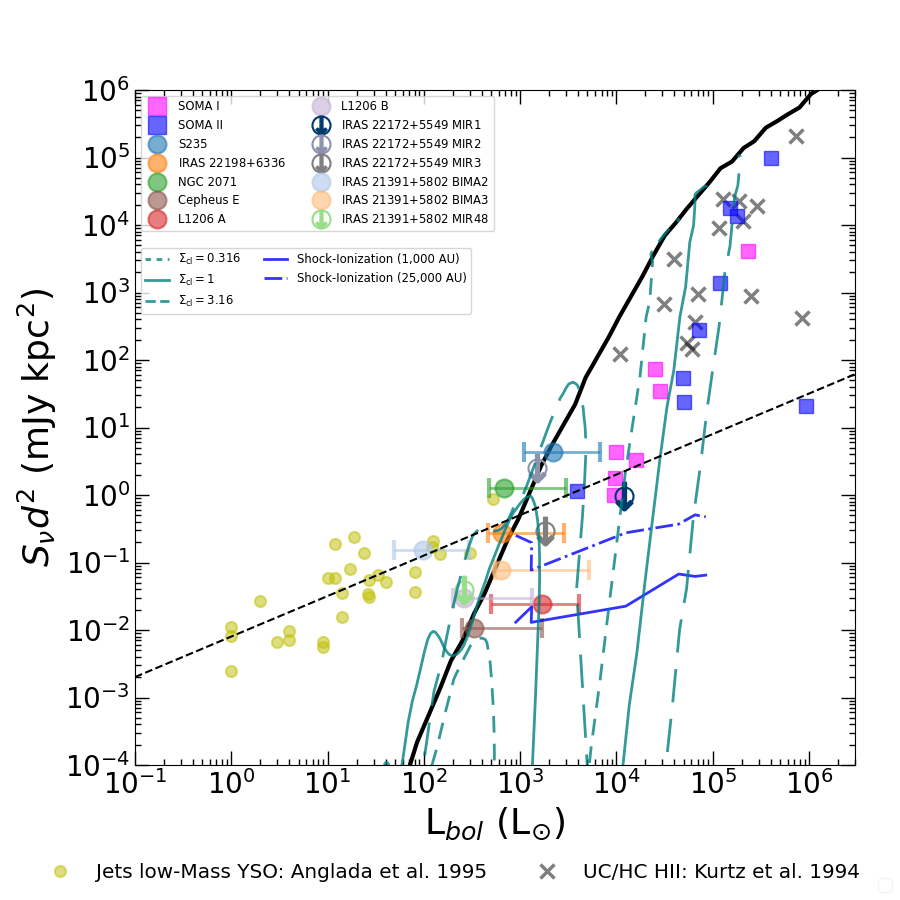}
\caption{As Fig.~\ref{fig:Anglada_plot}, but now plotting $L_{\rm bol,iso}$ for the SOMA sources (left: inner scale, right: SOMA scale).
\label{fig:Anglada_plot2}}
\end{center}
\end{figure*}

Figure \ref{fig:Anglada_plot2} shows a similar plot to Figure \ref{fig:Anglada_plot}, but in this case we used the average of the good models of the isotropic bolometric luminosity (Table C.1 of \citetalias{Telkamp_2025}). Everything else is the same as Figure \ref{fig:Anglada_plot}.

\section{Evolutionary tracks from radio luminosities}\label{sec:appB}

\begin{figure*}[ht!]
\figurenum{C1}
\begin{center}
\includegraphics[width=0.32\linewidth]{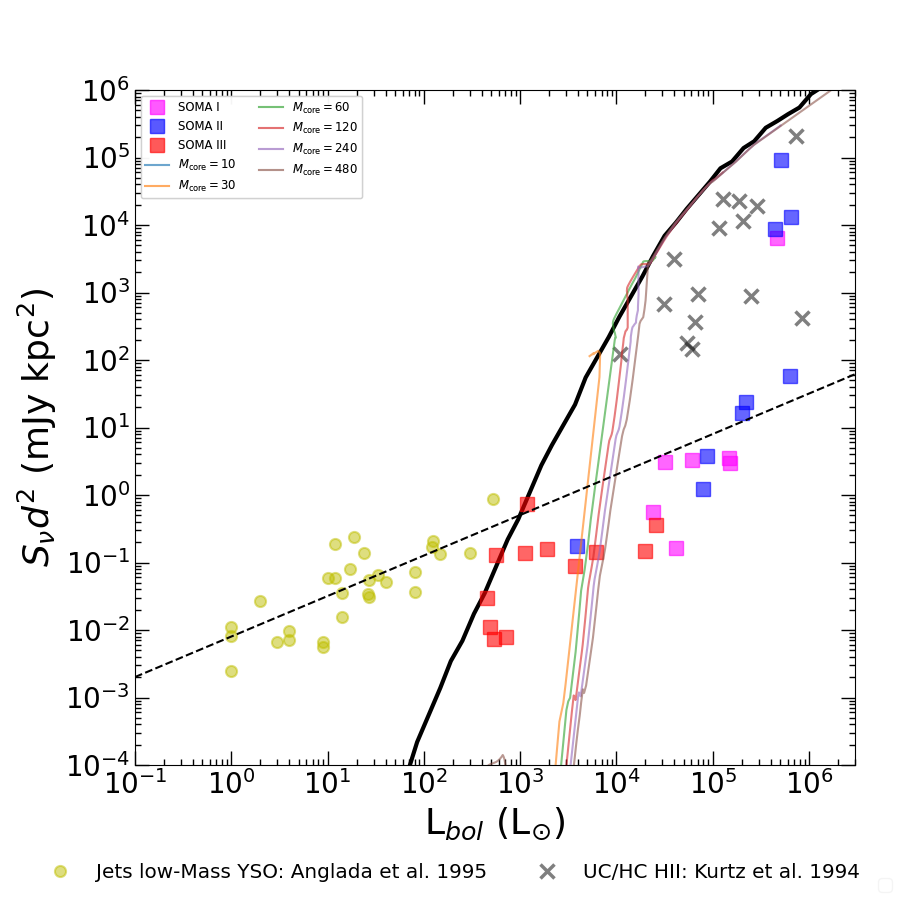}\quad\includegraphics[width=0.32\linewidth]{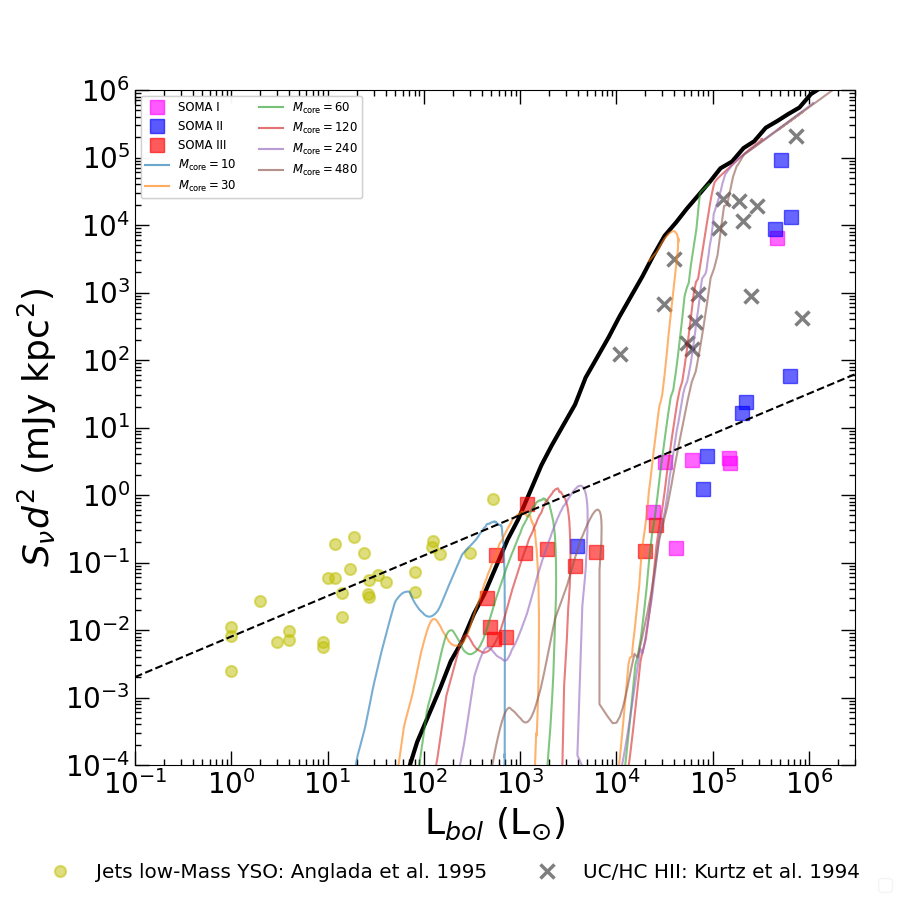}\quad\includegraphics[width=0.32\linewidth]{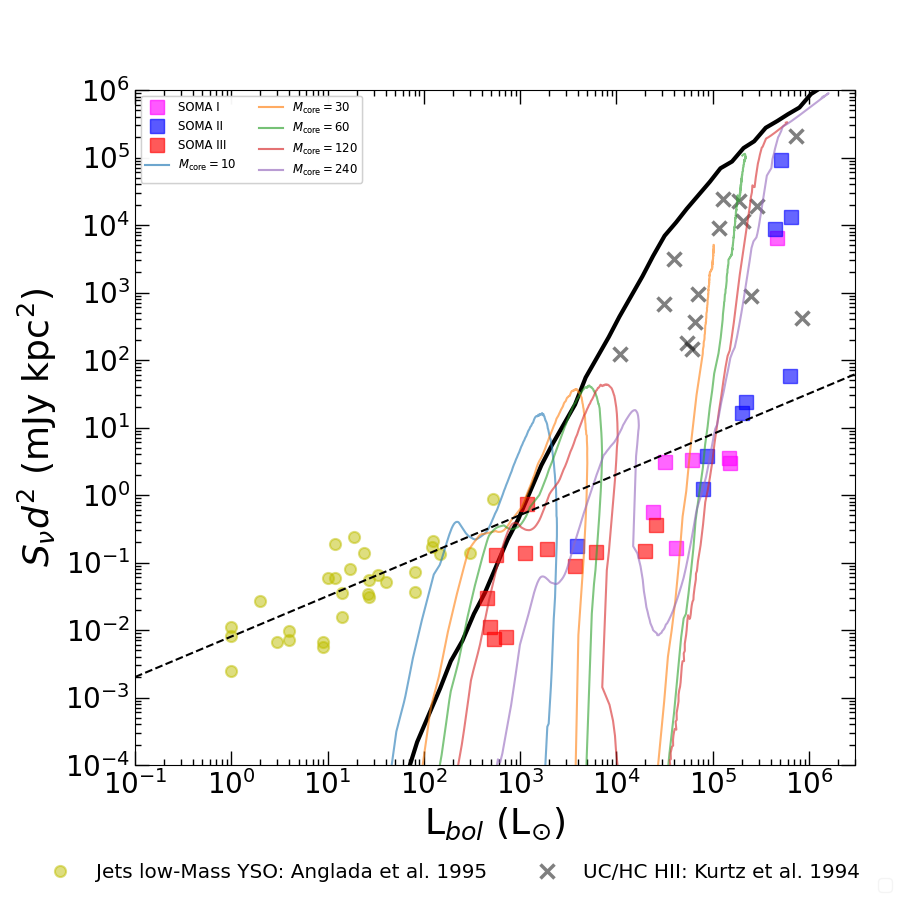}\\
\includegraphics[width=0.32\linewidth]{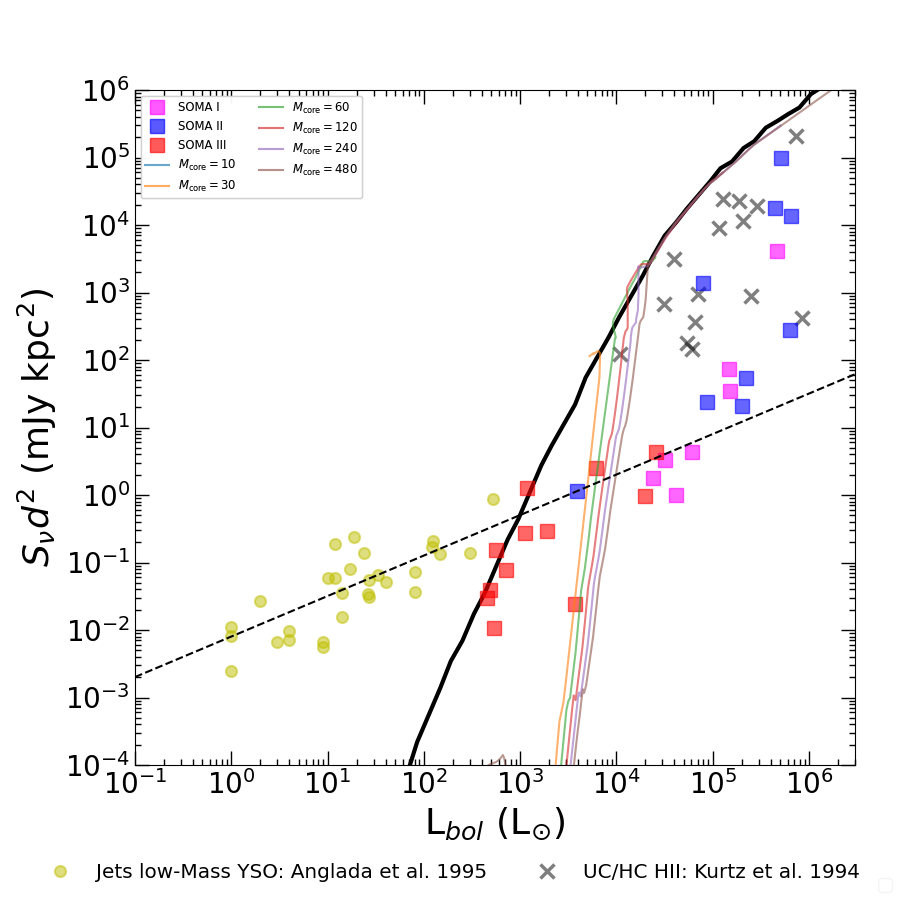}\quad\includegraphics[width=0.32\linewidth]{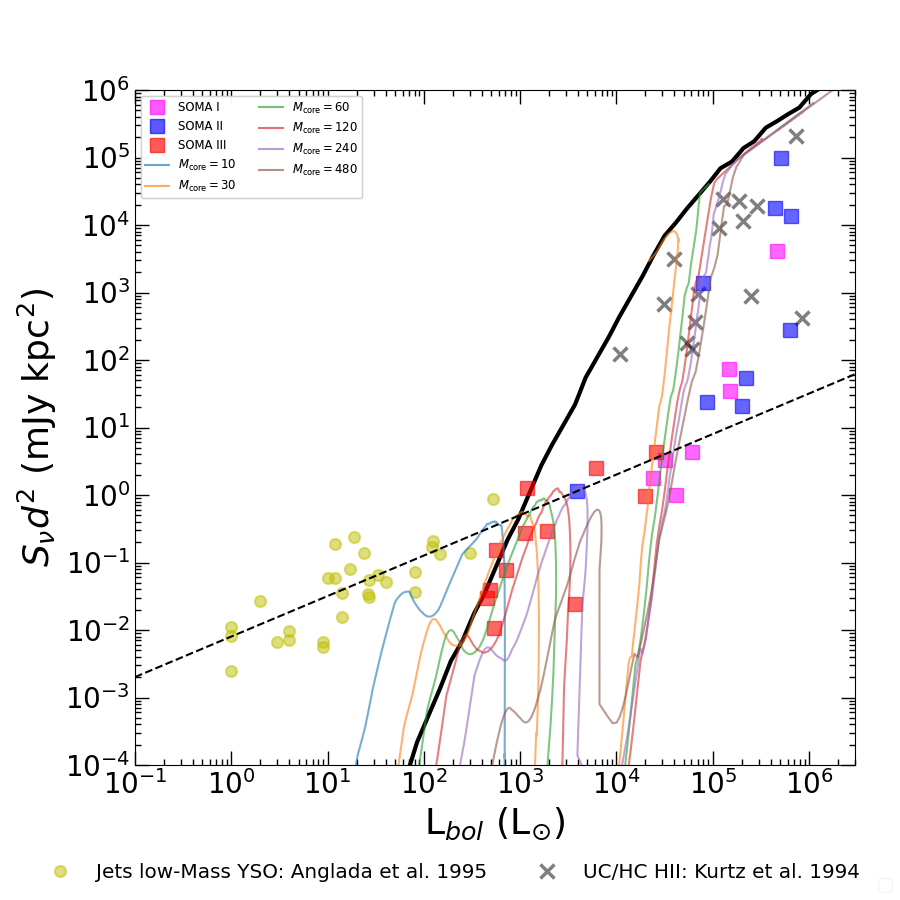}\quad\includegraphics[width=0.32\linewidth]{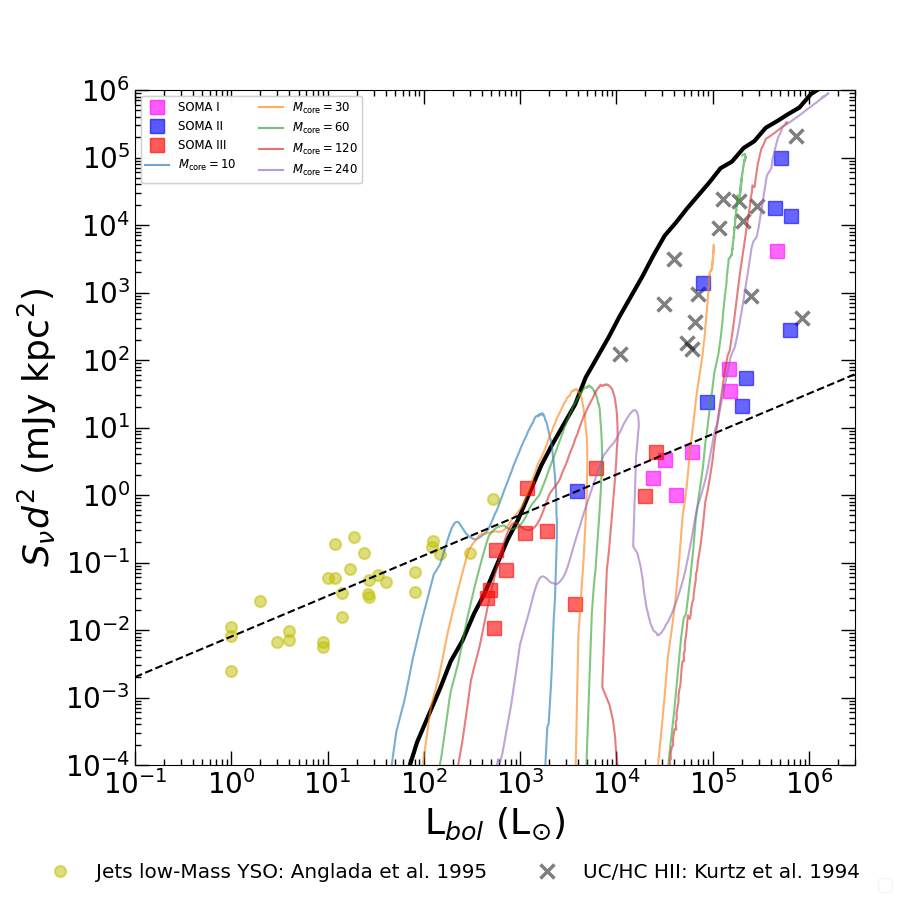}
\caption{As Fig.~\ref{fig:Anglada_plot}, but now showing protostellar evolutionary tracks from \citet{Zhang_2018}, under the assumption of optically thin radio emission. These models are for $M_{c} =$ 10, 30, 60, 120, 240 and 480 $M_{\odot}$ with $\Sigma_{cl} =$ 0.316 g cm$^{-2}$ (left), 1 g cm$^{-2}$ (center) and 3.16 g cm$^{-2}$ (right). Top plots are with the radio luminosity at the inner scale and bottom plots are with the radio luminosity at the SOMA scale.
\label{fig:YichenModels}}
\end{center}
\end{figure*}

\begin{figure*}[ht!]
\figurenum{C2}
\begin{center}
\includegraphics[width=0.32\linewidth]{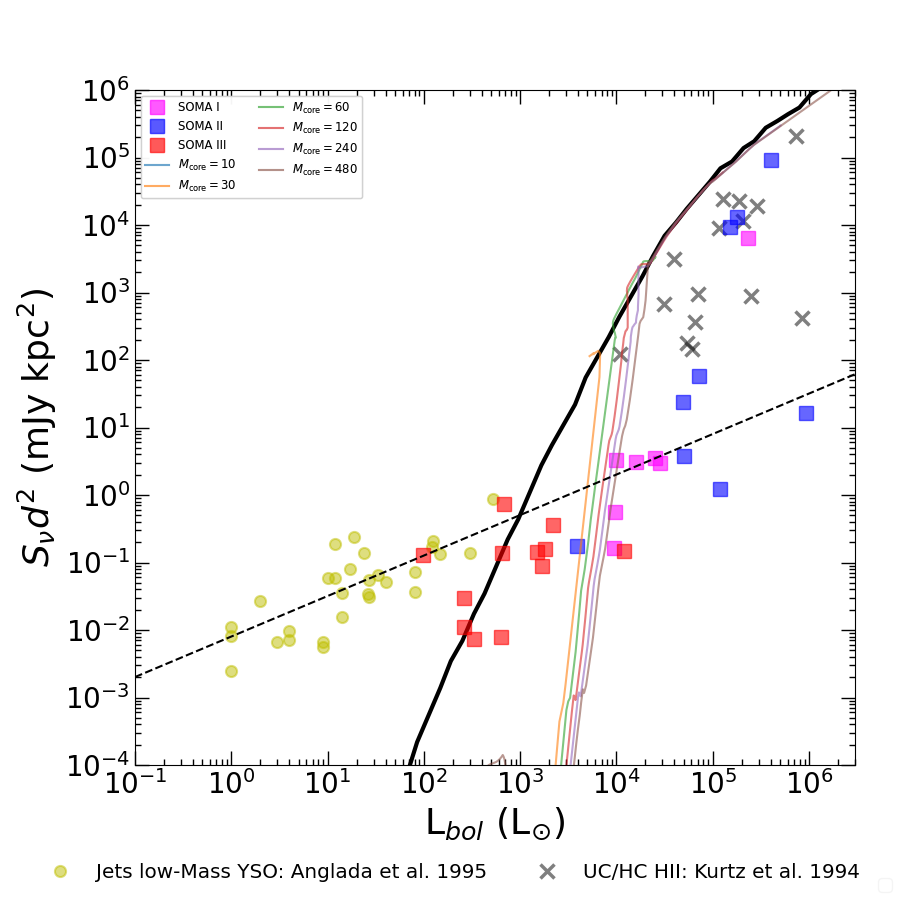}\quad\includegraphics[width=0.32\linewidth]{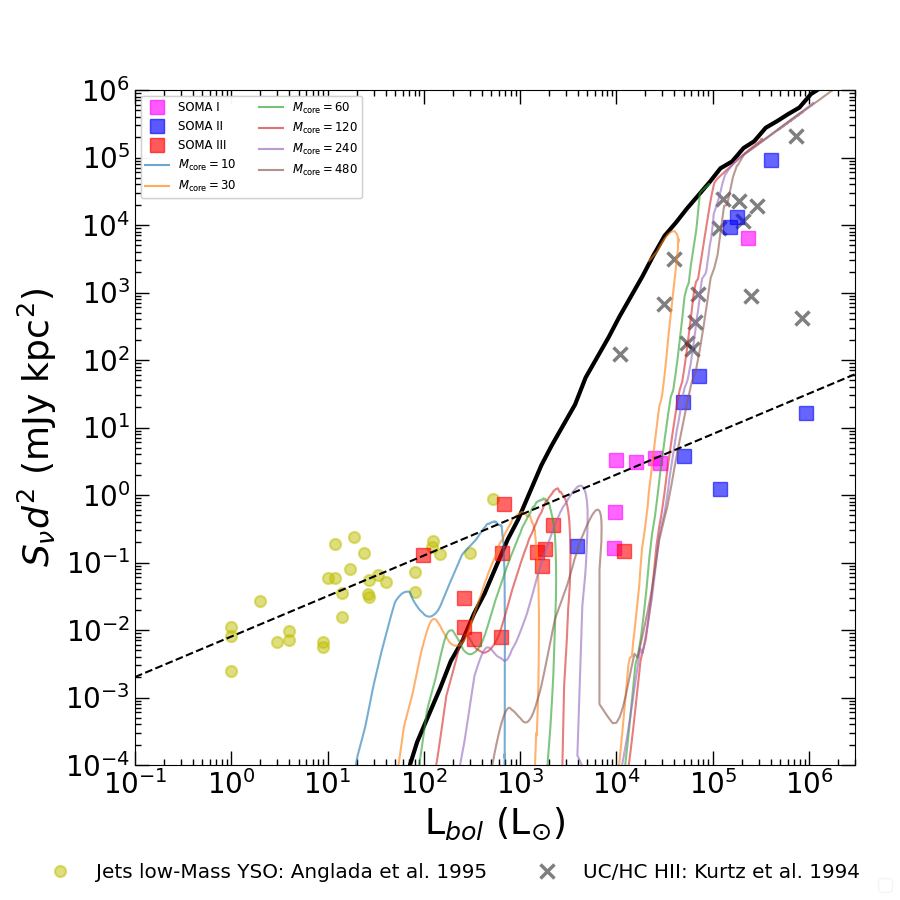}\quad\includegraphics[width=0.32\linewidth]{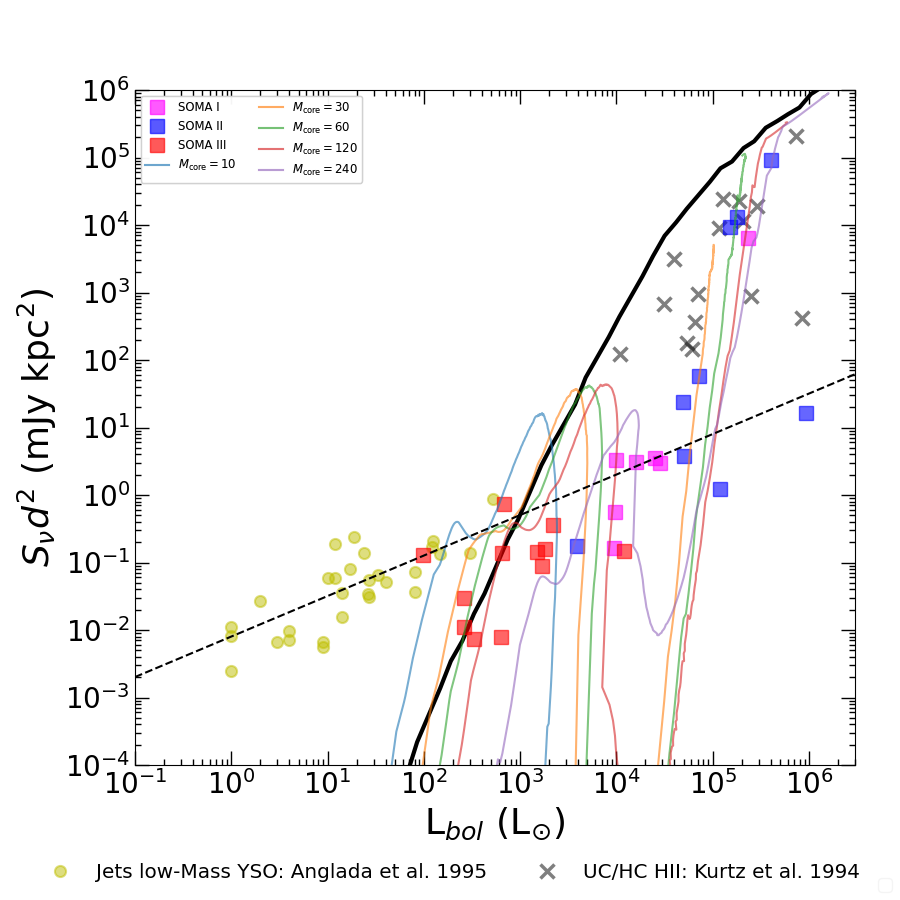}\\
\includegraphics[width=0.32\linewidth]{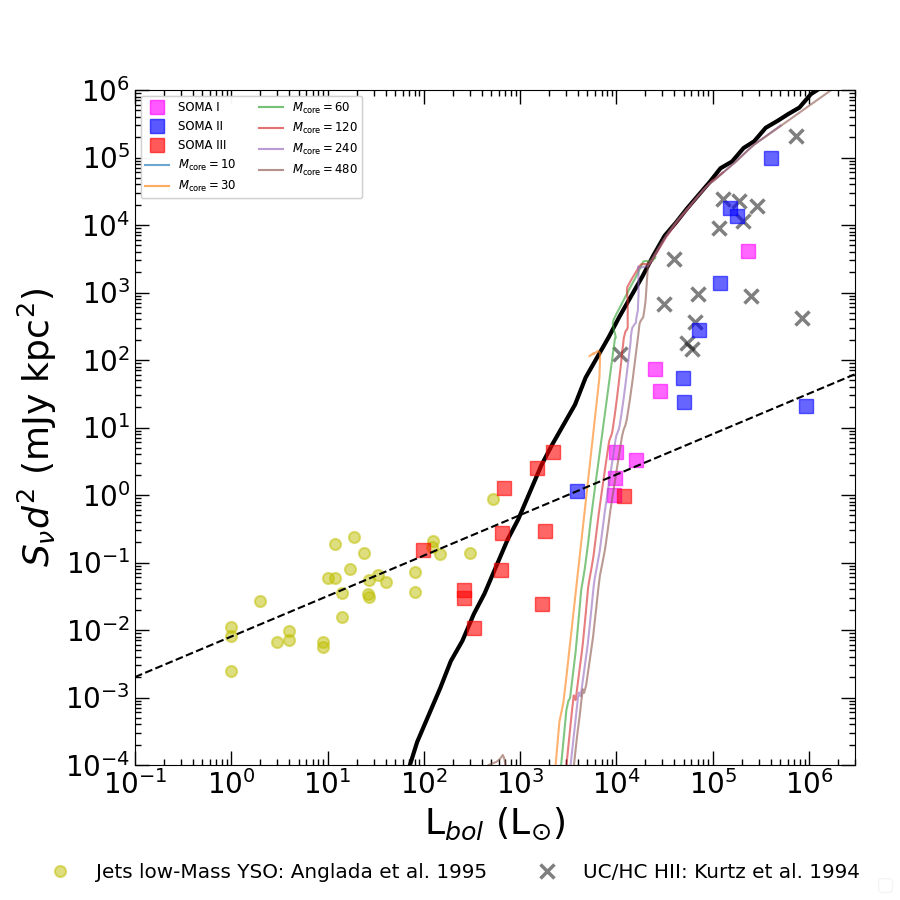}\quad\includegraphics[width=0.32\linewidth]{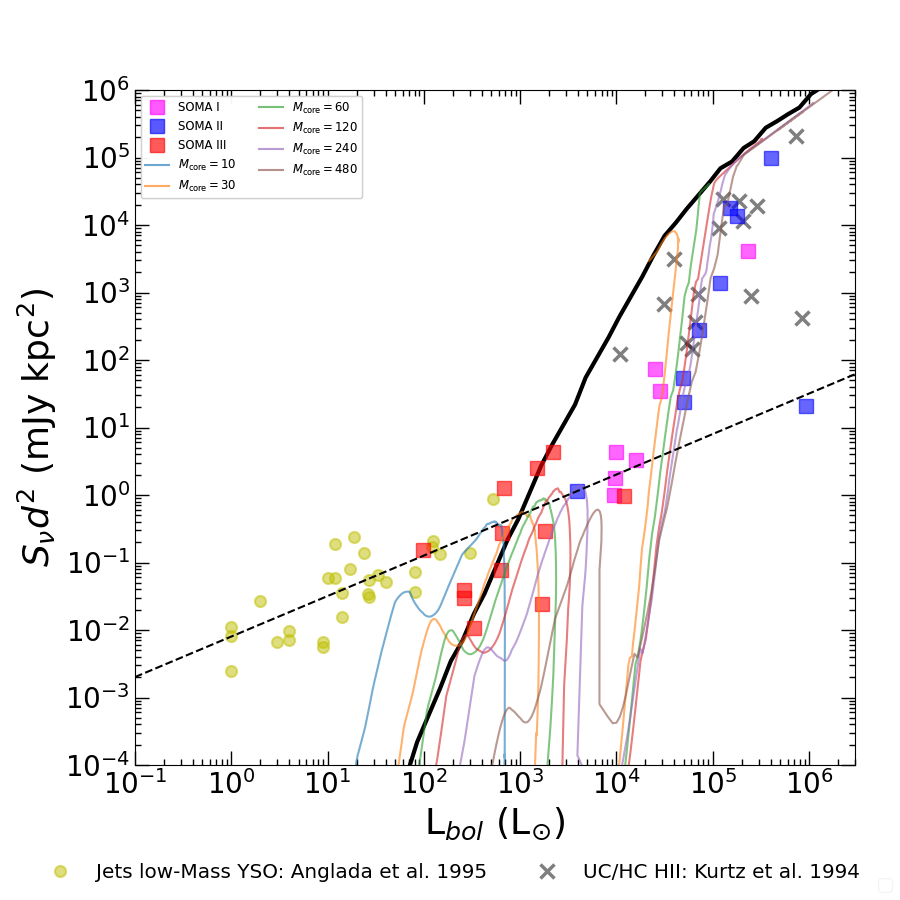}\quad\includegraphics[width=0.32\linewidth]{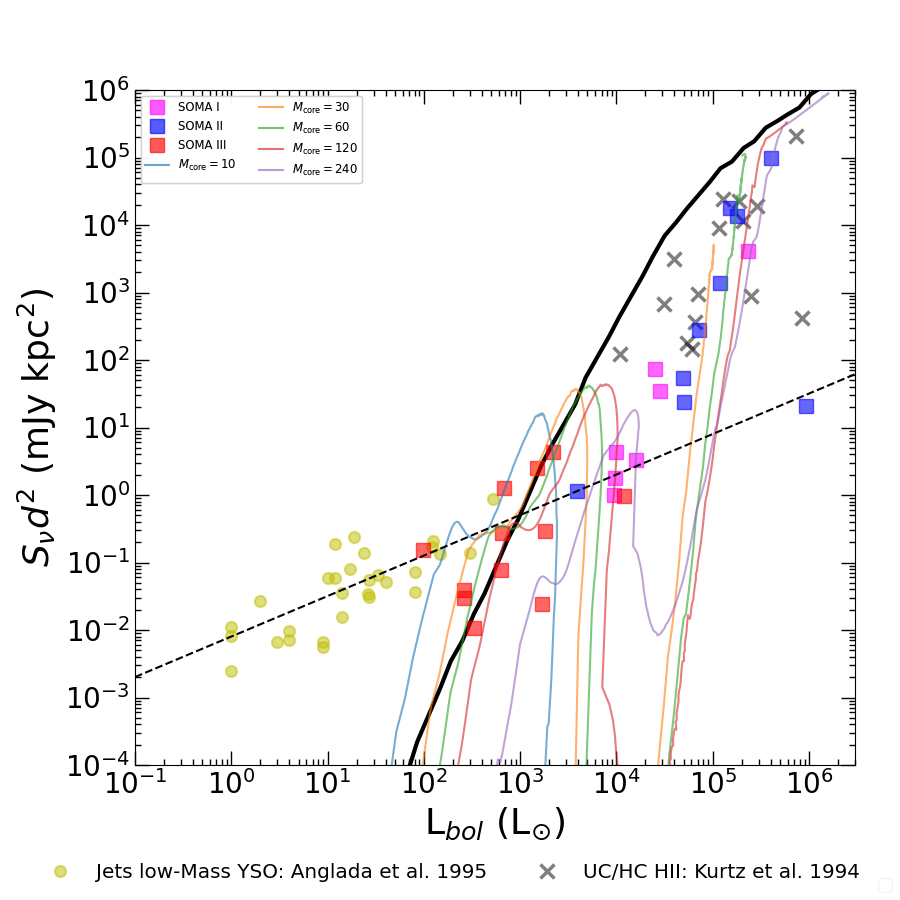}
\caption{As Fig.~\ref{fig:YichenModels}, but now with the isotropic bolometric luminosity. Top plots are with the radio luminosity at the inner scale and bottom plots are with the radio luminosity at the SOMA scale.
\label{fig:YichenModels2}}
\end{center}
\end{figure*}

Figure \ref{fig:YichenModels} makes a comparison of the SOMA Radio data at the inner (top) and SOMA (bottom) scale, with the full set of protostellar evolutionary tracks from \citet{Zhang_2018}, i.e., with $M_{c} =$ 10, 30, 60, 120, 240 and 480 $M_{\odot}$ with $\Sigma_{cl} =$ 0.316 g cm$^{-2}$ (left), 1 g cm$^{-2}$ (center) and 3.16 g cm$^{-2}$ (right). In these models the radio luminosities are not set from radiation transfer like the tracks from \citep{Tanaka_2016}, but instead assumes a simple spherical HII region. Meanwhile, Figure \ref{fig:YichenModels2} shows the same information but with the average of the good models of the isotropic bolometric luminosity (Table C.1 of \citetalias{Telkamp_2025}).

\clearpage
\bibliography{sample701}{}

@article{astropy:2013,
Adsurl = {http://adsabs.harvard.edu/abs/2013A%26A...558A..33A},
Archiveprefix = {arXiv},
Author = {{Astropy Collaboration} and {Robitaille}, T.~P. and {Tollerud}, E.~J. and {Greenfield}, P. and {Droettboom}, M. and {Bray}, E. and {Aldcroft}, T. and {Davis}, M. and {Ginsburg}, A. and {Price-Whelan}, A.~M. and {Kerzendorf}, W.~E. and {Conley}, A. and {Crighton}, N. and {Barbary}, K. and {Muna}, D. and {Ferguson}, H. and {Grollier}, F. and {Parikh}, M.~M. and {Nair}, P.~H. and {Unther}, H.~M. and {Deil}, C. and {Woillez}, J. and {Conseil}, S. and {Kramer}, R. and {Turner}, J.~E.~H. and {Singer}, L. and {Fox}, R. and {Weaver}, B.~A. and {Zabalza}, V. and {Edwards}, Z.~I. and {Azalee Bostroem}, K. and {Burke}, D.~J. and {Casey}, A.~R. and {Crawford}, S.~M. and {Dencheva}, N. and {Ely}, J. and {Jenness}, T. and {Labrie}, K. and {Lim}, P.~L. and {Pierfederici}, F. and {Pontzen}, A. and {Ptak}, A. and {Refsdal}, B. and {Servillat}, M. and {Streicher}, O.},
Doi = {10.1051/0004-6361/201322068},
Eid = {A33},
Eprint = {1307.6212},
Journal = {\aap},
Month = oct,
Pages = {A33},
Primaryclass = {astro-ph.IM},
Title = {{Astropy: A community Python package for astronomy}},
Volume = 558,
Year = 2013}

@ARTICLE{astropy:2018,
       author = {{Astropy Collaboration} and {Price-Whelan}, A.~M. and
         {Sip{\H{o}}cz}, B.~M. and {G{\"u}nther}, H.~M. and {Lim}, P.~L. and
         {Crawford}, S.~M. and {Conseil}, S. and {Shupe}, D.~L. and
         {Craig}, M.~W. and {Dencheva}, N. and {Ginsburg}, A. and {Vand
        erPlas}, J.~T. and {Bradley}, L.~D. and {P{\'e}rez-Su{\'a}rez}, D. and
         {de Val-Borro}, M. and {Aldcroft}, T.~L. and {Cruz}, K.~L. and
         {Robitaille}, T.~P. and {Tollerud}, E.~J. and {Ardelean}, C. and
         {Babej}, T. and {Bach}, Y.~P. and {Bachetti}, M. and {Bakanov}, A.~V. and
         {Bamford}, S.~P. and {Barentsen}, G. and {Barmby}, P. and
         {Baumbach}, A. and {Berry}, K.~L. and {Biscani}, F. and {Boquien}, M. and
         {Bostroem}, K.~A. and {Bouma}, L.~G. and {Brammer}, G.~B. and
         {Bray}, E.~M. and {Breytenbach}, H. and {Buddelmeijer}, H. and
         {Burke}, D.~J. and {Calderone}, G. and {Cano Rodr{\'\i}guez}, J.~L. and
         {Cara}, M. and {Cardoso}, J.~V.~M. and {Cheedella}, S. and {Copin}, Y. and
         {Corrales}, L. and {Crichton}, D. and {D'Avella}, D. and {Deil}, C. and
         {Depagne}, {\'E}. and {Dietrich}, J.~P. and {Donath}, A. and
         {Droettboom}, M. and {Earl}, N. and {Erben}, T. and {Fabbro}, S. and
         {Ferreira}, L.~A. and {Finethy}, T. and {Fox}, R.~T. and
         {Garrison}, L.~H. and {Gibbons}, S.~L.~J. and {Goldstein}, D.~A. and
         {Gommers}, R. and {Greco}, J.~P. and {Greenfield}, P. and
         {Groener}, A.~M. and {Grollier}, F. and {Hagen}, A. and {Hirst}, P. and
         {Homeier}, D. and {Horton}, A.~J. and {Hosseinzadeh}, G. and {Hu}, L. and
         {Hunkeler}, J.~S. and {Ivezi{\'c}}, {\v{Z}}. and {Jain}, A. and
         {Jenness}, T. and {Kanarek}, G. and {Kendrew}, S. and {Kern}, N.~S. and
         {Kerzendorf}, W.~E. and {Khvalko}, A. and {King}, J. and {Kirkby}, D. and
         {Kulkarni}, A.~M. and {Kumar}, A. and {Lee}, A. and {Lenz}, D. and
         {Littlefair}, S.~P. and {Ma}, Z. and {Macleod}, D.~M. and
         {Mastropietro}, M. and {McCully}, C. and {Montagnac}, S. and
         {Morris}, B.~M. and {Mueller}, M. and {Mumford}, S.~J. and {Muna}, D. and
         {Murphy}, N.~A. and {Nelson}, S. and {Nguyen}, G.~H. and
         {Ninan}, J.~P. and {N{\"o}the}, M. and {Ogaz}, S. and {Oh}, S. and
         {Parejko}, J.~K. and {Parley}, N. and {Pascual}, S. and {Patil}, R. and
         {Patil}, A.~A. and {Plunkett}, A.~L. and {Prochaska}, J.~X. and
         {Rastogi}, T. and {Reddy Janga}, V. and {Sabater}, J. and
         {Sakurikar}, P. and {Seifert}, M. and {Sherbert}, L.~E. and
         {Sherwood-Taylor}, H. and {Shih}, A.~Y. and {Sick}, J. and
         {Silbiger}, M.~T. and {Singanamalla}, S. and {Singer}, L.~P. and
         {Sladen}, P.~H. and {Sooley}, K.~A. and {Sornarajah}, S. and
         {Streicher}, O. and {Teuben}, P. and {Thomas}, S.~W. and
         {Tremblay}, G.~R. and {Turner}, J.~E.~H. and {Terr{\'o}n}, V. and
         {van Kerkwijk}, M.~H. and {de la Vega}, A. and {Watkins}, L.~L. and
         {Weaver}, B.~A. and {Whitmore}, J.~B. and {Woillez}, J. and
         {Zabalza}, V. and {Astropy Contributors}},
        title = "{The Astropy Project: Building an Open-science Project and Status of the v2.0 Core Package}",
      journal = {\aj},
         year = 2018,
        month = sep,
       volume = {156},
       number = {3},
          eid = {123},
        pages = {123},
          doi = {10.3847/1538-3881/aabc4f},
archivePrefix = {arXiv},
       eprint = {1801.02634},
 primaryClass = {astro-ph.IM},
       adsurl = {https://ui.adsabs.harvard.edu/abs/2018AJ....156..123A}
}

@ARTICLE{astropy:2022,
       author = {{Astropy Collaboration} and {Price-Whelan}, Adrian M. and {Lim}, Pey Lian and {Earl}, Nicholas and {Starkman}, Nathaniel and {Bradley}, Larry and {Shupe}, David L. and {Patil}, Aarya A. and {Corrales}, Lia and {Brasseur}, C.~E. and {N{"o}the}, Maximilian and {Donath}, Axel and {Tollerud}, Erik and {Morris}, Brett M. and {Ginsburg}, Adam and {Vaher}, Eero and {Weaver}, Benjamin A. and {Tocknell}, James and {Jamieson}, William and {van Kerkwijk}, Marten H. and {Robitaille}, Thomas P. and {Merry}, Bruce and {Bachetti}, Matteo and {G{"u}nther}, H. Moritz and {Aldcroft}, Thomas L. and {Alvarado-Montes}, Jaime A. and {Archibald}, Anne M. and {B{'o}di}, Attila and {Bapat}, Shreyas and {Barentsen}, Geert and {Baz{'a}n}, Juanjo and {Biswas}, Manish and {Boquien}, M{'e}d{'e}ric and {Burke}, D.~J. and {Cara}, Daria and {Cara}, Mihai and {Conroy}, Kyle E. and {Conseil}, Simon and {Craig}, Matthew W. and {Cross}, Robert M. and {Cruz}, Kelle L. and {D'Eugenio}, Francesco and {Dencheva}, Nadia and {Devillepoix}, Hadrien A.~R. and {Dietrich}, J{"o}rg P. and {Eigenbrot}, Arthur Davis and {Erben}, Thomas and {Ferreira}, Leonardo and {Foreman-Mackey}, Daniel and {Fox}, Ryan and {Freij}, Nabil and {Garg}, Suyog and {Geda}, Robel and {Glattly}, Lauren and {Gondhalekar}, Yash and {Gordon}, Karl D. and {Grant}, David and {Greenfield}, Perry and {Groener}, Austen M. and {Guest}, Steve and {Gurovich}, Sebastian and {Handberg}, Rasmus and {Hart}, Akeem and {Hatfield-Dodds}, Zac and {Homeier}, Derek and {Hosseinzadeh}, Griffin and {Jenness}, Tim and {Jones}, Craig K. and {Joseph}, Prajwel and {Kalmbach}, J. Bryce and {Karamehmetoglu}, Emir and {Ka{l}uszy{'n}ski}, Miko{l}aj and {Kelley}, Michael S.~P. and {Kern}, Nicholas and {Kerzendorf}, Wolfgang E. and {Koch}, Eric W. and {Kulumani}, Shankar and {Lee}, Antony and {Ly}, Chun and {Ma}, Zhiyuan and {MacBride}, Conor and {Maljaars}, Jakob M. and {Muna}, Demitri and {Murphy}, N.~A. and {Norman}, Henrik and {O'Steen}, Richard and {Oman}, Kyle A. and {Pacifici}, Camilla and {Pascual}, Sergio and {Pascual-Granado}, J. and {Patil}, Rohit R. and {Perren}, Gabriel I. and {Pickering}, Timothy E. and {Rastogi}, Tanuj and {Roulston}, Benjamin R. and {Ryan}, Daniel F. and {Rykoff}, Eli S. and {Sabater}, Jose and {Sakurikar}, Parikshit and {Salgado}, Jes{'u}s and {Sanghi}, Aniket and {Saunders}, Nicholas and {Savchenko}, Volodymyr and {Schwardt}, Ludwig and {Seifert-Eckert}, Michael and {Shih}, Albert Y. and {Jain}, Anany Shrey and {Shukla}, Gyanendra and {Sick}, Jonathan and {Simpson}, Chris and {Singanamalla}, Sudheesh and {Singer}, Leo P. and {Singhal}, Jaladh and {Sinha}, Manodeep and {Sip{H{o}}cz}, Brigitta M. and {Spitler}, Lee R. and {Stansby}, David and {Streicher}, Ole and {{{S}}umak}, Jani and {Swinbank}, John D. and {Taranu}, Dan S. and {Tewary}, Nikita and {Tremblay}, Grant R. and {Val-Borro}, Miguel de and {Van Kooten}, Samuel J. and {Vasovi{'c}}, Zlatan and {Verma}, Shresth and {de Miranda Cardoso}, Jos{'e} Vin{'i}cius and {Williams}, Peter K.~G. and {Wilson}, Tom J. and {Winkel}, Benjamin and {Wood-Vasey}, W.~M. and {Xue}, Rui and {Yoachim}, Peter and {Zhang}, Chen and {Zonca}, Andrea and {Astropy Project Contributors}},
        title = "{The Astropy Project: Sustaining and Growing a Community-oriented Open-source Project and the Latest Major Release (v5.0) of the Core Package}",
      journal = {\apj},
         year = 2022,
        month = aug,
       volume = {935},
       number = {2},
          eid = {167},
        pages = {167},
          doi = {10.3847/1538-4357/ac7c74},
archivePrefix = {arXiv},
       eprint = {2206.14220},
 primaryClass = {astro-ph.IM},
       adsurl = {https://ui.adsabs.harvard.edu/abs/2022ApJ...935..167A}
}

@ARTICLE{2022PASP..134k4501C,
       author = {{CASA Team} and {Bean}, Ben and {Bhatnagar}, Sanjay and {Castro}, Sandra and {Donovan Meyer}, Jennifer and {Emonts}, Bjorn and {Garcia}, Enrique and {Garwood}, Robert and {Golap}, Kumar and {Gonzalez Villalba}, Justo and {Harris}, Pamela and {Hayashi}, Yohei and {Hoskins}, Josh and {Hsieh}, Mingyu and {Jagannathan}, Preshanth and {Kawasaki}, Wataru and {Keimpema}, Aard and {Kettenis}, Mark and {Lopez}, Jorge and {Marvil}, Joshua and {Masters}, Joseph and {McNichols}, Andrew and {Mehringer}, David and {Miel}, Renaud and {Moellenbrock}, George and {Montesino}, Federico and {Nakazato}, Takeshi and {Ott}, Juergen and {Petry}, Dirk and {Pokorny}, Martin and {Raba}, Ryan and {Rau}, Urvashi and {Schiebel}, Darrell and {Schweighart}, Neal and {Sekhar}, Srikrishna and {Shimada}, Kazuhiko and {Small}, Des and {Steeb}, Jan-Willem and {Sugimoto}, Kanako and {Suoranta}, Ville and {Tsutsumi}, Takahiro and {van Bemmel}, Ilse M. and {Verkouter}, Marjolein and {Wells}, Akeem and {Xiong}, Wei and {Szomoru}, Arpad and {Griffith}, Morgan and {Glendenning}, Brian and {Kern}, Jeff},
        title = "{CASA, the Common Astronomy Software Applications for Radio Astronomy}",
      journal = {\pasp},
         year = 2022,
        month = nov,
       volume = {134},
       number = {1041},
          eid = {114501},
        pages = {114501},
          doi = {10.1088/1538-3873/ac9642},
archivePrefix = {arXiv},
       eprint = {2210.02276},
 primaryClass = {astro-ph.IM},
       adsurl = {https://ui.adsabs.harvard.edu/abs/2022PASP..134k4501C}
}

@misc{aplpy2012,
      author        = {{Robitaille}, T. and {Bressert}, E.},
      title         = "{APLpy: Astronomical Plotting Library in Python}",
      howpublished  = {Astrophysics Source Code Library},
      year          = 2012,
      month         = aug,
      archivePrefix = "ascl",
      eprint        = {1208.017},
      adsurl        = {http://adsabs.harvard.edu/abs/2012ascl.soft08017R}
}

@misc{aplpy2019,
      author       = {Robitaille, Thomas},
      title        = {{APLpy v2.0: The Astronomical Plotting Library in Python}},
      month        = feb,
      year         = 2019,
      doi          = {10.5281/zenodo.2567476},
      url          = {https://doi.org/10.5281/zenodo.2567476}
}

@ARTICLE{2015Ap&SS.355..283B,
       author = {{Beltr{\'a}n}, Maria T.},
        title = "{Observational perspective of the youngest phases of intermediate-mass stars}",
      journal = {\apss},
         year = 2015,
        month = feb,
       volume = {355},
       number = {2},
        pages = {283-290},
          doi = {10.1007/s10509-014-2151-0},
       adsurl = {https://ui.adsabs.harvard.edu/abs/2015Ap&SS.355..283B}
}

@ARTICLE{2019ApJ...880...99R,
       author = {{Rosero}, V. and {Hofner}, P. and {Kurtz}, S. and {Cesaroni}, R. and {Carrasco-Gonz{\'a}lez}, C. and {Araya}, E.~D. and {Rodr{\'\i}guez}, L.~F. and {Menten}, K.~M. and {Wyrowski}, F. and {Loinard}, L. and {Ellingsen}, S.~P. and {Molinari}, S.},
        title = "{Weak and Compact Radio Emission in Early High-mass Star-forming Regions. II. The Nature of the Radio Sources}",
      journal = {\apj},
         year = 2019,
        month = aug,
       volume = {880},
       number = {2},
          eid = {99},
        pages = {99},
          doi = {10.3847/1538-4357/ab2595},
archivePrefix = {arXiv},
       eprint = {1905.12089},
 primaryClass = {astro-ph.GA},
       adsurl = {https://ui.adsabs.harvard.edu/abs/2019ApJ...880...99R}
}

@ARTICLE{Bruizer_2017,
       author = {{De Buizer}, James M. and {Liu}, Mengyao and {Tan}, Jonathan C. and {Zhang}, Yichen and {Beltr{\'a}n}, Maria T. and {Shuping}, Ralph and {Staff}, Jan E. and {Tanaka}, Kei E.~I. and {Whitney}, Barbara},
        title = "{The SOFIA Massive (SOMA) Star Formation Survey. I. Overview and First Results}",
      journal = {\apj},
         year = 2017,
        month = jul,
       volume = {843},
       number = {1},
          eid = {33},
        pages = {33},
          doi = {10.3847/1538-4357/aa74c8},
archivePrefix = {arXiv},
       eprint = {1610.05373},
 primaryClass = {astro-ph.GA},
       adsurl = {https://ui.adsabs.harvard.edu/abs/2017ApJ...843...33D}
}

@ARTICLE{Tanaka_2016,
       author = {{Tanaka}, Kei E.~I. and {Tan}, Jonathan C. and {Zhang}, Yichen},
        title = "{Outflow-confined HII Regions. I. First Signposts of Massive Star Formation}",
      journal = {\apj},
         year = 2016,
        month = feb,
       volume = {818},
       number = {1},
          eid = {52},
        pages = {52},
          doi = {10.3847/0004-637X/818/1/52},
archivePrefix = {arXiv},
       eprint = {1509.06754},
 primaryClass = {astro-ph.SR},
       adsurl = {https://ui.adsabs.harvard.edu/abs/2016ApJ...818...52T}
}

@ARTICLE{Zhang_2018,
       author = {{Zhang}, Yichen and {Tan}, Jonathan C.},
        title = "{Radiation Transfer of Models of Massive Star Formation. IV. The Model Grid and Spectral Energy Distribution Fitting}",
      journal = {\apj},
         year = 2018,
        month = jan,
       volume = {853},
       number = {1},
          eid = {18},
        pages = {18},
          doi = {10.3847/1538-4357/aaa24a},
archivePrefix = {arXiv},
       eprint = {1708.08853},
 primaryClass = {astro-ph.GA},
       adsurl = {https://ui.adsabs.harvard.edu/abs/2018ApJ...853...18Z}
}

@ARTICLE{Rosero_2019,
       author = {{Rosero}, V. and {Tanaka}, K.~E.~I. and {Tan}, J.~C. and {Marvil}, J. and {Liu}, M. and {Zhang}, Y. and {De Buizer}, J.~M. and {Beltr{\'a}n}, M.~T.},
        title = "{The SOMA Radio Survey. I. Comprehensive SEDs of High-mass Protostars from Infrared to Radio and the Emergence of Ionization Feedback}",
      journal = {\apj},
         year = 2019,
        month = mar,
       volume = {873},
       number = {1},
          eid = {20},
        pages = {20},
          doi = {10.3847/1538-4357/ab0209},
archivePrefix = {arXiv},
       eprint = {1809.01264},
 primaryClass = {astro-ph.SR},
       adsurl = {https://ui.adsabs.harvard.edu/abs/2019ApJ...873...20R}
}

@ARTICLE{Liu_2019,
       author = {{Liu}, Mengyao and {Tan}, Jonathan C. and {De Buizer}, James M. and {Zhang}, Yichen and {Beltr{\'a}n}, Maria T. and {Staff}, Jan E. and {Tanaka}, Kei E.~I. and {Whitney}, Barbara and {Rosero}, Viviana},
        title = "{The SOFIA Massive (SOMA) Star Formation Survey. II. High Luminosity Protostars}",
      journal = {\apj},
         year = 2019,
        month = mar,
       volume = {874},
       number = {1},
          eid = {16},
        pages = {16},
          doi = {10.3847/1538-4357/ab07b7},
archivePrefix = {arXiv},
       eprint = {1901.01958},
 primaryClass = {astro-ph.GA},
       adsurl = {https://ui.adsabs.harvard.edu/abs/2019ApJ...874...16L}
}

@ARTICLE{Liu_2020,
       author = {{Liu}, Mengyao and {Tan}, Jonathan C. and {De Buizer}, James M. and {Zhang}, Yichen and {Moser}, Emily and {Beltr{\'a}n}, Maria T. and {Staff}, Jan E. and {Tanaka}, Kei E.~I. and {Whitney}, Barbara and {Rosero}, Viviana and {Yang}, Yao-Lun and {Fedriani}, Rub{\'e}n},
        title = "{The SOFIA Massive (SOMA) Star Formation Survey. III. From Intermediate- to High-mass Protostars}",
      journal = {\apj},
         year = 2020,
        month = nov,
       volume = {904},
       number = {1},
          eid = {75},
        pages = {75},
          doi = {10.3847/1538-4357/abbefb},
archivePrefix = {arXiv},
       eprint = {2006.06424},
 primaryClass = {astro-ph.GA},
       adsurl = {https://ui.adsabs.harvard.edu/abs/2020ApJ...904...75L}
}

@ARTICLE{Fedriani_2023,
       author = {{Fedriani}, Rub{\'e}n and {Tan}, Jonathan C. and {Telkamp}, Zoie and {Zhang}, Yichen and {Yang}, Yao-Lun and {Liu}, Mengyao and {De Buizer}, James M. and {Law}, Chi-Yan and {Beltran}, Maria T. and {Rosero}, Viviana and {Tanaka}, Kei E.~I. and {Cosentino}, Giuliana and {Gorai}, Prasanta and {Farias}, Juan and {Staff}, Jan E. and {Whitney}, Barbara},
        title = "{The SOFIA Massive (SOMA) Star Formation Survey. IV. Isolated Protostars}",
      journal = {\apj},
         year = 2023,
        month = jan,
       volume = {942},
       number = {1},
          eid = {7},
        pages = {7},
          doi = {10.3847/1538-4357/aca4cf},
archivePrefix = {arXiv},
       eprint = {2205.11422},
 primaryClass = {astro-ph.GA},
       adsurl = {https://ui.adsabs.harvard.edu/abs/2023ApJ...942....7F}
}

@article{Telkamp_2025,
doi = {10.3847/1538-4357/adcd79},
url = {https://dx.doi.org/10.3847/1538-4357/adcd79},
year = {2025},
month = {jun},
publisher = {The American Astronomical Society},
volume = {986},
number = {1},
pages = {15},
author = {Telkamp, Zoie and Fedriani, Rubén and Tan, Jonathan C. and Law, Chi-Yan and Zhang, Yichen and Plunkett, Adele and Crowe, Samuel and Yang, Yao-Lun and De Buizer, James M. and Beltran, Maria T. and Bonfand, Mélisse and Boyden, Ryan and Cosentino, Giuliana and Gorai, Prasanta and Liu, Mengyao and Rosero, Viviana and Taniguchi, Kotomi and Tanaka, Kei E. I. and Rodríguez, Tatiana M.},
title = {The SOFIA Massive (SOMA) Star Formation Survey. V. Clustered Protostars},
journal = {The Astrophysical Journal},
}

@ARTICLE{Sequeira-Murillo_2025,
       author = {{Sequeira-Murillo}, Francisco and {Rosero}, Viviana and {Marvil}, Joshua and {Tan}, Jonathan C. and {Fedriani}, Ruben and {Zhang}, Yichen and {Robinson}, Azia and {Gorai}, Prasanta and {Tanaka}, Kei E.~I. and {De Buizer}, James M. and et al.},
        title = "{The SOFIA Massive (SOMA) Radio Survey. II. Radio Emission from High-luminosity Protostars}",
      journal = {\apj},
         year = 2026,
        month = jan,
       volume = {996},
       number = {2},
          eid = {119},
        pages = {119},
          doi = {10.3847/1538-4357/ae10bb},
archivePrefix = {arXiv},
       eprint = {2507.16775},
 primaryClass = {astro-ph.SR},
       adsurl = {https://ui.adsabs.harvard.edu/abs/2026ApJ...996..119S}
}

@ARTICLE{2024ApJ...967..145G,
       author = {{Gardiner}, Emiko C. and {Tan}, Jonathan C. and {Staff}, Jan E. and {Ramsey}, Jon P. and {Zhang}, Yichen and {Tanaka}, Kei E.~I.},
        title = "{Disk Wind Feedback from High-mass Protostars. IV. Shock-ionized Jets}",
      journal = {\apj},
         year = 2024,
        month = jun,
       volume = {967},
       number = {2},
          eid = {145},
        pages = {145},
          doi = {10.3847/1538-4357/ad39e1},
archivePrefix = {arXiv},
       eprint = {2309.03887},
 primaryClass = {astro-ph.GA},
       adsurl = {https://ui.adsabs.harvard.edu/abs/2024ApJ...967..145G}
}

@ARTICLE{2016ApJS..227...25R,
       author = {{Rosero}, V. and {Hofner}, P. and {Claussen}, M. and {Kurtz}, S. and {Cesaroni}, R. and {Araya}, E.~D. and {Carrasco-Gonz{\'a}lez}, C. and {Rodr{\'\i}guez}, L.~F. and {Menten}, K.~M. and {Wyrowski}, F. and {Loinard}, L. and {Ellingsen}, S.~P.},
        title = "{Weak and Compact Radio Emission in Early High-mass Star-forming Regions. I. VLA Observations}",
      journal = {\apjs},
         year = 2016,
        month = dec,
       volume = {227},
       number = {2},
          eid = {25},
        pages = {25},
          doi = {10.3847/1538-4365/227/2/25},
archivePrefix = {arXiv},
       eprint = {1609.03269},
 primaryClass = {astro-ph.GA},
       adsurl = {https://ui.adsabs.harvard.edu/abs/2016ApJS..227...25R}
}

@PHDTHESIS{1995PhDT.......238B,
       author = {{Briggs}, Daniel Shenon},
        title = "{High fidelity deconvolution of moderately resolved sources}",
       school = {New Mexico Institute of Mining and Technology},
         year = 1995,
        month = jan,
       adsurl = {https://ui.adsabs.harvard.edu/abs/1995PhDT.......238B}
}

@ARTICLE{2018A&ARv..26....3A,
       author = {{Anglada}, Guillem and {Rodr{\'\i}guez}, Luis F. and {Carrasco-Gonz{\'a}lez}, Carlos},
        title = "{Radio jets from young stellar objects}",
      journal = {\aapr},
         year = 2018,
        month = jun,
       volume = {26},
       number = {1},
          eid = {3},
        pages = {3},
          doi = {10.1007/s00159-018-0107-z},
archivePrefix = {arXiv},
       eprint = {1806.06444},
 primaryClass = {astro-ph.SR},
       adsurl = {https://ui.adsabs.harvard.edu/abs/2018A&ARv..26....3A}
}

@ARTICLE{2013ApJ...762..120P,
       author = {{Palau}, Aina and {Fuente}, Asunci{\'o}n and {Girart}, Josep M. and {Estalella}, Robert and {Ho}, Paul T.~P. and {S{\'a}nchez-Monge}, {\'A}lvaro and {Fontani}, Francesco and {Busquet}, Gemma and {Commer{\c{c}}on}, Benoit and {Hennebelle}, Patrick and {Boissier}, J{\'e}r{\'e}mie and {Zhang}, Qizhou and {Cesaroni}, Riccardo and {Zapata}, Luis A.},
        title = "{Early Stages of Cluster Formation: Fragmentation of Massive Dense Cores down to <\raisebox{-0.5ex}\textasciitilde 1000 AU}",
      journal = {\apj},
         year = 2013,
        month = jan,
       volume = {762},
       number = {2},
          eid = {120},
        pages = {120},
          doi = {10.1088/0004-637X/762/2/120},
archivePrefix = {arXiv},
       eprint = {1211.2666},
 primaryClass = {astro-ph.GA},
       adsurl = {https://ui.adsabs.harvard.edu/abs/2013ApJ...762..120P}
}

@ARTICLE{Sanchez-Monge2010IRASCore,
       author = {{S{\'a}nchez-Monge}, {\'A}lvaro and {Palau}, Aina and {Estalella}, Robert and {Kurtz}, Stan and {Zhang}, Qizhou and {Di Francesco}, James and {Shepherd}, Debra},
        title = "{IRAS 22198+6336: Discovery of an Intermediate-mass Hot Core}",
      journal = {\apjl},
         year = 2010,
        month = oct,
       volume = {721},
       number = {2},
        pages = {L107-L111},
          doi = {10.1088/2041-8205/721/2/L107},
archivePrefix = {arXiv},
       eprint = {1007.5258},
 primaryClass = {astro-ph.GA},
       adsurl = {https://ui.adsabs.harvard.edu/abs/2010ApJ...721L.107S}
}

@ARTICLE{Sanchez-Monge2008SurveyWavelengths,
       author = {{S{\'a}nchez-Monge}, {\'A}. and {Palau}, Aina and {Estalella}, R. and {Beltr{\'a}n}, M.~T. and {Girart}, J.~M.},
        title = "{Survey of intermediate/high mass star-forming regions at centimeter and millimeter wavelengths}",
      journal = {\aap},
         year = 2008,
        month = jul,
       volume = {485},
       number = {2},
        pages = {497-515},
          doi = {10.1051/0004-6361:20078406},
archivePrefix = {arXiv},
       eprint = {0802.3132},
 primaryClass = {astro-ph},
       adsurl = {https://ui.adsabs.harvard.edu/abs/2008A&A...485..497S}
}

@ARTICLE{Hirota2008Astrometry1204G,
       author = {{Hirota}, Tomoya and {Ando}, Kazuma and {Bushimata}, Takeshi and {Choi}, Yoon Kyung and {Honma}, Mareki and {Imai}, Hiroshi and {Iwadate}, Kenzaburo and {Jike}, Takaaki and {Kameno}, Seiji and {Kameya}, Osamu and {Kamohara}, Ryuichi and {Kan-Ya}, Yukitoshi and {Kawaguchi}, Noriyuki and {Kijima}, Masachika and {Kim}, Mi Kyoung and {Kobayashi}, Hideyuki and {Kuji}, Seisuke and {Kurayama}, Tomoharu and {Manabe}, Seiji and {Matsui}, Makoto and {Matsumoto}, Naoko and {Miyaji}, Takeshi and {Miyazaki}, Atsushi and {Nagayama}, Takumi and {Nakagawa}, Akiharu and {Namikawa}, Daichi and {Nyu}, Daisuke and {Oh}, Chung Sik and {Omodaka}, Toshihiro and {Oyama}, Tomoaki and {Sakai}, Satoshi and {Sasao}, Tetsuo and {Sato}, Katsuhisa and {Sato}, Mayumi and {Shibata}, Katsunori M. and {Tamura}, Yoshiaki and {Ueda}, Kosuke and {Yamashita}, Kazuyoshi},
        title = "{Astrometry of H$_{2}$O Masers in Nearby Star-Forming Regions with VERA III. IRAS 22198+6336 in Lynds1204G}",
      journal = {\pasj},
         year = 2008,
        month = oct,
       volume = {60},
        pages = {961},
          doi = {10.1093/pasj/60.5.961},
archivePrefix = {arXiv},
       eprint = {0808.0546},
 primaryClass = {astro-ph},
       adsurl = {https://ui.adsabs.harvard.edu/abs/2008PASJ...60..961H}
}

@ARTICLE{Anglada1998SpectralSources,
       author = {{Anglada}, Guillem and {Villuendas}, Eva and {Estalella}, Robert and {Beltr{\'a}n}, Maria T. and {Rodr{\'\i}guez}, Luis F. and {Torrelles}, Jos{\'e} M. and {Curiel}, Salvador},
        title = "{Spectral Indices of Centimeter Continuum Sources in Star-forming Regions: Implications on the Nature of the Outflow Exciting Sources}",
      journal = {\aj},
         year = 1998,
        month = dec,
       volume = {116},
       number = {6},
        pages = {2953-2964},
          doi = {10.1086/300637},
       adsurl = {https://ui.adsabs.harvard.edu/abs/1998AJ....116.2953A}
}

@ARTICLE{Jenness1995Embedded,
       author = {{Jenness}, T. and {Scott}, P.~F. and {Padman}, R.},
        title = "{Studies of Embedded Far Infrared Sources in the Vicinity of H2O Masers - I. Observations}",
      journal = {\mnras},
         year = 1995,
        month = oct,
       volume = {276},
        pages = {1024},
          doi = {10.1093/mnras/276.3.1024},
       adsurl = {https://ui.adsabs.harvard.edu/abs/1995MNRAS.276.1024J}
}

@ARTICLE{Reynolds1986Continuum,
       author = {{Reynolds}, S.~P.},
        title = "{Continuum Spectra of Collimated, Ionized Stellar Winds}",
      journal = {\apj},
         year = 1986,
        month = may,
       volume = {304},
        pages = {713},
          doi = {10.1086/164209},
       adsurl = {https://ui.adsabs.harvard.edu/abs/1986ApJ...304..713R}
}

@ARTICLE{1960ApJS....4..337H,
       author = {{Herbig}, George H.},
        title = "{The Spectra of Be- and Ae-Type Stars Associated with Nebulosity}",
      journal = {\apjs},
         year = 1960,
        month = mar,
       volume = {4},
        pages = {337},
          doi = {10.1086/190050},
       adsurl = {https://ui.adsabs.harvard.edu/abs/1960ApJS....4..337H}
}

@ARTICLE{2009MNRAS.399..778B,
       author = {{Boley}, Paul A. and {Sobolev}, Andrey M. and {Krushinsky}, Vadim V. and {van Boekel}, Roy and {Henning}, Thomas and {Moiseev}, Aleksei V. and {Yushkin}, Maksim V.},
        title = "{S 235 B explained: an accreting Herbig Be star surrounded by reflection nebulosity}",
      journal = {\mnras},
         year = 2009,
        month = oct,
       volume = {399},
       number = {2},
        pages = {778-782},
          doi = {10.1111/j.1365-2966.2009.15308.x},
       adsurl = {https://ui.adsabs.harvard.edu/abs/2009MNRAS.399..778B}
}

@ARTICLE{1996ApJ...463..630H,
       author = {{Heyer}, Mark H. and {Carpenter}, John M. and {Ladd}, E.~F.},
        title = "{Giant Molecular Cloud Complexes with Optical H II Regions: 12CO and 13CO Observations and Global Cloud Properties}",
      journal = {\apj},
         year = 1996,
        month = jun,
       volume = {463},
        pages = {630},
          doi = {10.1086/177277},
       adsurl = {https://ui.adsabs.harvard.edu/abs/1996ApJ...463..630H}
}

@ARTICLE{2006A&A...453..911F,
       author = {{Felli}, M. and {Massi}, F. and {Robberto}, M. and {Cesaroni}, R.},
        title = "{New signposts of massive star formation in the S235A-B region}",
      journal = {\aap},
         year = 2006,
        month = jul,
       volume = {453},
       number = {3},
        pages = {911-922},
          doi = {10.1051/0004-6361:20054646},
archivePrefix = {arXiv},
       eprint = {astro-ph/0603818},
 primaryClass = {astro-ph},
       adsurl = {https://ui.adsabs.harvard.edu/abs/2006A&A...453..911F}
}

@ARTICLE{2016ApJ...819...66D,
       author = {{Dewangan}, L.~K. and {Ojha}, D.~K. and {Luna}, A. and {Anandarao}, B.~G. and {Ninan}, J.~P. and {Mallick}, K.~K. and {Mayya}, Y.~D.},
        title = "{A Multi-wavelength Study of Star Formation Activity in the S235 Complex}",
      journal = {\apj},
         year = 2016,
        month = mar,
       volume = {819},
       number = {1},
          eid = {66},
        pages = {66},
          doi = {10.3847/0004-637X/819/1/66},
archivePrefix = {arXiv},
       eprint = {1601.04488},
 primaryClass = {astro-ph.GA},
       adsurl = {https://ui.adsabs.harvard.edu/abs/2016ApJ...819...66D}
}

@ARTICLE{1986PASJ...38..531N,
       author = {{Nakano}, Makoto and {Yoshida}, Shigeomi},
        title = "{Molecular line observations of the S235B region.}",
      journal = {\pasj},
         year = 1986,
        month = jan,
       volume = {38},
        pages = {531-545},
       adsurl = {https://ui.adsabs.harvard.edu/abs/1986PASJ...38..531N}
}

@ARTICLE{2004Fontani,
       author = {{Fontani}, F. and {Cesaroni}, R. and {Testi}, L. and {Molinari}, S. and {Zhang}, Q. and {Brand}, J. and {Walmsley}, C.~M.},
        title = "{Nature of two massive protostellar candidates:  IRAS 21307+5049 and IRAS 22172+5549}",
      journal = {\aap},
         year = 2004,
        month = sep,
       volume = {424},
        pages = {179-195},
          doi = {10.1051/0004-6361:20035848},
archivePrefix = {arXiv},
       eprint = {astro-ph/0406023},
 primaryClass = {astro-ph},
       adsurl = {https://ui.adsabs.harvard.edu/abs/2004A&A...424..179F}
}

@ARTICLE{2021Purser,
       author = {{Purser}, S.~J.~D. and {Lumsden}, S.~L. and {Hoare}, M.~G. and {Kurtz}, S.},
        title = "{A Galactic survey of radio jets from massive protostars}",
      journal = {\mnras},
         year = 2021,
        month = jun,
       volume = {504},
       number = {1},
        pages = {338-355},
          doi = {10.1093/mnras/stab747},
archivePrefix = {arXiv},
       eprint = {2103.08990},
 primaryClass = {astro-ph.GA},
       adsurl = {https://ui.adsabs.harvard.edu/abs/2021MNRAS.504..338P}
}

@ARTICLE{2002Molinari,
       author = {{Molinari}, Sergio and {Testi}, Leonardo and {Rodr{\'\i}guez}, Luis F. and {Zhang}, Qizhou},
        title = "{The Formation of Massive Stars. I. High-Resolution Millimeter and Radio Studies of High-Mass Protostellar Candidates}",
      journal = {\apj},
         year = 2002,
        month = may,
       volume = {570},
       number = {2},
        pages = {758-778},
          doi = {10.1086/339630},
       adsurl = {https://ui.adsabs.harvard.edu/abs/2002ApJ...570..758M}
}

@ARTICLE{1977ApJ...218..736S,
       author = {{Sargent}, A.~I.},
        title = "{Molecular clouds and star formation. I. Observations of the Cepheus OB3 molecular cloud.}",
      journal = {\apj},
         year = 1977,
        month = dec,
       volume = {218},
        pages = {736-748},
          doi = {10.1086/155729},
       adsurl = {https://ui.adsabs.harvard.edu/abs/1977ApJ...218..736S}
}

@ARTICLE{1996A&A...313L..17L,
       author = {{Lefloch}, B. and {Eisloeffel}, J. and {Lazareff}, B.},
        title = "{The remarkable Class 0 source CEP E}",
      journal = {\aap},
         year = 1996,
        month = sep,
       volume = {313},
        pages = {L17-L20},
       adsurl = {https://ui.adsabs.harvard.edu/abs/1996A&A...313L..17L}
}

@ARTICLE{2001ApJ...555..146M,
       author = {{Moro-Mart{\'\i}n}, Amaya and {Noriega-Crespo}, Alberto and {Molinari}, Sergio and {Testi}, Leonardo and {Cernicharo}, Jos{\'e} and {Sargent}, Anneila},
        title = "{Infrared and Millimetric Study of the Young Outflow Cepheus E}",
      journal = {\apj},
         year = 2001,
        month = jul,
       volume = {555},
       number = {1},
        pages = {146-159},
          doi = {10.1086/321443},
archivePrefix = {arXiv},
       eprint = {astro-ph/0103065},
 primaryClass = {astro-ph},
       adsurl = {https://ui.adsabs.harvard.edu/abs/2001ApJ...555..146M}
}

@ARTICLE{Gusdorf_CepE,
       author = {{Gusdorf}, A. and {Anderl}, S. and {Lefloch}, B. and {Leurini}, S. and {Wiesemeyer}, H. and {G{\"u}sten}, R. and {Benedettini}, M. and {Codella}, C. and {Godard}, B. and {G{\'o}mez-Ruiz}, A.~I. and {Jacobs}, K. and {Kristensen}, L.~E. and {Lesaffre}, P. and {Pineau des For{\^e}ts}, G. and {Lis}, D.~C.},
        title = "{Nature of shocks revealed by SOFIA OI observations in the Cepheus E protostellar outflow}",
      journal = {\aap},
         year = 2017,
        month = jun,
       volume = {602},
          eid = {A8},
        pages = {A8},
          doi = {10.1051/0004-6361/201730454},
archivePrefix = {arXiv},
       eprint = {1704.03796},
 primaryClass = {astro-ph.GA},
       adsurl = {https://ui.adsabs.harvard.edu/abs/2017A&A...602A...8G}
}

@ARTICLE{2018A&A...618A.145O,
       author = {{Ospina-Zamudio}, J. and {Lefloch}, B. and {Ceccarelli}, C. and {Kahane}, C. and {Favre}, C. and {L{\'o}pez-Sepulcre}, A. and {Montarges}, M.},
        title = "{First hot corino detected around an isolated intermediate-mass protostar: Cep E-mm}",
      journal = {\aap},
         year = 2018,
        month = oct,
       volume = {618},
          eid = {A145},
        pages = {A145},
          doi = {10.1051/0004-6361/201832857},
archivePrefix = {arXiv},
       eprint = {1807.11278},
 primaryClass = {astro-ph.SR},
       adsurl = {https://ui.adsabs.harvard.edu/abs/2018A&A...618A.145O}
}

@ARTICLE{2019MNRAS.490.2679O,
       author = {{Ospina-Zamudio}, J. and {Lefloch}, B. and {Favre}, C. and {L{\'o}pez-Sepulcre}, A. and {Bianchi}, E. and {Ceccarelli}, C. and {De Simone}, M. and {Bouvier}, M. and {Kahane}, C.},
        title = "{Molecules in the Cep E-mm jet: evidence for shock-driven photochemistry?}",
      journal = {\mnras},
         year = 2019,
        month = dec,
       volume = {490},
       number = {2},
        pages = {2679-2691},
          doi = {10.1093/mnras/stz2733},
archivePrefix = {arXiv},
       eprint = {1909.12002},
 primaryClass = {astro-ph.GA},
       adsurl = {https://ui.adsabs.harvard.edu/abs/2019MNRAS.490.2679O}
}

@ARTICLE{LeflochCepE1,
       author = {{Lefloch}, B. and {Cernicharo}, J. and {Pacheco}, S. and {Ceccarelli}, C.},
        title = "{Shocked water in the Cepheus E protostellar outflow}",
      journal = {\aap},
         year = 2011,
        month = mar,
       volume = {527},
          eid = {L3},
        pages = {L3},
          doi = {10.1051/0004-6361/201016247},
archivePrefix = {arXiv},
       eprint = {1101.2327},
 primaryClass = {astro-ph.GA},
       adsurl = {https://ui.adsabs.harvard.edu/abs/2011A&A...527L...3L}
}

@ARTICLE{LeflochCepE2,
       author = {{Lefloch}, B. and {Gusdorf}, A. and {Codella}, C. and {Eisl{\"o}ffel}, J. and {Neri}, R. and {G{\'o}mez-Ruiz}, A.~I. and {G{\"u}sten}, R. and {Leurini}, S. and {Risacher}, C. and {Benedettini}, M.},
        title = "{The structure of the Cepheus E protostellar outflow: The jet, the bowshock, and the cavity}",
      journal = {\aap},
         year = 2015,
        month = sep,
       volume = {581},
          eid = {A4},
        pages = {A4},
          doi = {10.1051/0004-6361/201425521},
       adsurl = {https://ui.adsabs.harvard.edu/abs/2015A&A...581A...4L}
}

@ARTICLE{2000AJ....120..909A,
       author = {{Ayala}, S. and {Noriega-Crespo}, A. and {Garnavich}, P.~M. and {Curiel}, S. and {Raga}, A.~C. and {B{\"o}hm}, K. -H. and {Raymond}, J.},
        title = "{Optical and Near-Infrared Study of the Cepheus E Outflow, A Very Low-Excitation Object}",
      journal = {\aj},
         year = 2000,
        month = aug,
       volume = {120},
       number = {2},
        pages = {909-919},
          doi = {10.1086/301500},
archivePrefix = {arXiv},
       eprint = {astro-ph/0004298},
 primaryClass = {astro-ph},
       adsurl = {https://ui.adsabs.harvard.edu/abs/2000AJ....120..909A}
}

@ARTICLE{Tobin2020,
       author = {{Tobin}, John J. and {Sheehan}, Patrick D. and {Megeath}, S. Thomas and {D{\'\i}az-Rodr{\'\i}guez}, Ana Karla and {Offner}, Stella S.~R. and {Murillo}, Nadia M. and {van 't Hoff}, Merel L.~R. and {van Dishoeck}, Ewine F. and {Osorio}, Mayra and {Anglada}, Guillem and {Furlan}, Elise and {Stutz}, Amelia M. and {Reynolds}, Nickalas and {Karnath}, Nicole and {Fischer}, William J. and {Persson}, Magnus and {Looney}, Leslie W. and {Li}, Zhi-Yun and {Stephens}, Ian and {Chandler}, Claire J. and {Cox}, Erin and {Dunham}, Michael M. and {Tychoniec}, {\L}ukasz and {Kama}, Mihkel and {Kratter}, Kaitlin and {Kounkel}, Marina and {Mazur}, Brian and {Maud}, Luke and {Patel}, Lisa and {Perez}, Laura and {Sadavoy}, Sarah I. and {Segura-Cox}, Dominique and {Sharma}, Rajeeb and {Stephenson}, Brian and {Watson}, Dan M. and {Wyrowski}, Friedrich},
        title = "{The VLA/ALMA Nascent Disk and Multiplicity (VANDAM) Survey of Orion Protostars. II. A Statistical Characterization of Class 0 and Class I Protostellar Disks}",
      journal = {\apj},
         year = 2020,
        month = feb,
       volume = {890},
       number = {2},
          eid = {130},
        pages = {130},
          doi = {10.3847/1538-4357/ab6f64},
archivePrefix = {arXiv},
       eprint = {2001.04468},
 primaryClass = {astro-ph.GA},
       adsurl = {https://ui.adsabs.harvard.edu/abs/2020ApJ...890..130T}
}

@ARTICLE{Trinidad2009,
       author = {{Trinidad}, M.~A. and {Rodr{\'\i}guez}, T. and {Rodr{\'\i}guez}, L.~F.},
        title = "{Radio Jets and Disks in the Intermediate-Mass Star-Forming Region NGC2071IR}",
      journal = {\apj},
         year = 2009,
        month = nov,
       volume = {706},
       number = {1},
        pages = {244-251},
          doi = {10.1088/0004-637X/706/1/244},
       adsurl = {https://ui.adsabs.harvard.edu/abs/2009ApJ...706..244T}
}

@ARTICLE{Cheng2022,
       author = {{Cheng}, Yu and {Tobin}, John J. and {Yang}, Yao-Lun and {van't Hoff}, Merel L.~R. and {Sadavoy}, Sarah I. and {Osorio}, Mayra and {D{\'\i}az-Rodr{\'\i}guez}, Ana Karla and {Anglada}, Guillem and {Karnath}, Nicole and {Sheehan}, Patrick D. and {Li}, Zhi-Yun and {Reynolds}, Nickalas and {Murillo}, Nadia M. and {Zhang}, Yichen and {Megeath}, S. Thomas and {Tychoniec}, {\L}ukasz},
        title = "{Disks and Outflows in the Intermediate-mass Star-forming Region NGC 2071 IR}",
      journal = {\apj},
         year = 2022,
        month = jul,
       volume = {933},
       number = {2},
          eid = {178},
        pages = {178},
          doi = {10.3847/1538-4357/ac7464},
archivePrefix = {arXiv},
       eprint = {2205.15108},
 primaryClass = {astro-ph.GA},
       adsurl = {https://ui.adsabs.harvard.edu/abs/2022ApJ...933..178C}
}

@ARTICLE{Carrasco2012,
       author = {{Carrasco-Gonz{\'a}lez}, Carlos and {Osorio}, Mayra and {Anglada}, Guillem and {D'Alessio}, Paola and {Rodr{\'\i}guez}, Luis F. and {G{\'o}mez}, Jos{\'e} F. and {Torrelles}, Jos{\'e} M.},
        title = "{Multiplicity, Disks, and Jets in the NGC 2071 Star-forming Region}",
      journal = {\apj},
         year = 2012,
        month = feb,
       volume = {746},
       number = {1},
          eid = {71},
        pages = {71},
          doi = {10.1088/0004-637X/746/1/71},
archivePrefix = {arXiv},
       eprint = {1111.5469},
 primaryClass = {astro-ph.GA},
       adsurl = {https://ui.adsabs.harvard.edu/abs/2012ApJ...746...71C}
}

@ARTICLE{1982ApJ...261..558B,
       author = {{Bally}, J.},
        title = "{Energetic activity in a star-forming molecular cloud core: a disk constrained bipolar outflow in NGC 2071.}",
      journal = {\apj},
         year = 1982,
        month = oct,
       volume = {261},
        pages = {558-568},
          doi = {10.1086/160366},
       adsurl = {https://ui.adsabs.harvard.edu/abs/1982ApJ...261..558B}
}

@ARTICLE{1983ApJ...266L..61B,
       author = {{Bally}, J. and {Stark}, A.~A.},
        title = "{Atomic hydrogen associated with the high-velocity flow in NGC 2071.}",
      journal = {\apjl},
         year = 1983,
        month = mar,
       volume = {266},
        pages = {L61-L64},
          doi = {10.1086/183978},
       adsurl = {https://ui.adsabs.harvard.edu/abs/1983ApJ...266L..61B}
}

@ARTICLE{1992ApJ...398..139R,
       author = {{Ruiz}, A. and {Rodriguez}, L.~F. and {Canto}, J. and {Mirabel}, I.~F.},
        title = "{High-Velocity OH in Absorption: A New Tracer of Shocked Gas in Outflows?}",
      journal = {\apj},
         year = 1992,
        month = oct,
       volume = {398},
        pages = {139},
          doi = {10.1086/171843},
       adsurl = {https://ui.adsabs.harvard.edu/abs/1992ApJ...398..139R}
}

@ARTICLE{2000ApJ...545..861G,
       author = {{Garay}, Guido and {Mardones}, Diego and {Rodr{\'\i}guez}, L.~F.},
        title = "{Silicon Monoxide and Methanol Emission from the NGC 2071 Molecular Outflow}",
      journal = {\apj},
         year = 2000,
        month = dec,
       volume = {545},
       number = {2},
        pages = {861-873},
          doi = {10.1086/317853},
       adsurl = {https://ui.adsabs.harvard.edu/abs/2000ApJ...545..861G}
}

@ARTICLE{1998ApJ...505..756T,
       author = {{Torrelles}, Jos{\'e} M. and {G{\'o}mez}, Jos{\'e} F. and {Rodr{\'\i}guez}, Luis F. and {Curiel}, Salvador and {Anglada}, Guillem and {Ho}, Paul T.~P.},
        title = "{Radio Continuum-H$_{2}$O Maser Systems in NGC 2071: H$_{2}$O Masers Tracing a Jet (IRS 1) and a Rotating Proto-planetary Disk of Radius 20 AU (IRS 3)}",
      journal = {\apj},
         year = 1998,
        month = oct,
       volume = {505},
       number = {2},
        pages = {756-765},
          doi = {10.1086/306205},
       adsurl = {https://ui.adsabs.harvard.edu/abs/1998ApJ...505..756T}
}

@ARTICLE{2010Rygl,
       author = {{Rygl}, K.~L.~J. and {Brunthaler}, A. and {Reid}, M.~J. and {Menten}, K.~M. and {van Langevelde}, H.~J. and {Xu}, Y.},
        title = "{Trigonometric parallaxes of 6.7 GHz methanol masers}",
      journal = {\aap},
         year = 2010,
        month = feb,
       volume = {511},
          eid = {A2},
        pages = {A2},
          doi = {10.1051/0004-6361/200913135},
archivePrefix = {arXiv},
       eprint = {0910.0150},
 primaryClass = {astro-ph.GA},
       adsurl = {https://ui.adsabs.harvard.edu/abs/2010A&A...511A...2R}
}

@ARTICLE{Beltran2006,
       author = {{Beltr{\'a}n}, M.~T. and {Girart}, J.~M. and {Estalella}, R.},
        title = "{Who is eating the outflow? High-angular resolution study of an intermediate-mass protostar in L1206}",
      journal = {\aap},
         year = 2006,
        month = oct,
       volume = {457},
       number = {3},
        pages = {865-876},
          doi = {10.1051/0004-6361:20065132},
archivePrefix = {arXiv},
       eprint = {astro-ph/0604028},
 primaryClass = {astro-ph},
       adsurl = {https://ui.adsabs.harvard.edu/abs/2006A&A...457..865B}
}

@ARTICLE{Surcis2013II,
       author = {{Surcis}, G. and {Vlemmings}, W.~H.~T. and {van Langevelde}, H.~J. and {Hutawarakorn Kramer}, B. and {Quiroga-Nu{\~n}ez}, L.~H.},
        title = "{EVN observations of 6.7 GHz methanol maser polarization in massive star-forming regions. II. First statistical results}",
      journal = {\aap},
         year = 2013,
        month = aug,
       volume = {556},
          eid = {A73},
        pages = {A73},
          doi = {10.1051/0004-6361/201321501},
archivePrefix = {arXiv},
       eprint = {1306.6335},
 primaryClass = {astro-ph.SR},
       adsurl = {https://ui.adsabs.harvard.edu/abs/2013A&A...556A..73S}
}

@ARTICLE{1991AJ....102.1398R,
       author = {{Ressler}, Michael E. and {Shure}, Mark},
        title = "{Two protostar candidates in the bright-rimmed dark cloud LDN 1206.}",
      journal = {\aj},
         year = 1991,
        month = oct,
       volume = {102},
        pages = {1398},
          doi = {10.1086/115965},
       adsurl = {https://ui.adsabs.harvard.edu/abs/1991AJ....102.1398R}
}

@ARTICLE{2000AJ....119..323S,
       author = {{Sugitani}, K. and {Matsuo}, H. and {Nakano}, M. and {Tamura}, M. and {Ogura}, K.},
        title = "{2 Millimeter Observations of Bright-rimmed Clouds with IRAS Point Sources}",
      journal = {\aj},
         year = 2000,
        month = jan,
       volume = {119},
       number = {1},
        pages = {323-334},
          doi = {10.1086/301164},
       adsurl = {https://ui.adsabs.harvard.edu/abs/2000AJ....119..323S}
}

@ARTICLE{2021ApJS..253...15L,
       author = {{Liu}, De-Jian and {Xu}, Ye and {Li}, Ying-Jie and {Zheng}, Sheng and {Lu}, Deng-Rong and {Hao}, Chao-Jie and {Lin}, Ze-Hao and {Bian}, Shuai-Bo and {Liu}, Li-Ming},
        title = "{High-sensitivity Millimeter Imaging of Molecular Outflows in Nine Nearby High-mass Star-forming Regions}",
      journal = {\apjs},
         year = 2021,
        month = mar,
       volume = {253},
       number = {1},
          eid = {15},
        pages = {15},
          doi = {10.3847/1538-4365/abcece},
archivePrefix = {arXiv},
       eprint = {2012.03226},
 primaryClass = {astro-ph.GA},
       adsurl = {https://ui.adsabs.harvard.edu/abs/2021ApJS..253...15L}
}

@ARTICLE{1993AJ....106..250W,
       author = {{Wilking}, Bruce and {Mundy}, Lee and {McMullin}, Joseph and {Hezel}, Thomas and {Keene}, Jocelyn},
        title = "{IRAS 21391+5802: A Study in Intermediate Mass Star Formation}",
      journal = {\aj},
         year = 1993,
        month = jul,
       volume = {106},
        pages = {250},
          doi = {10.1086/116633},
       adsurl = {https://ui.adsabs.harvard.edu/abs/1993AJ....106..250W}
}

@ARTICLE{1979A&A....75..345M,
       author = {{Matthews}, H.~I.},
        title = "{High resolution radio observations of bright rims in IC 1396.}",
      journal = {\aap},
         year = 1979,
        month = jun,
       volume = {75},
        pages = {345-350},
       adsurl = {https://ui.adsabs.harvard.edu/abs/1979A&A....75..345M}
}

@ARTICLE{2001A&A...376..271C,
       author = {{Codella}, C. and {Bachiller}, R. and {Nisini}, B. and {Saraceno}, P. and {Testi}, L.},
        title = "{Star formation in the bright rimmed globule IC 1396N}",
      journal = {\aap},
         year = 2001,
        month = sep,
       volume = {376},
        pages = {271-287},
          doi = {10.1051/0004-6361:20010963},
       adsurl = {https://ui.adsabs.harvard.edu/abs/2001A&A...376..271C}
}

@ARTICLE{2005A&A...443..535V,
       author = {{Valdettaro}, R. and {Palla}, F. and {Brand}, J. and {Cesaroni}, R.},
        title = "{H\{2\}O maser emission from bright rimmed clouds in the northern hemisphere}",
      journal = {\aap},
         year = 2005,
        month = nov,
       volume = {443},
       number = {2},
        pages = {535-540},
          doi = {10.1051/0004-6361:20053731},
archivePrefix = {arXiv},
       eprint = {astro-ph/0508446},
 primaryClass = {astro-ph},
       adsurl = {https://ui.adsabs.harvard.edu/abs/2005A&A...443..535V}
}

@ARTICLE{2010ApJ...717.1067C,
       author = {{Choudhury}, Rumpa and {Mookerjea}, Bhaswati and {Bhatt}, H.~C.},
        title = "{Triggered Star Formation and Young Stellar Population in Bright-rimmed Cloud SFO 38}",
      journal = {\apj},
         year = 2010,
        month = jul,
       volume = {717},
       number = {2},
        pages = {1067-1083},
          doi = {10.1088/0004-637X/717/2/1067},
archivePrefix = {arXiv},
       eprint = {1005.1841},
 primaryClass = {astro-ph.GA},
       adsurl = {https://ui.adsabs.harvard.edu/abs/2010ApJ...717.1067C}
}

@ARTICLE{Beltran2002,
       author = {{Beltr{\'a}n}, Maria T. and {Girart}, Jos{\'e} M. and {Estalella}, Robert and {Ho}, Paul T.~P. and {Palau}, Aina},
        title = "{IRAS 21391+5802: The Molecular Outflow and Its Exciting Source}",
      journal = {\apj},
         year = 2002,
        month = jul,
       volume = {573},
       number = {1},
        pages = {246-259},
          doi = {10.1086/340592},
archivePrefix = {arXiv},
       eprint = {astro-ph/0203206},
 primaryClass = {astro-ph},
       adsurl = {https://ui.adsabs.harvard.edu/abs/2002ApJ...573..246B}
}

@ARTICLE{2007A&A...468L..33N,
       author = {{Neri}, R. and {Fuente}, A. and {Ceccarelli}, C. and {Caselli}, P. and {Johnstone}, D. and {van Dishoeck}, E.~F. and {Wyrowski}, F. and {Tafalla}, M. and {Lefloch}, B. and {Plume}, R.},
        title = "{The IC1396N proto-cluster at a scale of \raisebox{-0.5ex}\textasciitilde250 AU}",
      journal = {\aap},
         year = 2007,
        month = jun,
       volume = {468},
       number = {3},
        pages = {L33-L36},
          doi = {10.1051/0004-6361:20077320},
archivePrefix = {arXiv},
       eprint = {0705.2663},
 primaryClass = {astro-ph},
       adsurl = {https://ui.adsabs.harvard.edu/abs/2007A&A...468L..33N}
}

@INPROCEEDINGS{2015aska.confE.121A,
       author = {{Anglada}, G. and {Rodr{\'\i}guez}, L.~F. and {Carrasco-Gonzalez}, C.},
        title = "{Radio Jets in Young Stellar Objects with the SKA}",
    booktitle = {Advancing Astrophysics with the Square Kilometre Array (AASKA14)},
         year = 2015,
        month = apr,
          eid = {121},
        pages = {121},
          doi = {10.22323/1.215.0121},
archivePrefix = {arXiv},
       eprint = {1412.6409},
 primaryClass = {astro-ph.SR},
       adsurl = {https://ui.adsabs.harvard.edu/abs/2015aska.confE.121A}
}

@INPROCEEDINGS{1995RMxAC...1...67A,
       author = {{Anglada}, G.},
        title = "{Centimeter Continuum Emission from Outflow Sources}",
    booktitle = {Revista Mexicana de Astronomia y Astrofisica Conference Series},
         year = 1995,
       editor = {{Lizano}, S. and {Torrelles}, J.~M.},
       series = {Revista Mexicana de Astronomia y Astrofisica Conference Series},
       volume = {1},
        month = apr,
        pages = {67},
       adsurl = {https://ui.adsabs.harvard.edu/abs/1995RMxAC...1...67A}
}

@ARTICLE{1994ApJS...91..659K,
       author = {{Kurtz}, S. and {Churchwell}, E. and {Wood}, D.~O.~S.},
        title = "{Ultracompact H II Regions. II. New High-Resolution Radio Images}",
      journal = {\apjs},
         year = 1994,
        month = apr,
       volume = {91},
        pages = {659},
          doi = {10.1086/191952},
       adsurl = {https://ui.adsabs.harvard.edu/abs/1994ApJS...91..659K}
}

@ARTICLE{1984ApJ...283..165T,
       author = {{Thompson}, R.~I.},
        title = "{Lyman and Balmer continuum ionization in zero-age main-sequence stars: applications to the line excess phenomenon.}",
      journal = {\apj},
         year = 1984,
        month = aug,
       volume = {283},
        pages = {165-168},
          doi = {10.1086/162287},
       adsurl = {https://ui.adsabs.harvard.edu/abs/1984ApJ...283..165T}
}

@ARTICLE{1973AJ.....78..929P,
       author = {{Panagia}, Nino},
        title = "{Some Physical parameters of early-type stars}",
      journal = {\aj},
         year = 1973,
        month = nov,
       volume = {78},
        pages = {929-934},
          doi = {10.1086/111498},
       adsurl = {https://ui.adsabs.harvard.edu/abs/1973AJ.....78..929P}
}

@ARTICLE{Panagia1978,
  author  = {Panagia, N. and Walmsley, C.~M.},
  title   = {Radio Source Angular Diameters},
  journal = {Astronomy and Astrophysics},
  year    = {1978},
  month   = {nov},
  volume  = {70},
  pages   = {411},
  note    = {\url{https://ui.adsabs.harvard.edu/abs/1978A&A....70..411P}}
}

@ARTICLE{2013ApJS..208....9P,
       author = {{Pecaut}, Mark J. and {Mamajek}, Eric E.},
        title = "{Intrinsic Colors, Temperatures, and Bolometric Corrections of Pre-main-sequence Stars}",
      journal = {\apjs},
         year = 2013,
        month = sep,
       volume = {208},
       number = {1},
          eid = {9},
        pages = {9},
          doi = {10.1088/0067-0049/208/1/9},
archivePrefix = {arXiv},
       eprint = {1307.2657},
 primaryClass = {astro-ph.SR},
       adsurl = {https://ui.adsabs.harvard.edu/abs/2013ApJS..208....9P}
}

@ARTICLE{2021A&A...645A..29K,
       author = {{Kavak}, {\"U}. and {S{\'a}nchez-Monge}, {\'A}. and {L{\'o}pez-Sepulcre}, A. and {Cesaroni}, R. and {van der Tak}, F.~F.~S. and {Moscadelli}, L. and {Beltr{\'a}n}, M.~T. and {Schilke}, P.},
        title = "{Search for radio jets from massive young stellar objects. Association of radio jets with H$_{2}$O and CH$_{3}$OH masers}",
      journal = {\aap},
         year = 2021,
        month = jan,
       volume = {645},
          eid = {A29},
        pages = {A29},
          doi = {10.1051/0004-6361/202037652},
archivePrefix = {arXiv},
       eprint = {2011.14729},
 primaryClass = {astro-ph.GA},
       adsurl = {https://ui.adsabs.harvard.edu/abs/2021A&A...645A..29K}
}
\bibliographystyle{aasjournalv7}

\end{document}